# Input and Horizon Dependent Structural Fidelity in Chunk Coupled Spiking Networks

Xinjun Li
NEXT BRAIN LTD

## Abstract

A fixed structural edit can preserve a spike sequence under one input and change it under another. We study this dependence using statistics-preserving edge swaps, controlled inputs and finite-horizon parameter certificates. The original 125-chunk numerical model is tested on ten artificial graphs from three connection rules. Despite identical total external drive at every slot, global and focal inputs change the visibility of 4–13 of nineteen fixed swaps per graph. A paired homogeneous and heterogeneous leaky integrate-and-fire validation in brian2-atlas reuses three originals and extends observation from 30 to 300 ms under a constant background drive sufficient for isolated-cell firing. Heterogeneous short-window switching persists, but by 300 ms every tested swap differs from its reference, saturating the binary visibility readout. Fixed-support weight families provide a complementary regional guarantee. Positive certified intervals cover 4,308 of 4,320 LIF direction conditions and include distinct floating-point configurations. A post hoc comparison finds a long-window failure inside the short-window certified interval in 710 of 720 heterogeneous conditions; 394 of 829 direction pairs ordered under both matched-drive inputs change order at 300 ms. Input-specific filtering and threshold-margin scores retain useful ordering but degrade with horizon and do not consistently beat activity controls. Earlier temporal-error, paired-dose and unsuccessful structural-selection results delimit what these findings imply. Together, the evidence links input-conditioned structural effects to finite-horizon fidelity regions in two deterministic kernels, without establishing universal topology rankings, model compression or asymptotic robustness.

## 1 Introduction

Which changes of a network's endpoints matter for its trajectory after node-specific connection statistics are preserved? A structural intervention need not change the effective input under every source activity pattern. Even when inputs change, thresholded spike outputs can remain equal, and temporal or population aggregation can conceal differences that are present in fine trajectories. These distinctions make the input protocol and observation rule part of a fidelity claim.

Efficient neural simulation has long used the consolidation of convergent additive inputs [1,2], with communication and accumulation organized by channel and delay [3]. We use this principle as an execution control in a scheduler where arrivals can activate dormant chunks. Embedded spiking networks and population descriptions address the further task of reducing represented dynamics [4–7], while brian2-atlas describes execution contracts for heterogeneous simulation [8]. Our control retains adjacency and unit states, allowing subsequent comparisons to change the graph without conflating them with a different input-delivery rule.

Structural effects on neural dynamics are well established. Motif analysis [9], network downscaling [10], equitable partitions and cluster synchronization [11,12], degree-preserving rewiring [13], and structure–activity and structure–correlation studies [14–16] describe complementary aspects of this relationship. Our question concerns cross-run trajectories of the same identified chunks under fixed node-specific degree and strength summaries. It does not infer a new structural principle from the fact that rewiring changes activity.

The temporal resolution of a spike comparison is also substantive. Metric-space and filtered-spike approaches explicitly parameterize sensitivity to timing and cumulative count differences [17,18]. We therefore examine both exact slot counts and readouts that retain different amounts of temporal and node-identity information. Triangle inequalities guarantee that certain aggregations cannot increase L1 discrepancy; they do not determine how much of a particular intervention's discrepancy will disappear.

We distinguish endpoint replacement from fixed-support weight perturbation. Statistics-matched rewiring and single legal double-edge swaps test which identified spike effects depend on input; reference-activity graph selection [19] provides a negative control on inferring closed-loop fidelity from an open-loop score. Temporal-error composition, paired doses and initial-state controls characterize the observed discrepancies. The discovery input/readout analyses are post hoc and use seed 99 over 30 ms; prospectively fixed graph and LIF comparisons test the input effect beyond that discovery setting.

Sensitivity and finite-time spike preservation have established context. In vivo experiments demonstrate propagation of a single extra cortical spike [20]. Distance to threshold sets has been used to characterize spiking dynamics [21], prescribed spike trains can be expressed as parameter constraints [22], and learning precisely timed spikes links synaptic weights, threshold crossings and reset history [23]. Stability regions around trajectories have also been studied in recurrent spiking models [24]. These studies motivate the distinction between sensitivity and preservation; they do not establish the mechanism of the present deterministic networks. Our contribution concerns auditable region guarantees under a specified numerical execution contract and their input and horizon dependence, rather than the first observation that threshold margins or stable regions exist.

Neural-network verification provides related methods for guarantees over sets of perturbations. Symbolic interval propagation has been used to bound outputs of ReLU networks [25], while temporal encodings and SMT constraints have been used to verify SNN classification robustness under input-spike perturbations [26]. Our verification target is preservation of the complete labeled finite spike record in a recurrent simulation under fixed-support joint weight perturbations and a specified floating-point execution contract. The enclosure principle is established; the present study develops and audits its model-specific realization and examines the resulting guarantees across inputs and observation horizons.

The article addresses two connected questions: whether matched input summaries determine a fixed structural edit's spike effect, and which nonzero weight perturbations are guaranteed to preserve a finite reference sequence. Ten original graphs from three connection rules test the first question. Paired homogeneous and heterogeneous LIF networks in brian2-atlas reuse three originals and examine both questions over nested 30/100/300 ms prefixes. Certified lower bounds and actual replay failures distinguish regional preservation from observed equality and locate horizon-dependent losses. Supporting execution, readout and selection controls remain part of the evidence; full proofs and the separate LIF enclosure derivation appear in the article and its appendices. Edge swaps and fixed-support weight families are different interventions; the latter are not a complete mechanism explanation of the former. Results first present swap visibility and its graph and LIF validations, then regional guarantees, horizon failures and direction comparisons, followed by supporting temporal-readout, dose and selection controls.

## 2 Models and fidelity conditions

### 2.1 Chunk graph and unit dynamics

The original execution benchmark is a directed graph of C chunks. Each chunk contains U = 100 allocated units, giving N = CU. Chunks have either excitatory or inhibitory source identity, with 80% excitatory chunks in every workload. A projection connects a source chunk to a destination chunk and carries a nonnegative weight. When source chunk j emits $S_j$[n] spikes in slot n, explicit routing generates one message of magnitude $w_e$ $S_j$[n] for each outgoing edge e. The destination broadcasts the increment to all of its units in the appropriate excitatory or inhibitory input channel. The graph therefore specifies chunk-to-chunk coupling.

Except for the initial-voltage calibration in Section 3.5, each unit starts at v = −65, with other dynamical variables zero. Outside its refractory interval, the numerical kernel updates voltage and three input states as follows:

$$v_i^+ = v_i + dt(-v_i + g_{Ei} - g_{Ii} + g_{Li}) \quad \text{(1a)}$$

$$g_{qi}^+ = (1 - dt)g_{qi}, \qquad q \in \{E, I, L\} \quad \text{(1b)}$$

Here q indexes the excitatory, inhibitory and long-range input states. A threshold crossing at $v \geq 30$ emits a spike, resets v to −65 and sets a refractory deadline two slots ahead. The next slot skips the voltage and input-decay updates, although delivered inputs still accumulate. The stored recovery variable is unused. These equations define a simplified numerical spiking model without a biophysical calibration of voltage or conductance units. The timestep is dt = 0.1 ms; the code uses its numerical value in the Euler update. With uninterrupted updates and no input, voltage leaks toward zero, not the reset value. The kernel has no membrane capacitance, reversal potentials or lower voltage bound; its input states enter as current-like numerical quantities. The shortest permitted interval between spikes is two slots, or 0.2 ms at this timestep.

Before each active chunk update, local fields contribute 0.01 $F_E$ and 0.01 $F_I$ to the corresponding unit input states. Afterward the fields decay with a 10 ms time constant, and the chunk spike count is added to $F_E$. This update also applies to inhibitory chunks; $F_I$ starts at zero and has no spike-driven source, so it remains identically zero in these workloads. Identical initial conditions, parameters and broadcast inputs keep units within a chunk synchronized. N thus describes the allocated workload, while the number of distinct unit trajectories is at most C.

## 2.2 Scheduling and delivery time

All routing variants use the same active, pending and dormant chunk scheduler. Due messages wake their destination. A stepped chunk remains active when it spikes, receives an event, or has a local field or input state above the scheduler tolerance of $10^{-6}$. Future queued events also retain scheduler activity. Skipped unit states are frozen; voltage alone does not trigger an update. The explicit reference is defined by this scheduler together with the numerical kernel.

Routing occurs after the harness advances its current slot counter. For the uniform stored edge delay of one used here, a spike emitted in slot n produces explicit delivery in slot n + 2. The raw-sum, static-gain and history-modulated paths use the same arrival slot. The two-channel and filtered four-channel paths deliver in slot n + 1. These are comparisons between complete execution rules: the filtered and history-modulated methods differ in arrival time as well as their input transformations (Figure 1).

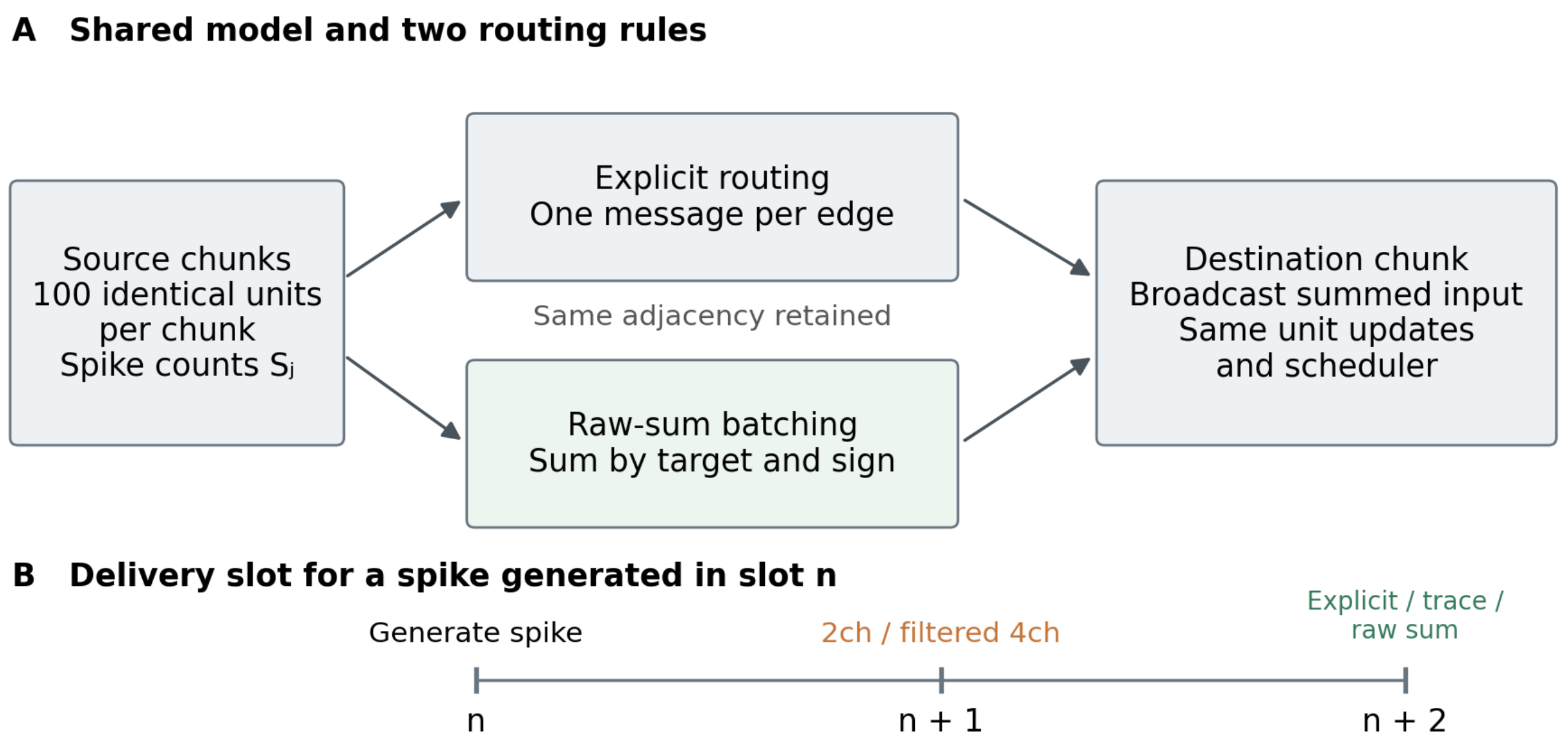

Figure 1. Chunk coupling and delivery semantics. Explicit routing and raw-sum aggregation use the same projection graph, unit updates and scheduler. With stored edge delay one, both deliver a contribution from slot n at n + 2. Normalized two-channel and filtered four-channel proxies deliver at n + 1. Excitatory and inhibitory contributions are grouped separately.

### 2.3 Delay aligned raw sums

Let $P_{kq}$ be the set of aggregated incoming edges to destination k with source polarity q. Their raw contribution, with j = src(e) denoting the source of edge e, is

$$R_{kq}[n] = \sum_{e \in P_{kq}} w_e \; S_j[n] \quad (2)$$

Raw-sum aggregation emits one message with magnitude $D_{kq}$[n] = $R_{kq}$[n] when $R_{kq}$[n] > 0. Excitatory and inhibitory contributions remain separate. Explicitly retained edges are excluded from $P_{kq}$ and are routed individually, so every edge belongs to exactly one path. Both paths deliver at the same slot. The current implementation uses a uniform delay; aggregation over heterogeneous delays would require a separate group for each arrival slot.

All recurrent edge weights in the E1, E3, E4 and E5 workloads are strictly positive. For zero-weight recurrent edges, explicit routing can enqueue zero-amplitude messages that still activate a target. Preserving their activation would require separate wake-up markers, which the current raw-sum path does not generate. External input events use the same harness path in both executions; separately supplied wake-up events must likewise be preserved outside recurrent aggregation.

The following proposition applies the established superposition principle for additive inputs [1,2] to discrete routing and activation. It specifies a sufficient execution condition, including preservation of events that wake dormant chunks. Linearity is required of input accumulation, not of the complete neuronal dynamics.

**Proposition 1 (Sufficient condition for grouped delivery).** Consider a deterministic discrete-time simulation with nonnegative message magnitudes and separate excitatory and inhibitory input channels. At each routing step, group positive-amplitude messages only when destination, input channel and arrival slot agree, replacing each nonempty group by a message carrying its summed amplitude. Assume that each delivery only adds the message magnitude to its designated input accumulator, apart from activation bookkeeping, and that all messages due in a slot are accumulated before a common neuronal update. Assume identical initial states, external inputs, integration rules and routing phases, and scheduling that depends on state and message existence and arrival time, rather than message multiplicity, identity or ordering. Preserve separately any zero-amplitude events that have activation effects. Then explicit and grouped routing produce identical neural dynamical states, spikes and activation decisions in exact arithmetic.

**Proof.** For fixed source spike counts, each group delivers the same input increment by associativity and commutativity of addition. Positive amplitudes and separate polarity channels preserve the existence of a delivery event. This preserves the set of pending destination–channel–arrival-slot triples after each routing step, as well as event existence at delivery. The identical states and fully accumulated inputs therefore give the same subsequent integration and activation decisions. The resulting spikes generate the same aggregate inputs at the next routing step. Induction over slots completes the argument. Queue representations and message counters may differ.

This sufficient condition describes an algebraic property of the execution rule. Equivalence is relative to the fixed discrete reference; it does not establish exact integration of a continuous-time model. Floating-point addition can depend on accumulation order, especially near a firing threshold. The experiments therefore measure actual population-count agreement separately; they do not assume that all floating-point states are identical.

## 2.4 Approximation controls

Filtered and history-modulated outputs are heuristic implementation controls. They alter delivered magnitudes and, for the filtered variants, arrival time. A static-gain ablation and raw sums remove successive output transformations. Appendix H specifies all equations, parameters and earlier comparisons; these variants do not represent the space of local circuit reductions.

## 2.5 Execution cost

All methods retain the unit-state arrays and projection adjacency. At each slot, the aggregation index scans source slots and sums the spike flags of each source with replaced outgoing edges. It traverses those edges only when that source has a nonzero spike count. The raw-sum and history-modulated paths also scan all target slots to clear accumulators, update internal states and emit messages. With uniform delay, these paths emit at most two aggregated messages per target receiving positive aggregate input, one for each polarity. This bounds the output message count, while source scans, edge traversal and neural-state integration remain.

## 2.6 Statistical preservation and the input map

We distinguish changes of execution on a fixed graph from changes of the graph itself. For one common routing and arrival slot, let $W^q$ and $\widetilde{W}^q$ be C × C nonnegative weight matrices for channel $q \in \{E, I\}$, with target rows and source columns; edges outside that source channel have zero entries. Let $\Delta W^q = \widetilde{W}^q - W^q$. The fixed partition has nonempty E and I groups, and P is its C × 2 indicator matrix: $P_{kg} = 1$ when chunk k belongs to group g and zero otherwise. The vector s contains the same source-chunk spike counts for both input maps. No neuronal linearity is assumed; the statements concern additive input increments in exact arithmetic.

**Proposition 2 (Input redistribution under matched group strengths).** Suppose, separately for each channel, that rewiring preserves every target's incoming weight sum from each source group and every source's outgoing weight sum to each target group. Equivalently,

$$\Delta W^q P = 0, \qquad P^T \Delta W^q = 0 \quad (3)$$

Define the group means and within-group residual by

$$\bar{s} = \left(P^T P\right)^{-1} P^T s, \qquad r = s - P\bar{s} \quad (4a)$$

Then the input difference and its Euclidean norm obey

$$\Delta I^q = \Delta W^q s = \Delta W^q r, \qquad P^T \Delta I^q = 0 \quad (4b)$$

$$\| \Delta I^q \|_2 \leq \| \Delta W^q \|_2 \| r \|_2 \quad (5)$$

Here the matrix norm in Eq. (5) is the induced Euclidean norm. In particular, activity that is constant within each source group gives zero input difference; for arbitrary shared source activity, the input differences sum to zero separately within each target group.

**Proof.** The (k,g) entry of $\Delta W^q$ P is the change in the weight sum from source group g to target k, so preservation of the incoming group strengths gives the first equality in Eq. (3). The (g,j) entry of $P^T \Delta W^q$ is the change in the weight sum from source j to target group g, giving the second equality. Since all groups are nonempty, $P^T$P is diagonal with positive group sizes and is invertible. The definition of the group mean yields s = P $\bar{s}$ + r and $P^T$r = 0. Substitution gives $\Delta W^q$ s = $\Delta W^q$ P $\bar{s}$ + $\Delta W^q$ r = $\Delta W^q$ r. Left multiplication by $P^T$ gives zero, proving both identities in Eq. (4b). Submultiplicativity of the induced norm gives Eq. (5). If r = 0, the input difference vanishes. More generally it vanishes whenever r lies in the null space of $\Delta W^q$; nonuniform activity alone does not guarantee an input difference. This proves the proposition.

These conditions hold for the homogeneous rewiring controls in Section 3.4. For the heterogeneous controls, source weights are preserved but target incoming strengths are not: the second equality in Eq. (3) still holds, whereas the first need not. The full residual identity therefore does not apply to that auxiliary family. For multiple delays the same reasoning would require matching strengths separately by arrival slot, rather than matching weights after pooling delays.

Equitable-partition and cluster-synchronization theory studies graph partitions compatible with invariant collective dynamics [11,12]. Proposition 2 uses related partition-matrix algebra to compare two input operators. Its condition $\Delta W^q$ P = 0 is not the equitable-partition condition $W^q$ P = P $B^q$ for some quotient matrix $B^q$. Matching strengths between graphs does not require equal incoming strengths across nodes of a group within either graph. Consequently, it does not establish an invariant synchronized manifold or an exact closed population model. Within-chunk synchrony in Section 2.1 is also distinct from synchrony across all E or I chunks.

The shared vector s is essential. In two autonomous simulations, source histories may differ, and their input difference is $\widetilde{W}^q(\tilde{s} - s) + \Delta W^q$ s. Proposition 2 constrains only the latter term. Equal group input totals for a shared source vector do not imply equal subsequent population spike counts, since local states and threshold responses may differ. Equation (5) is a one-step upper bound, not a lower bound or a prediction of the observed closed-loop trajectory errors.

**Proposition 3 (Insufficiency of degree and strength summaries for universal input reconstruction).** Let T(W) record the partition, per-node group-specific in/out degrees and incoming/outgoing weight multisets, but not edge endpoints. Consider any deterministic input predictor f(T(W),s) with no additional graph-identifying information. If $W_1$ and $W_2$ have the same summary, the predictor returns a common estimate $\hat{I}$ for a shared source vector s, and for any norm

$$max\{\|\hat{I} - W_1 s\|, \|\hat{I} - W_2 s\|\} \geq \frac{1}{2}\|(W_1 - W_2)s\| \quad (6)$$

There exist positive-weight, loop-free directed graphs with identical such summaries and a nonnegative source vector for which the right-hand side is strictly positive. Thus these summaries cannot support exact reconstruction of every node's input for all admissible graphs and source vectors.

**Proof.** Identical arguments to the deterministic predictor give the same $\hat{I}$. The triangle inequality gives $\|W_1 s - W_2 s\| \leq \|W_1 s - \hat{I}\| + \|\hat{I} - W_2 s\|$, and this sum is at most twice the larger error, proving Eq. (6). For a concrete pair, take four distinct nodes a, b, c, d in one source-type group. In $W_1$ use edges a→b and c→d, each with weight w > 0; in $W_2$ use a→d and c→b with the same weights and delay. The per-node in/out degrees and weight multisets agree, there are no loops or duplicate edges, and both edges belong to the same projection class. Any other nodes and edges can be chosen identically where they do not interfere with these edges. The input difference is confined to b and d, with opposite values $w(s_a - s_c)$. Hence

$$\|(W_1 - W_2)s\|_1 = 2w\ |s_a - s_c| \quad (7)$$

Choosing $s_a = 1$, $s_c = 0$ and all other source counts zero gives L1 input separation 2w and worst-case predictor error at least w. In the synchronized U = 100 implementation, the choice $s_a = 100$ scales the separation and bound by 100 without changing the argument. This completes the constructive proof.

The positive lower bound is an existence result over the stated graph class. It applies when both graphs supply the predictor with identical information, and need not be attained by every E3 realization. A predictor with graph-dependent fitted parameters, initialization, or informative history may distinguish the graphs and is outside the common-information premise. The result concerns inputs: distinct inputs can still give identical finite spike sequences. Neither proposition proves that history is useless or that a particular rewiring must change spikes.

### 2.7 A finite horizon sufficient condition

**Corollary 1 (Trajectory agreement on reference activity).** Consider two fixed graphs on the same labeled chunks. Graph dependence enters the common deterministic kernel only through additive routed inputs and their activation bookkeeping. Use the channel separation and common execution rules of Proposition 1, with nonnegative edge weights. Let $W_d^{\,q}$ and $\widetilde{W}_d^{\,q}$ denote their input matrices separately for each channel q and emission-to-arrival delay class d, and let $s_{ref}[n]$ be the reference graph's emitted source-count vector in exact arithmetic. Suppose initial neuronal and local states, refractory deadlines, active/dormant status and external inputs agree. Initial pending messages must have the same aggregate increment for every destination, channel and arrival slot and the same activation effects, although their representations may differ. If

$$\Delta W_d^{\,q}\ s_{ref}[n] = 0, \qquad \forall q,\ d,\ 0 \leq n < T \quad (8)$$

then the neural states, emitted spikes and activation decisions agree through slots 0 to T − 1 in exact arithmetic. The condition includes every delay group emitted before T, including arrivals after T when pending events can affect scheduling before arrival. Any zero-amplitude events with activation effects must be preserved separately.

**Proof.** At the initial prefix, equal dynamic states and equivalent pending increments and activation information give the same stepping decisions and due inputs. Their common deterministic update therefore gives identical new states and emitted spikes. Suppose this holds through the start of slot n. The common source vector is $s_{ref}$[n]; the displayed condition gives the same newly scheduled aggregate increment for every destination, channel and delay group. A positive sum of nonnegative messages is present in one graph exactly when it is present in the other. Together with the separately matched zero-amplitude wake-ups, this preserves event-existence and arrival information, including pending events queried before their arrival. The previously pending increments also agree by the induction hypothesis. Thus all scheduling-relevant pending summaries remain equal at the next prefix, and so do the common update's states and spikes. Induction over n = 0,…,T − 1 proves the claim. Queue entries and message counters need not agree.

For the equal-weight swap in Proposition 3, $\Delta W^q\ s_{ref}$[n] = w($s_c$[n] − $s_a$[n])($e_b$ − $e_d$), where $e_b$ and $e_d$ are coordinate vectors. Equal activity at the exchanged sources therefore makes this swap invisible at the input interface, even if the two sources are active. This is a sufficient condition, not a necessary condition for equal spikes: changed inputs can leave threshold outcomes unchanged. The corollary concerns the specified finite reference trajectory and exact arithmetic; it does not extend equality to other inputs, longer windows or floating-point state bits.

## 2.8 Finite horizon certification with nonzero input differences

The exact-input corollary guarantees a stronger equality than is needed for identical spike outputs. We additionally enclose candidate continuous states under the reference branch history, routing the reference source spikes through the candidate graph in actual source order. Each affected target must preserve execution or skipping, threshold decisions, reset and refractory updates, status labels and next-active membership, including pending arrivals. Local fields can be reused only after the common spike and execution history has been established. The sufficient threshold checks are:

$$V_{hi} < \theta \quad (reference\ \ silent), \qquad V_{lo} \geq \theta \quad (reference\ \ spike) \quad (9)$$

Here V = [$V_{lo}$, $V_{hi}$] encloses the candidate pre-reset voltage and θ = 30. The first condition applies to a silent reference update and the second to a reference spike, only at refractory-eligible executed updates. Failed or unresolved checks are not predictions of spike divergence. Soundness is a joint induction over slots and affected targets, not an assumption that one changed target may be treated as correct without checking it. Appendix J supplies the statement, proof and numerical enclosure contract. The guarantee permits unequal continuous states and ends with the declared window.

The singleton checker uses conservative binary32 rounding bounds [27]. The optimized implementation computes outward binary64 bounds and falls back to exact rational arithmetic when a projected binary32 endpoint is unresolved. The family checker retains affine dependence on fixed parameters plus a uniform residual enclosure; exact support bounds and a rigorous corner-envelope implementation are compared on the same boxes. Ordinary endpoint or grid replay is a different baseline: it checks specified points and does not certify the unsampled interior. For a single specified candidate, however, a complete local consistency replay can itself establish the common-branch property by the same induction. We therefore do not claim a logical advantage over local replay for singleton verification.

## 2.9 Paired homogeneous and heterogeneous LIF model

The LIF validation reuses seeds 100, 201 and 301, the first existing graph from each rule. It does not add three independent original graphs. Each chunk has four actual neurons; each block edge expands to all sixteen ordered unit pairs with unit weight float64(archived float32 weight)×0.001/4 V. The homogeneous control has membrane time constant 20 ms and threshold −50 mV. In the heterogeneous model each chunk has four evenly spaced time constants from 18 to 22 ms and four thresholds from −51 to −49 mV, independently permuted with PCG64 seed 20261109+graph_seed. All other parameters are paired. The initial/reset voltage is −60 mV, rest potential −65 mV, constant background equivalent drive 18 mV, synaptic time constants 5/10 ms, and refractory duration 2 ms.

The background alone is suprathreshold for every isolated unit: with synaptic and task inputs set to zero, the subthreshold equilibrium is −65 + 18 = −47 mV, above the −51 to −49 mV thresholds. Thus isolated cells fire periodically under this constant drive. For the homogeneous unit, the continuous-time approximation gives an interspike interval of 2 + 20 ln(13/3) ≈ 31.33 ms, or 31.9 Hz; this is an analytic reference, not a new simulation or a measured network rate. The study does not test a fluctuation-driven regime with subthreshold background.

$$v_i^+ = v_i + \frac{dt}{\tau_i}\left(E_L + I_b + g_{Ei} - g_{Ii} + d_i - v_i\right) \quad (10a)$$

$$g_{qi}^+ = \left(1 - dt/\tau_q\right)g_{qi}, \quad q \in \{E, I\} \quad (10b)$$

Equation (10) is an Euler update with dt=0.1 ms; $g_E$ and $g_I$ are current-equivalent voltage contributions, not reversal-potential conductances. Voltage is clamped during refractoriness, while both input states continue to decay and receive events. Thresholding uses v≥θ. The native brian2-atlas 0.2.0 reference-f64 backend executes the exported model; no upstream Brian2 distribution or fallback is used. A spike is delivered after one tick and first affects the voltage update two ticks after emission. This differs from the legacy numerical scale and scheduler, so absolute cross-kernel error or radius differences are not a pure heterogeneity effect.

The two models serve different roles. Sections 2.1–2.8 specify the original execution benchmark and its sufficient conditions; Section 2.9 supplies an independently audited deterministic LIF instance of the common-branch condition. Identical units in the homogeneous LIF control and parameter-diverse units in its heterogeneous counterpart share the same graph, input masks, integration rule and recording contract. This paired comparison isolates the specified parameter heterogeneity within LIF, whereas differences between LIF and the original kernel also include units, reset, refractory handling, local fields and stepping.

## 3 Experimental design

### 3.1 Workloads and retention policies

We use sustained-drive, retention, weight-regime and pulse workloads, denoted E1, E3, E4 and E5 to link the manuscript to the accompanying implementation (Table 1). Each legacy workload in Table 1 uses dt = 0.1 ms and 100 units per chunk. Sustained-drive experiments deliver an external input of amplitude 10 to every excitatory chunk in every slot. E5 supplies amplitude 150 for 8 ms, followed by 22 ms without external drive. Initial states and graph realizations are shared between compared routing rules.

Table 1. Workload dimensions and observation windows. Each chunk has 100 units and dt = 0.1 ms.

| Workload | Units N | Chunks C | Slots | Time / ms |
|---|---|---|---|---|
| E1 principal | 50,000 | 500 | 500 | 50 |
| E1 gain scan | 12,500 | 125 | 200 | 20 |
| E3 heterogeneous | 12,500 | 125 | 200 | 20 |
| E3 homogeneous | 12,500 | 125 | 300 | 30 |
| E4 weight settings | 5,000 | 50 | 200 | 20 |
| E5 pulse and recovery | 5,000 | 50 | 300 | 30 |
| E3 rewiring extension | 12,500 | 125 | 300 | 30 |
| Phase diagnostic archive | 12,500 | 125 | 300 | 30 |
| Perturbation response | 12,500 | 125 | 1000 | 100 |

| Workload | Units N | Chunks C | Slots | Time / ms |
|---|---|---|---|---|
| Sensitivity calibration | 12,500 | 125 | 300 | 30 |

The E1 graph has connection probabilities ($p_{EE}$, $p_{EI}$, $p_{IE}$, $p_{II}$) = (0.10, 0.20, 0.20, 0.15) and weights (0.8, 0.8, 1.2, 0.6). E3 uses probabilities (0.15, 0.25, 0.25, 0.20) and base weights (0.6, 0.8, 1.4, 0.5). Its heterogeneous realization multiplies each accepted edge weight by a uniform factor between 0.5 and 1.5. Self-chunk edges are excluded in the graph constructor. Appendix A specifies seeds, all workload parameters and policy cutoffs. Homogeneous and heterogeneous E3 runs use different seeds; the additional weight draws also alter the generator's subsequent edge draws. They are separate configurations rather than a paired isolation of weight heterogeneity.

Hybrid routing retains edges selected by an inhibitory-source weight threshold, an I-to-I threshold, an E-to-I threshold, or the union of the first two rules. These membership tests do not inspect reciprocal edges or cycles. The E3 sweep includes zero retention, implemented by an empty selection policy, and inhibitory-source thresholds retaining 2.0%, 5.8%, 11.2%, 15.9% and 20.5% of edges. The zero-retention extension uses the same graph, input and method-specific gains; the 2.0% setting is rerun to check agreement with the existing scan. The edge-class comparison uses different budgets for its four policies. These comparisons characterize policy behavior at the observed budgets rather than identifying a budget-matched structural optimum.

### 3.2 Measurements and analysis scope

The population activity in slot n is y[n] = S[n]/(N $dt_s$), where S[n] is the total emitted spike count and $dt_s$ is the timestep in seconds. Temporal agreement is measured by zero-lag Pearson correlation r between the reference and approximate per-slot series, without smoothing, trial averaging or lag optimization. Mean-rate error is $\delta_{FR}$ = 100(mean($y_{\text{approx}}$)/mean($y_{ref}$) − 1). We also count slots in which the two total population spike counts differ. This count tests equality of the measured series more directly than a correlation rounded to one.

Message cost is Q = $M_{\text{approx}}/M_{ref}$, where M counts recurrent messages enqueued after simulation steps. It excludes externally injected drive and includes messages scheduled beyond the observation window. In hybrid runs, M includes both individually routed and aggregated messages. Q measures queue insertions; it does not measure runtime or peak memory. It depends on the generated spike activity as well as the routing rule, so comparing Q across different activity trajectories does not isolate routing efficiency. Retained-edge fraction describes the graph partition and is distinct from the fraction of queued messages using explicit routing.

The gain study evaluates a 3 × 3 grid and a second 3 × 3 neighborhood, with one overlapping pair: 18 evaluations and 17 distinct gain pairs per method. We use r ≥ 0.70 and $|\delta_{FR}|$ ≤ 20% as descriptive engineering screening criteria. Each execution-benchmark graph setting has one deterministic realization, and gain pairs are parameter settings rather than statistical replicates. The raw-sum and static-gain controls are post-hoc exploratory analyses; there is no held-out graph set or preregistered inference. Integer totals, mismatch counts, numerical summaries and reproducible source are archived.

For the historical workloads, deterministic means a fixed graph and input protocol without additional process noise. Explicit routing uses an unspecified HashMap traversal order, so floating-point accumulation order and bitwise results are not guaranteed to match across runs.

### 3.3 Controlled delivery fixtures

Eight deterministic fixtures, B1–B8, exercise the frozen Scheduler and Chunk implementation with two chunks of four units, dt = 0.1 ms and 12 slots. Both chunks are explicitly initialized as dormant; events are preloaded at slot zero. These are open-loop delivery tests, not recurrent network workloads. Correct grouping is tested with homogeneous and heterogeneous initial voltages, preserved zero-amplitude wake-ups, separate polarity channels and separate arrival slots. Three variants deliberately drop a wake-up, combine opposite channels, or move inhibitory input to an earlier slot. The latter two retain wake-up markers at the original arrival slots to distinguish input transformation from direct event removal. Subsequent activation may still change because the dynamical states change.

Candidate events are constructed by a separate grouping helper or prescribed event lists and injected directly into Scheduler/Chunk. These tests cover receiver delivery conditions, but do not exercise online aggregation in LpuTopologyIndex or recurrent feedback. The suite uses exactly representable f32 amplitudes and is not designed as a general rounding-stress test.

For every unit and slot we compare voltage, input states, local fields, recovery state, refractory deadline and spike flag; we also compare stepping decisions, post-step chunk status and earliest pending arrival. Floating states are compared by their f32 bit patterns, with absolute differences recorded separately. Queue multiplicity and message counters may differ. Appendix D gives the prescribed events. These fixtures do not extend the state-level validation of E1–E5.

### 3.4 Statistically matched rewiring controls

We extend the two existing E3 graph families to a paired structural control. The homogeneous graph (seed 99) has 2,845 edges; the heterogeneous graph (seed 200, weight spread 50%) has 2,840. Both have 100 E and 25 I chunks, each with 100 units. Their topology is already generated by E/I block connection probabilities. This is a test of particular edge assignments conditional on specified statistics, not a biological-connectome or motif-enrichment test.

For each original, 19 independent chains start from that same graph, with seeds 91001–91019. Within each EE, EI, IE or II arm, a proposal replaces a→b and c→d by a→d and c→b, retaining source, weight, delay and edge-table row. Proposals with shared sources or targets, self-loops or duplicate edges are rejected. Each arm makes 100 proposals per edge, including rejected proposals. No uniform-sampling or mixing guarantee is established; control ranges and ranks are descriptive, not p-values.

All controls preserve each node's per-arm in/out degree, source outgoing weight/delay multisets, node types, group sizes and delay one. In the homogeneous primary family, equal weights within each arm additionally preserve every target's incoming weight multiset and weighted strength exactly. Both identities of Proposition 2 therefore hold channel by channel. In the heterogeneous auxiliary family target weighted strengths can change; it cannot isolate topology from incoming-strength changes. Independent Counter-based audits verify the preserved statistics without relying on the rewiring generator.

Each graph is tested with three fixed inputs for 300 slots (30 ms): tonic amplitude 10 to all 100 E chunks throughout; global burst amplitude 150 to all E chunks in slots 0–79; and focal burst with the same amplitude and timing to 20 E chunks numbered 0, 5, …, 95. Global and focal bursts differ in total drive as well as spatial extent. Every graph/input pair uses explicit routing, raw_sum and the unchanged history output (trace), with zero explicit retention and fixed gains (2.5, 0.4). There are 2 × 20 × 3 × 3 = 360 simulations, without per-graph fitting. Configuration and primary analysis hashes were locked before these runs; the experiment is exploratory, not publicly preregistered.

We compare rewired explicit execution against original explicit execution to measure structural effects, and each proxy against its own graph's explicit reference to measure approximation error. For 1 ms bins, let $B_{tk}$ denote chunk k's spike count in bin t. Chunk L1 is the sum of $|B_{test} - B_{ref}|$ over bins and identified chunks divided by the reference total; population L1 sums chunks before taking the absolute difference. Percent values multiply these ratios by 100 and may exceed 100%. The population measure is no larger by the triangle inequality; that inequality is not a new empirical result. Medians and full ranges summarize the 19 controls. Mean-rate changes and late-window activity provide additional readouts.

This extension sorts explicit source chunks by ChunkId to fix the previous HashMap accumulation order. It adds observation code and a raw-output ablation, without changing the neuronal equations, scheduler or trace rule. A bridge reproduces the archived heterogeneous E3 first 200 slots, including explicit and trace population spikes and messages. Every recorded slot checks equality of the 100 units within each chunk before saving the representative state and chunk count. Technical repetition is assessed separately from structural variation. Appendix E specifies the supplementary diagnostics and archived evidence.

### 3.5 Local sensitivity calibration

We use the homogeneous seed-99 original, its 125 chunks, 2,845 edges, explicit canonical-order routing and the three 300-slot inputs from Section 3.4. Nineteen local structural controls independently start from that original. Each accepts exactly one legal same-arm target exchange a→b, c→d to a→d, c→b, with seeds 93001–93019 and a fixed EE/EI/IE/II allocation of 5/5/5/4. Weight bits, delays, sources and edge-table rows are retained; proposals with shared sources or targets, self-loops or duplicate edges are rejected. Each control removes two original adjacency pairs (0.07030%) and preserves the same per-node degree, weight-multiset and strength constraints as the extensive homogeneous controls. This arm-stratified local intervention is not a uniform random-graph sample.

A separate control changes only initial voltage. Nineteen seeded directions, 94001–94019, assign balanced signs within E and I: 50 positive and 50 negative E chunks, and 12 positive, 12 negative and one unchanged I chunk. All 100 units in a chunk share its value. Each direction is reused at nominal amplitudes 0.0001, 0.001 and 0.01 around −65; these are uncalibrated model coordinates. Every other initial state and the dormant initialization remain unchanged, with no added wake-up. The supplied voltage bits and the voltage immediately before each chunk's first executed update are recorded. The protocol, analysis and inputs are hashed before simulation.

There are 3 original, 57 single-swap and 171 initial-voltage conditions: 231 new scientific runs. Three zero-perturbation and three inverse-swap technical shams are reported separately. The 57 archived extensive controls and three archived originals are reanalyzed, not counted as new runs. New originals must reproduce all ten recorded fields and count records of their archived counterparts exactly.

The primary errors use the same 1 ms population and identified-chunk L1 definition as Section 3.4. For each individual comparison we also compute the fraction of chunk error canceled by population aggregation, one minus population L1 divided by chunk L1; it is undefined when chunk L1 is zero. Raw-slot differences distinguish spike equality from equality after binning. State RMS uses the original trajectory's V, E-input and I-input standard deviations, separately for each input and fixed across all controls and windows. All 19 points, medians, ranges and zero counts are reported without confidence intervals. Appendix G describes shared-source input diagnostics, state recoincidence and time-resolved readouts.

### 3.6 Post hoc input switching and readout analysis

This analysis reuses the 57 explicit single-swap trajectories, their three originals and 57 extensive homogeneous controls. It adds no simulation. The questions and readouts were selected after inspecting the archived switching and cumulative-count results; the saved analysis plan fixes the subsequent computation and robustness checks, not a prospective discovery test. The 19 swaps are paired across the three inputs. A swap is classified as spike-identical under all inputs, spike-different under all inputs, or switching between these outcomes. Four mutually exclusive case categories distinguish zero shared-reference-input difference with state agreement, changed input with equal spikes, changed spikes with equal 1 ms bin counts, and changed 1 ms counts. The first diagnostic evaluates the archived source history for arrivals within the recorded window; it is not a certification of every assumption of the exact-arithmetic corollary.

Let $\delta_{kn}$ be the control minus reference count for chunk k in raw slot n. Define the raw-slot discrepancy, the discrepancy in each identified chunk's full-window count, and the discrepancy in the total count by

$$D_{slot} = \sum_{n,k} |\delta_{kn}|, \qquad D_{count} = \sum_k |\sum_n \delta_{kn}| \quad \text{(11a)}$$

$$D_{global} = |\sum_{n,k} \delta_{kn}|, \qquad R_{count} = \frac{D_{count}}{D_{slot}} \quad (D_{slot} > 0) \quad \text{(11b)}$$

The triangle inequality gives $D_{global} \leq D_{count} \leq D_{slot}$ and the exact temporal-first decomposition

$$D_{slot} - D_{count} = 2\sum_k \min\{\sum_n [\delta_{kn}]_+, \sum_n [-\delta_{kn}]_+\} \quad \text{(12a)}$$

$$D_{slot} = (D_{slot} - D_{count}) + (D_{count} - D_{global}) + D_{global} \quad \text{(12b)}$$

Here $[x]_+$ = max(x, 0). The first difference is discrepancy canceled by time accumulation within nodes. The second is the further discrepancy canceled by pooling nodes after time accumulation. The last term remains in the whole-network total. This order is fixed: pooling nodes first would give a different allocation. The identity is bookkeeping, not a novel inequality. For each comparison with $D_{slot} > 0$, we report $D_{count}/D_{slot}$ and the three component fractions; zero-denominator cases remain undefined. Group medians are computed from individual ratios. The archived per-comparison fractions sum to one; their three separately computed group medians need not. The main plot displays the retained fraction rather than stacking those medians. Relative L1 scores divide discrepancies by that input's original total count. Whole-window counts are finite-window quantities, not estimates of stationary firing rates.

For bin robustness, widths are 1, 2, 10, 50 and 300 raw slots, or 0.1, 0.2, 1, 5 and 30 ms. For every width b, all offsets o = 0,…,b − 1 are evaluated with partial first and last bins retained; every recorded slot contributes once. A shifted 30 ms partition can thus contain two partial bins. Identified-chunk, E/I-pooled and globally pooled readouts are compared separately. At each common width and offset, the range gap is the smallest extensive-control score minus the largest single-swap score. These are repeated readouts of the same trajectories, not independent observations.

A bin-free check uses the exponential-filter distance of van Rossum [18]. One representative spike per synchronized chunk is used, dividing its counts by U = 100. With $h_\tau(t) = \exp(-t/\tau)$ for $t \geq 0$ and zero otherwise, filter each representative spike train to obtain $f_{k,\tau}$ and $f_{ref,k,\tau}$, and calculate

$$V_\tau^2 = \frac{1}{\tau}\int_0^\infty \sum_k [f_{k,\tau}(t) - f_{ref,k,\tau}(t)]^2 \, dt \quad \text{(13a)}$$

$$R_\tau = \frac{2V_\tau^2}{N_{ref}} \quad \text{(13b)}$$

$N_{ref}$ is the total number of representative reference spikes. The integral includes the full analytic filter tail of spikes recorded within the 30 ms window and assumes no additional future spikes. We use τ = 0.1, 0.2, 1, 5 and 30 ms. $R_\tau$ is a normalized squared distance, not an L1 percentage, and need not be bounded by one or decrease with τ. The factor two makes a single isolated unmatched representative spike contribute one to its unnormalized numerator. This check asks whether the range separation persists under another timing-sensitive readout; it does not equate the different metrics.

### 3.7 Matched drive and temporally rearranged inputs

The controlled-input protocol is locked before new trajectories, with unchanged seed 99 graph, nineteen swap graphs, numerical kernel, horizon and direction pool. G_B stimulates all 100 excitatory chunks with amplitude 150 for slots 0–79; F_B stimulates chunks 0,5,...,95 with the same amplitude and duration. F_H uses the same twenty targets with amplitude 750. G_T and F_HT retain the respective G_B and F_H per-node 300-slot amplitude multisets but apply a common temporal permutation: PCG64 seed 20261108 permutes slots 0–159 and selects its first 80 indices; slots 160–299 remain zero. G_B/F_H and G_T/F_HT have identical total external amplitude at every slot and sum to 1,200,000; F_B sums to 240,000. Spatial allocation changes together with local amplitude. Temporal order, run length, final drive time and recovery duration can change together; no pure spatial-range or phase effect is isolated.

Each input uses its own original reference. Five inputs times twenty graphs give 100 network conditions, of which 40 bridge existing runs and 60 use new inputs; a second 100-run execution is technical repetition.

### 3.8 Validation on additional original graphs

The controlled-input original-graph validations use 125 labeled chunks, with 100 excitatory and 25 inhibitory source identities. The discovery graph is seed 99. Nine prospectively fixed follow-up graphs use seeds 100–102 with the original independent-edge E/I generator, seeds 201–203 with module affinity, and seeds 301–303 with ring-distance affinity. Independent-edge probabilities for EE/EI/IE/II are 0.15/0.25/0.25/0.20; positive arm weights are 0.6/0.8/1.4/0.5. Module and ring graphs use weighted sampling without replacement with fixed arm counts 1467/610/648/120, totaling 2845 edges. These two affinity rules share a sampler. The ten-original cohort concerns equal-arm-weight graphs; the earlier heterogeneous-weight seed200 control in Section 3.4 is a separate auxiliary comparison. Original graphs are not matched for each node's degree or strength, so between-family differences do not identify a geometric causal effect.

Each original has 19 legal double-edge swaps, allocated 5/5/5/4 across the four arms. A swap replaces a→b and c→d by a→d and c→b within the same equal-weight arm, rejecting self edges, duplicate edges and repeated endpoints. It preserves each node's arm-specific in/out degree, incoming/outgoing weight multiset and strength. Sampling takes the first legal proposal under the locked seed and attempt budget, without using activity outcomes. It is a stratified intervention design, not a uniform graph-null ensemble. Every swap is compared with its own original under the same input.

Each new original uses the five unchanged event tables of Section 3.7 and its own original trajectory under each input. All nineteen swaps are retained, including silent targets and zero effects. The three independent-edge originals contribute 300 scientific network conditions; the six affinity originals contribute 600. Appendix L gives the generators, sampling rules, per-family decision criteria and direction denominators.

### 3.9 LIF input controls and nested observation windows

Legacy amplitudes are 150 and 750 in numerical input units, observed for 30 ms. LIF amplitudes are fixed dyadic values 15.625 and 78.125 mV acting in the voltage equation. Its 30 ms pattern repeats ten times; 30/100/300 ms are prefixes of the same 300 ms trajectory, not independent runs. Total-drive matching holds at every prefix, but 100 ms includes a partial cycle and therefore does not preserve the per-node amplitude multiset under time rearrangement. Continuing stimulation and late activity are reported to distinguish prolonged observation from a silent tail.

A swap is visible when at least one labeled neuron's per-slot spike differs from its own reference; legacy within-chunk synchrony makes the corresponding chunk count comparison equivalent. A visibility switch means that this equality label changes between two inputs. We always retain all four combinations of equality labels. L1 slot discrepancy and within-neuron cumulative-count discrepancy supply secondary, nonbinary readouts. Neither a normalized L1 value near one nor a zero switch count establishes decorrelation or input independence. Raw switch counts also depend on the two marginal visibility rates. For N paired swaps and visible fractions $p_A$ and $p_B$, independent labels would have expected switch count $N[p_A(1-p_B)+(1-p_A)p_B]$. This is a descriptive benchmark, not an assumed sampling null or a significance test. Counts therefore cannot rank the strength of input dependence across contrasts; one observed switch establishes a counterexample to exact summary sufficiency without establishing excess switching above independence.

Three reused graphs × two parameter variants × five inputs × twenty graph versions produce 600 scientific 300 ms trajectories. The three horizons are repeated readouts, not 1,800 independent simulations. Thirty original references are rerun, and standalone 30 and 100 ms references are compared with their long-run prefixes. Full state observations cover every neuron, not a representative synchronized unit. Late activity, refractory behavior and within-chunk spike differences are recorded without selecting favorable trials. The fixed input masks, physical parameters, numerical expression order and technical validation are given in Appendices M and N.

### 3.10 Certificate and parameter family audits

All audits reuse the 57 archived single-swap/input cases, 125 chunks and 300 slots. The 17 input-changed, spike-equal cases are the target cohort; the four input-equal cases and 36 spike-different cases remain controls. Candidate trajectories are withheld from certificate computation and used afterward for validation, but these cases were already known to the investigators. The family prototype varies positive weights on the two affected targets' existing incoming support, with one or two fixed sign generators and seven predeclared radius levels through 25%. It is centred on each old candidate graph and tested against the original reference. It does not preserve strength or weight multisets relative to the original. Two-dimensional generators are each divided by two to bound the total edge budget; their coverage is not an equal-per-axis comparison with one dimension. Appendix J records coverage, distinct floating-point witnesses and the same-guarantee cost comparison.

### 3.11 Paired directions under original self references

The matched-direction comparison uses the original graph as the centre under each of tonic, global-burst and focal-burst input. The 19 distinct target pairs come from the earlier swaps; no swapped graph supplies a candidate centre. For each pair, incoming original edges are grouped by target, E/I channel and exact weight bits. Odd groups keep the final source-sorted edge fixed. Remaining signs contain equal positive and negative counts, randomly permuted with seed 20261108 plus pair index. Sixteen distinct directions are retained, identifying global sign reversals because intervals are symmetric. The same support, group sign counts and norm budgets are used in every direction:

$$w_e(t) = RN32[w_{e,0}(1 + t\ \sigma_e)], \qquad |t| \le \rho \le 0.25 \quad (14)$$

At a common t, all directions also have the same actual binary32 weight multiset within each group. This is cross-direction matching, not equality to the original weight multiset. Quantization may cause a common strength discrepancy. Support, polarity, delays, initial state and input remain fixed; the parameter is constant throughout the window.

Each symmetric radius is bracketed by L, certified for every t in [−L,L], and U, supported by an actual failure at +U or −U. Initial levels are 0, $2^{-16}$, $2^{-12}$, $2^{-8}$, $2^{-6}$, $2^{-4}$ and 0.25. After the first witnessed failure, at most eight midpoint steps target width max($2^{-20}$, 0.05U). A midpoint enclosure failure with both endpoints passing leaves a proof gap. Certification at the cap is censored, not a maximal radius. Independent local replay checks both targets separately and returns the earliest failed slot; Python replay and exact parameter-entry rounding validate the final witnesses.

For a direction, ΔI denotes its ideal unit-relative-weight change in arriving input along the corresponding unperturbed source history. A deliberately simplified score filters the net E-minus-I projection and compares it with the reference pre-reset threshold margin:

$$z_k[n] = d\ z_k[n - 1] + (1 - d)(\Delta I_{kE}[n] - \Delta I_{kI}[n]) \quad (15a)$$

$$J = max_{k,n\ eligible} \frac{|z_k[n]|}{max\{|v_{ref,k,pre}[n] - \theta|,\ 2^{-20}\}} \quad (15b)$$

Here d = binary32(1 − binary32(0.1)), the input-state decay factor and approximately the voltage-retention factor at this timestep; the maximum is over the two targets and originally executed, refractory-eligible slots. Lower score predicts a safer direction. The score does not propagate the complete conductance, reset or refractory dynamics. Mean activity and independently circular-shifted source activity use the same filter and original margins; random selection is candidate 0. All rules select on tonic only, break near ties by smallest ID and keep that ID under the other inputs. Pools, scores and selections are saved before perturbed outcomes. Full frozen-branch sensitivity is retained only as an algebraic ceiling, because it shares the certificate recurrence. Recomputed scores under other inputs use their complete unperturbed references, not test perturbation outcomes or an unseen-input forecast.

Only disjoint brackets order two directions. Concordance is computed within each target pair and averaged over pairs with comparisons; prediction ties receive half credit. Cross-input reversal requires opposite resolved order under both inputs. Overlapping selection brackets are unresolved; identical directions and two cap-certified choices are separate operational ties, not claims of equal true radii. All counts are descriptive correlated comparisons. The three inputs jointly vary drive, spatial support and timing, so this design does not isolate any one stimulus attribute.

### 3.12 Direction and score audits under controlled inputs and across originals

The fixed nineteen-by-sixteen weight directions give 1,520 conditions, including 608 old bridges and 912 new-input conditions. All scores and G_B selections are locked before certification. Mean and independent-shift controls preserve per-source counts over slots 0–297; the final two source slots remain unchanged. The new protocol fixes this delivered-window definition prospectively. Identical radius search, cap and stopping limits are used. Every finite upper witness is independently replayed; every lower interval is rechecked. Nine lower-bound cases are selected for exact-rational checks by pair IDs 0,9,18, direction 0 and the three new inputs. All failures, caps and unresolved comparisons are retained; no new original seed is introduced.

Target pairs are sorted and deduplicated before generating sixteen balanced directions; duplicates are not replaced to recover a preferred sample size. Graphs, swaps, inputs and pools are fixed before network simulation, and scores and G_B choices before radius outcomes.

The score-factor audit is post hoc and reuses all 912 direction/input conditions without new certificates or replay. The four scores combine the absolute net arriving projection or its existing one-pole filtered value with either no denominator or the original threshold-margin denominator. Every score maximizes over the same reference-executed, refractory-eligible slots. Original mean and independently shifted source controls already include both factors. We report target-level paired differences, coverage and concordance stratified by g = $L_{safer}/U_{less_safe}$ > 1, a conservative lower bound on the radius ratio. Small g need not mean that the true radii are close. Unknown orders remain unresolved; Appendix K gives their categories and loose score-completion ranges.

### 3.13 LIF direction families and reference conditioned scores

For each graph, target-pair indices 0, 9 and 18 are chosen before LIF outcomes; each supplies the sixteen existing directions. The same sign at a block edge applies to all sixteen expanded unit connections. For one common representable binary64 parameter α, weights use the nested rounding construction below; support, delays, other parameters and the input remain fixed. There are 3 × 2 × 3 × 16 × 5 × 3 = 4,320 direction conditions. These are one-parameter joint directions, not arbitrary simultaneous combinations of separately certified directions.

$$w_e(\alpha) = RN_{64}[w_{e,0} RN_{64}(1 + \sigma_e \alpha)], \quad |\alpha| \le \rho \le 0.25 \quad (16)$$

Radii refer to a centered symmetric interval contained in the safe set, not to its outermost isolated safe point. A lower bound L certifies every covered parameter with |α|≤L. An upper bound U is assigned only when point replay exhibits a different spike at an actual |α|=U. A failed enclosure with passing endpoint

replays is a proof gap, not a failure witness. Direction i is ordered safer than j only when $L_i > U_j$. Both directions may be capped at 25%, and overlapping brackets remain unresolved. The frozen-source replay is exact until its first discrepancy: before that event all sources still emit the reference spikes. Selected LIF witnesses are additionally tested in the full Atlas network.

Search levels are 0, $2^{-16}$, $2^{-12}$, $2^{-8}$, $2^{-6}$, $2^{-4}$ and 0.25, with at most eight refinements after a failure upper bound and width tolerance max($2^{-20}$, 0.05U). Searches stop at the fixed budget or a proof gap rather than refining until a desired answer. The retained records distinguish interval success, real point failures, zero lower bounds and numerical limits.

Scores use the input's own original reference. The joint score filters the signed input perturbation and normalizes by the pre-reset threshold margin, evaluated only on reference-eligible updates. In Eq. (17), $h_i = dt/\tau_i$, $c_q = 1 - dt/\tau_q$, and δI is indexed at its first possible voltage-update effect. The LIF projection includes separate 5/10 ms synaptic decay and each unit's membrane leak, without reset or refractory clamping in the projection itself. The legacy score uses its 0.9 single-stage filter. The full frozen-branch derivative includes those branches and is an algebraic comparator. These scores are related to the certificate dynamics; high agreement alone is not an independent causal explanation.

$$y_{iq}[n] = c_q y_{iq}[n-1] + \delta I_{iq}[n] \quad (17a)$$

$$z_i[n] = (1-h_i) z_i[n-1] + h_i (y_{iE}[n] - y_{iI}[n]) \quad (17b)$$

$$J = max_{i,n\ eligible} \frac{|z_i[n]|}{max\{|v_{ref,i,pre}[n] - \theta_i|,\ \varepsilon\}} \quad (17c)$$

The floor ε is fixed at $2^{-20}$ in the corresponding voltage units, hence $2^{-20}$ V in LIF. This is a denominator floor for the heuristic score, not a threshold tolerance or a rounding allowance in the certifier. In LIF it is approximately 0.954 μV, about $9.54 \times 10^{-5}$ of the homogeneous 10 mV threshold-to-reset distance. Reusing the same numerical value across uncalibrated legacy units and volts is not a unit-invariant normalization. The locked score and all reported radii remain unchanged. Ablations remove filtering or margin normalization; mean-activity and independent circular-shift controls replace source histories while preserving counts in slots that can affect the window, with the last two source slots unchanged. The same eligible mask and resolved direction set are used for each method. We average agreement within target pairs and then over nonempty pairs within each graph/input/horizon, reporting the possible-pair denominator. G_B scores and choices are frozen before evaluation under F_H. This is transfer of a reference-based diagnostic, not prediction without observing any reference trajectory.

In contrast to the legacy one-pole diagnostic, the LIF score uses its 5/10 ms synaptic constants and each neuron's membrane time constant. It does not clamp its projected voltage at reference resets or during refractoriness, although evaluation uses the reference-eligible mask. Full frozen-branch derivatives retain those operations and remain an algebraic comparator. All compared methods use the same bracket-resolved pairs. A target pair with no resolved comparison contributes no macro score; coverage and the number of contributing targets accompany every principal score. No accuracy is assigned to unresolved pairs.

### 3.14 Finite prefix nesting and post hoc bracket comparisons

For a fixed executable family, initial condition and one complete driving sequence, define $S_T$ as the representable parameter values whose labeled spike sequence equals the original through T. The observation horizons must be prefixes of that same run. Independent restarts with altered inputs or initial states are not covered by the following nesting statement.

$$S_{T_2} \subseteq S_{T_1}\ (T_2 \geq T_1), \quad U_{long} < L_{short} \quad (18)$$

**Corollary 2 (Nested spike-preserving sets).** If $T_2 \geq T_1$, then $S_{T_2} \subseteq S_{T_1}$. If a short-prefix certificate covers $[-L_{short}, L_{short}]$ and a long-prefix replay fails at an actual parameter with magnitude $U_{long} < L_{short}$, then an explicitly previously certified parameter is no longer spike-preserving at the longer horizon. Proof. Equality on a longer sequence implies equality on every prefix. The witnessed failing parameter belongs to the short certified interval and, by its actual long-run spike discrepancy, does not belong to the long safe set. This proves strict loss inside that interval. The full safe set need not be connected; the certified interval need not be maximal.

After completing the LIF study, we evaluate $U_{long} < L_{short}$ for every matched short/long direction condition. This compares a real replay failure with an interval guarantee rather than interpreting smaller numerical lower bounds alone. Separately, for every within-target direction pair, we compare its order under G_B and F_H only when both inputs have strictly disjoint certified-lower/failure-upper brackets. All candidate pairs and unresolved outcomes are retained. These two summaries, and the audit of clipping by the fixed score-margin floor, are post hoc; they were not the locked LIF success criteria. No new radius query or simulation is required. The source records and computations are specified in Appendix N.

### 3.15 Activity guided selection across nested budgets

Noise-driven covariance has been used to derive synaptic pruning rules with guarantees for restricted linear and rectified-linear networks [19]. Here every candidate retains all 2,845 edges; we test selection among statistically matched endpoint changes in the existing spiking model. The training input is the focal burst of Section 3.4. A first follow-up fixes ten candidate banks of 64 graphs, each removing 282 original adjacency pairs (9.9121%), with EE/EI/IE/II budgets 146/60/64/12. Independent legal same-arm exchanges use each original edge row at most once and never recreate an original pair. Node-specific degree, weight-multiset, strength and delay constraints remain exact. Each bank is a conditional graph search on one original, not an independent original network or a uniform graph sample.

Let $S_{ref}$ contain representative source counts $S_j$[n]/100 from the archived training trajectory, for n = 0,...,297, whose recurrent arrivals lie within the 300-slot window. Let L = 298 and μ be the mean across its L time columns. With the channel-separated weight differences of Section 2.6, the two deterministic scores are

$$J_{mean} = L \sum_q \| \Delta W^q \mu \|_2^2 \quad (19a)$$

$$J_{co} = \sum_q \| \Delta W^q S_{ref} \|_F^2 \quad (19b)$$

These scores include only emissions whose arrivals lie within the observation window. Even in exact arithmetic, a zero coactivity score would not by itself certify Corollary 1: its input-map condition must also hold for source slots 298 and 299, and all remaining hypotheses, including matching initial states and equivalent initial pending inputs and activation effects, must be satisfied.

Each bank supplies four choices: a uniform candidate index, the mean-activity minimum of Eq. (19a), the coactivity minimum of Eq. (19b), and the minimum mean coactivity score under 32 independent per-source time permutations. The same permutations are used for every candidate within a bank. They retain each source's count distribution while changing cross-source alignment. All scores use the full difference matrix, including interference between exchanges; ties use the smallest candidate index. The coactivity score depends on uncentered zero-lag second moments and is invariant to a common permutation of time. It does not encode complete event order or a memory mechanism. Shared graph choices retain all policy memberships but are simulated only once per input.

For the 9.9121% experiment, three evaluation inputs change the 20-target mask, split the burst into 0–4 and 8–12 ms segments, or change both. The new mask overlaps the training mask at two targets. Each input supplies 1,600 amplitude-150 events, totaling 240,000; splitting extends the final drive time from 8 to 12 ms. The primary endpoint is each bank's equal-weight mean of normalized raw-slot identified L1 over these three inputs, followed by paired differences between selectors. Training and individual-input outcomes are also reported. Design, code, inputs, candidate graphs, scores and selections were locked before any new simulation.

After examining that experiment, a second follow-up fixes budgets 28/2,845 (0.9842%; arm counts 14/6/6/2) and 56/2,845 (1.9684%; 30/12/12/2). Each lower-budget candidate is an exact per-arm prefix of its corresponding archived full-budget exchange sequence, with the nearest allowed even per-arm budget and a minimum of two changed original pairs. Selection is repeated within the same ten banks using the same training information and four rules. Lower-budget scores are evaluated from $\Delta W^q\ S_{ref}$ directly before squaring, avoiding cancellation in a Gram-matrix trace; the mathematical criteria are unchanged and the archived full-budget selections are retained. Budgets are nested, dependent comparisons; the discrete arm allocations are reported rather than treated as exactly proportional doses.

The second study first tests training-input raw-slot L1. Both budgets must have coactivity-minus-random paired mean and median below $-10^{-12}$ and at least seven wins with a paired difference below $-10^{-12}$ among ten banks. Against each mean and shuffled control, the paired mean must be at most $10^{-12}$ and wins must be no fewer than losses. Both budgets and an independent audit must pass before repeating the three changed inputs. Those inputs are already known evaluation conditions, not a fresh independent holdout set. This is a decision rule for continuing the study, not a statistical significance test. Failure stops further simulations under this protocol; secondary metrics cannot replace the gate.

Both follow-ups retain the frozen kernel, explicit canonical routing and default initial state. Secondary readouts are 1 ms identified L1, identified cumulative-count L1, the 1 ms exponential-filter score, and original-channel-standardized state RMS. Ratios with zero denominators remain undefined. All paired points, means, medians, ranges and ties are retained without population-level confidence intervals. The input-table driver must reproduce the archived training counts and all ten recorded fields exactly. Graph-search and scoring costs are offline work; the matched budget measures changed adjacency, not saved memory or execution time.

### 3.16 Paired doses and error composition

The paired-dose audit is also post hoc. Only ten archived random-policy chains have complete training trajectories at all three nested budgets: 28, 56 and 282 replaced original adjacencies, or 14, 28 and 141 legal double-edge exchanges. These thirty runs share one original focal-burst reference with 147,600 spikes and retain the same candidate identity within each chain. They are not 640 fully simulated candidates, independently selected best graphs or extensions of the nineteen earlier single swaps. Define X = $D_{slot}$/147600 and Y = $D_{count}$/147600 from the discrepancies in Eq. (11), and R = Y/X where X > 0. For two budgets a and b, the exact symmetric identity is $\Delta Y = ((R_a + R_b)/2)\Delta X + ((X_a + X_b)/2)\Delta R$. Contributions are calculated per chain before averaging. This separates magnitude from retained-fraction changes arithmetically, not the causal effects of edge number and identity.

### 3.17 Supporting phase and pulse response studies

Five controls in each of three constraint treatments preserve a matched adjacency-edit budget on the homogeneous original. The 48 archived trajectories from 16 graphs and three inputs are partitioned into stimulation, early recovery and late windows. Fixed-scale state errors and reference-subspace projections are descriptive. Appendix I.1 and I.3 retain the full design and results.

A separate 48-condition follow-up uses these 16 graphs over 100 ms, with a new conditioning location and zero or two nonzero pulse amplitudes. It distinguishes response to each graph's own baseline from fidelity to the original response, and tests fixed training-only activity features. Appendix I.2 and I.4 retain the complete design, equations and results; Appendix F supplies technical details.

# 4 Results

## 4.1 Execution control separates graph changes from delivery changes

Across the 120 graph/input pairs in the two-family rewiring study, raw sums match their own explicit reference at every recorded slot and chunk. Recorded floating-point states are not all identical. In E1, population counts also match over all 500 slots, with 84.30% fewer recurrent queue insertions. Active-source edge traversal, graph storage and unit states remain. These are execution checks and queue-traffic measurements, not a claim of graph compression or runtime acceleration. Appendix H retains all E1–E5 comparisons, gain and retention scans, and floating-point state discrepancies.

All five fixtures respecting the grouping conditions preserve every recorded dynamical state bit pattern, stepping decision, chunk status and earliest pending arrival (Table 2). The positive-input, heterogeneous-initial-state and separate-arrival controls produce 20, 22 and 8 spikes, respectively, so their agreement is not limited to silent trajectories. This is observed floating-point equality in the specified fixtures, not an extension of Proposition 1 to arbitrary floating-point executions.

Dropping the zero-amplitude wake-up first changes state and activation at slot 2. Combining equal excitatory and inhibitory contributions into a net zero retains a wake-up at slot 2, but loses both channel states; state differences begin at slot 2 and activation differences at slot 3. Moving inhibitory input from slot 4 to slot 2 changes state at slot 2 even though stepping and chunk status remain matched. All three violating variants retain the same per-unit spike sequences as their references over the 12-slot window. Thus, even spike-sequence agreement can miss internal-state and activation differences in these examples; the tests demonstrate possible failures of grouping outside the stated conditions, not that every condition violation must change spikes.

Table 2. Controlled delivery fixtures. First differences use zero-based slots; a dash denotes no difference in 12 slots. Message counts include prescribed wake-up events.

| Fixture | Messages ref / test | First state difference | First activation difference | First spike difference |
|---|---|---|---|---|
| B1 Positive aligned | 4 / 2 | — | — | — |
| B2 Heterogeneous initial state | 4 / 2 | — | — | — |
| B3 Drop zero wake-up | 1 / 0 | 2 | 2 | — |
| B4 Preserve zero wake-up | 1 / 1 | — | — | — |
| B5 Separate E/I channels | 4 / 2 | — | — | — |
| B6 Cancel E/I channels | 4 / 1 | 2 | 3 | — |
| B7 Separate arrival slots | 4 / 2 | — | — | — |
| B8 Move later input earlier | 4 / 3 | 2 | — | — |

State comparisons cover all recorded unit dynamical variables and chunk fields; activation compares stepping and post-step status. Spikes are compared per unit.

## 4.2 Matched statistics do not fix identified chunk trajectories

All primary controls exactly preserve the specified per-node degrees and weight multisets. They retain 21.4–23.7% of the original directed adjacency pairs. Although 97.7–98.4% of edge-table rows change target, that row-based measure is not the fraction of original adjacency pairs removed. Table 3 reports the primary homogeneous results; Figure 2 shows every control for both families.

Table 3. Homogeneous controls, normalized L1 errors (%). Rewiring columns give median (minimum–maximum) over 19 controls. Structural differences compare explicit runs; trace uses each graph's own explicit reference.

| Input | Rewired population | Rewired chunks | Original trace | Rewired trace |
|---|---|---|---|---|
| Tonic | 11.73 (4.89–47.49) | 69.01 (45.55–85.76) | 13.43 | 13.96 (9.51–24.80) |
| Global burst | 7.21 (3.56–16.68) | 28.99 (23.14–36.00) | 7.98 | 6.42 (5.13–9.00) |
| Focal burst | 15.31 (9.42–32.52) | 88.69 (73.64–115.58) | 10.50 | 14.99 (9.34–29.97) |

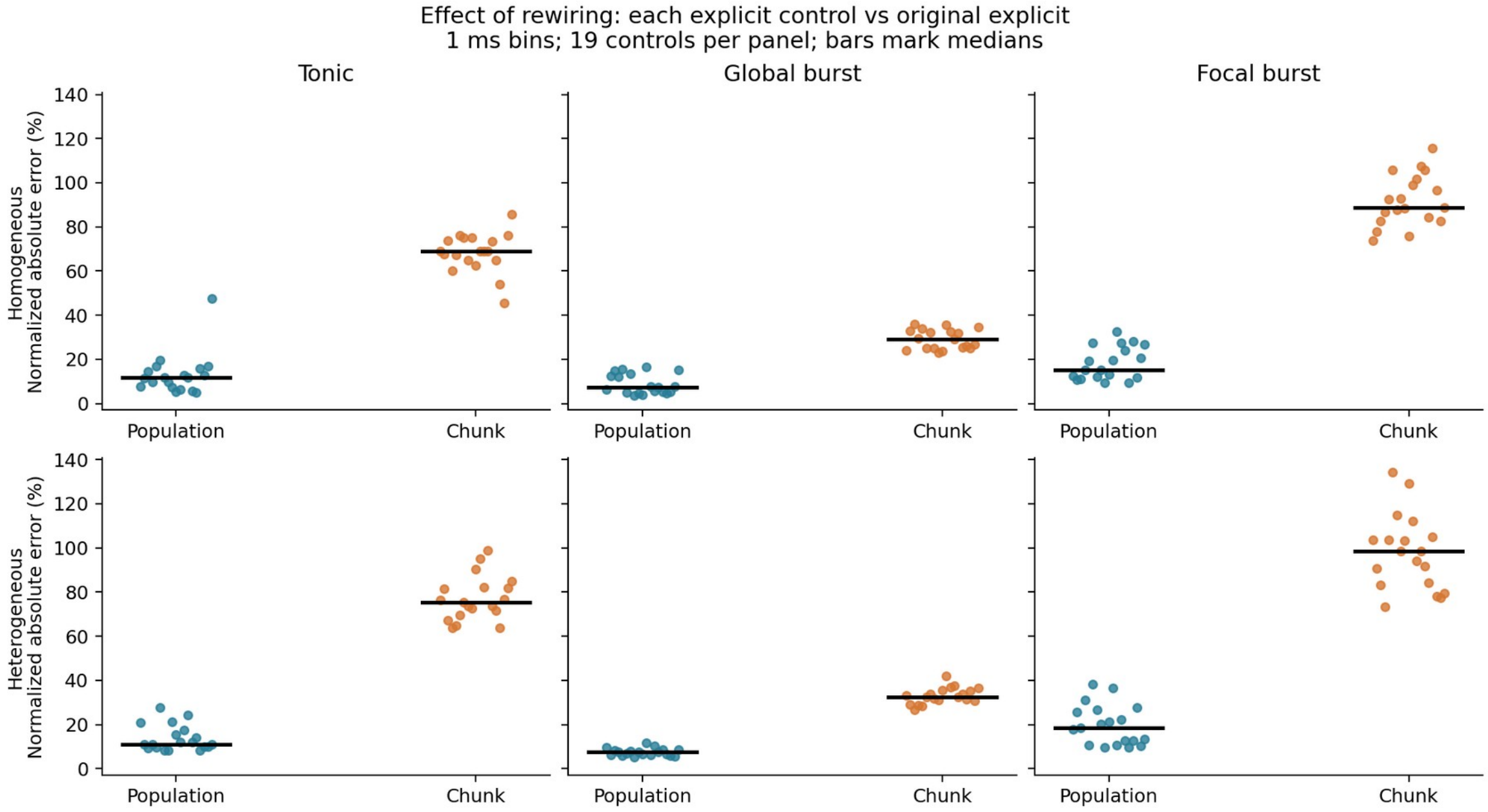

Figure 2. Explicit dynamics after constrained rewiring. Each point compares a rewired graph with its family's original under the same input, using 1 ms normalized L1 at population and identified-chunk resolution; bars mark medians of 19 controls. Homogeneous weights preserve per-node group strengths exactly; heterogeneous weights do not preserve target strengths. Ranges are descriptive, not confidence intervals.

For the homogeneous global burst, population L1 has median 7.21%, whereas identified-chunk L1 has median 28.99%. Whole-window mean rate changes by a median +2.79%, but the post-stimulus 22 ms window changes by +19.33%. Thus a relatively small change in an aggregate readout can coexist with appreciable redistribution across identified chunks and time. The focal-burst chunk difference is larger (median 88.69%), but the inputs also differ in total drive. These results are conditional on the tested graphs and protocols, not evidence that localization alone causes greater sensitivity.

Proposition 2 explains what the preserved statistics constrain for a shared source vector, and Proposition 3 shows why those statistics cannot guarantee universal input reconstruction. Neither predicts the numerical L1 values or makes spike divergence inevitable. The measured closed-loop differences supply the separate empirical evidence. In the heterogeneous controls the population/chunk L1 medians are 11.07%/75.31%, 7.44%/32.54% and 18.33%/98.52% for tonic, global and focal inputs, respectively. Their largest per-arm normalized L1 target-strength change is 16.53%; these are auxiliary sensitivity results, not a pure strength-matched structural comparison.

### 4.3 Extensive rewiring exceeds the local sensitivity controls

For each of the three inputs, every extensively rewired control has greater 1 ms identified-chunk error than every single-swap control (Table 4; Figure 3). The smallest extensive errors are 45.55%, 23.14% and 73.64% for tonic, global-burst and focal-burst inputs, respectively; the corresponding single-swap maxima are 39.15%, 9.83% and 22.70%. The comparison separates the observed error scales under these controls. It does not establish a causal law relating error to the number of changed edges: the interventions also differ in the sources, paths and endpoint assignments they affect.

Table 4. Identified-chunk L1 error in 1 ms bins (%). Broad rewiring and one-swap groups each contain 19 controls; entries are median (minimum–maximum). The final column tests finer 0.1 ms per-slot spike agreement.

| Input | Broad rewiring | One swap | Swap slot agreement |
|---|---|---|---|
| Tonic | 69.01 (45.55–85.76) | 0.00 (0.00–39.15) | 7/19 |
| Global burst | 28.99 (23.14–36.00) | 1.35 (0.00–9.83) | 7/19 |
| Focal burst | 88.69 (73.64–115.58) | 7.05 (0.00–22.70) | 7/19 |

All 171 runs across three small initial-voltage doses preserve per-slot spikes. One-swap bin-zero counts are 10, 7 and 8, respectively, and must be distinguished from per-slot agreement.

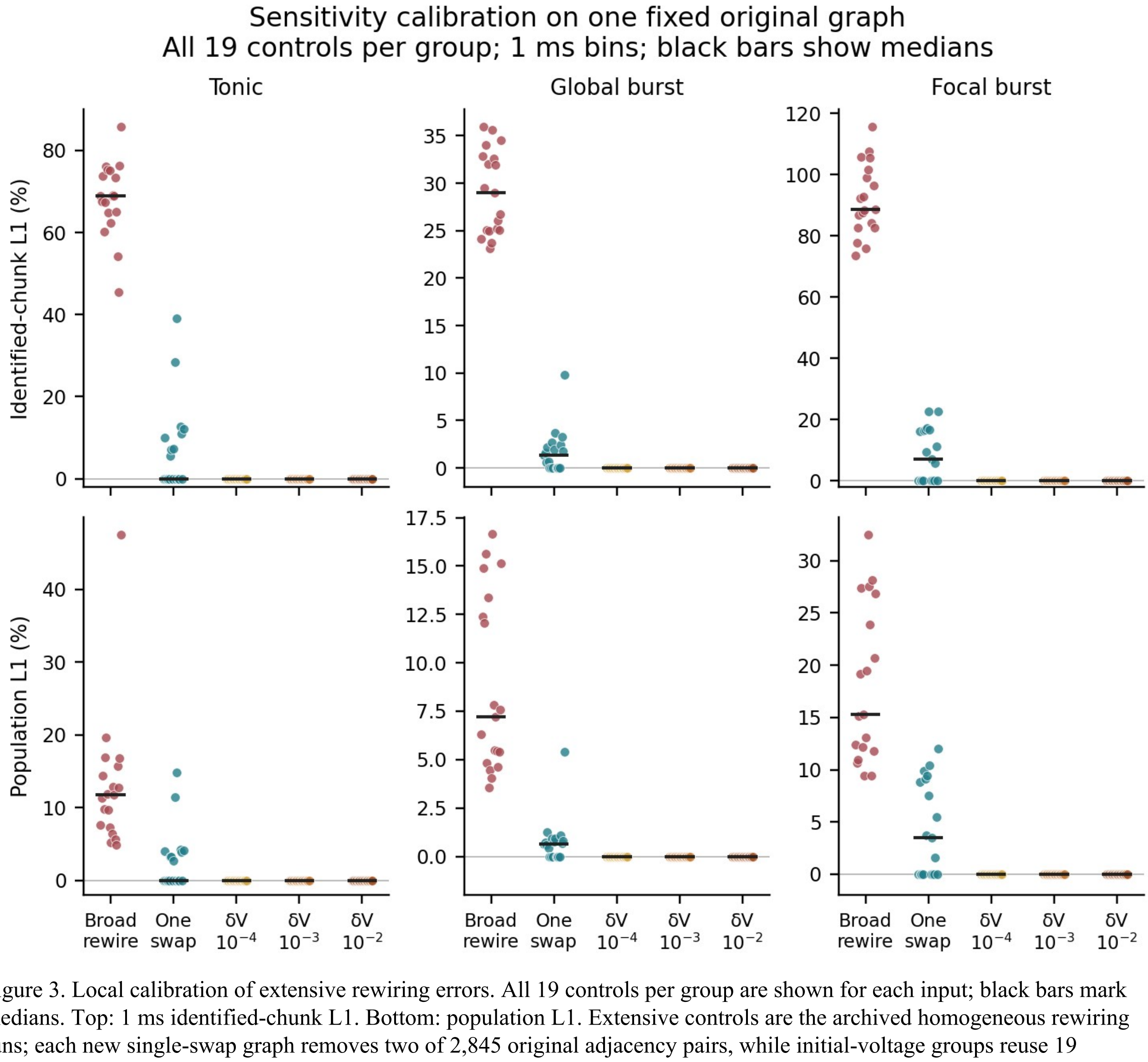

Figure 3. Local calibration of extensive rewiring errors. All 19 controls per group are shown for each input; black bars mark medians. Top: 1 ms identified-chunk L1. Bottom: population L1. Extensive controls are the archived homogeneous rewiring runs; each new single-swap graph removes two of 2,845 original adjacency pairs, while initial-voltage groups reuse 19 directions at the three displayed amplitudes. All comparisons use their input's original explicit trajectory. Points at zero can overlap. The groups are conditional interventions, not independent draws from a uniform graph ensemble.

Single swaps are not uniformly silent. For each input considered separately, seven of the nineteen controls preserve every raw-slot chunk spike count; their identities differ across inputs. The numbers with zero 1 ms binned error are 10, 7 and 8. The additional zero-bin cases contain two to eight mismatching chunk/slot cells. Thus a zero binned score can conceal timing changes. Shared-original-source input is unchanged within the observation window in four of the 57 swap/input cases, and all four reproduce the ten recorded fields bitwise. However, another 17 cases preserve spikes despite a nonzero shared-source input difference.

The input-map diagnostic explains some silent swaps but does not determine nonlinear spike divergence (Figure S6).

All 171 initial-voltage conditions preserve raw-slot chunk spike counts. Their early continuous states differ, but all 125 chunks' ten recorded fields agree with the original at the final slot. This outcome is consistent with the kernel's contraction between equal events, common spike resets and floating-point recoincidence. It is not evidence of general robustness to perturbations at other times or in other state variables. In particular, the separate online pulse experiment in Appendix I.4 can change spike sequences; the initial-voltage control does not contradict that result.

Population aggregation removes a substantial part of the extensive controls' identified-chunk discrepancy: the per-comparison cancellation fraction has medians 83.30%, 75.93% and 82.53% for the three inputs. These are medians of individual ratios, not ratios of the population and chunk medians. Figure S5 shows all defined ratios, and Figure S7 separates cumulative count error from state error over time. The tested small initial changes and single legal swaps do not reach the extensive controls' 1 ms error range. This calibrates the observed magnitudes without identifying an error law per changed edge.

### 4.4 A fixed swap can change its spike effect across inputs

Among the 19 fixed swap graphs, two preserve every raw-slot chunk spike count under all three inputs, seven differ under all three, and ten switch between equality and difference as the input protocol changes (Figure 4). The intervention itself is fixed within each row: the same endpoints, weights and edit count are used. This paired comparison demonstrates input-dependent visibility of the structural change in the measured spikes. It does not isolate spatial localization, because the protocols also differ in total drive and time course.

The 57 swap/input cases separate into four categories. Four have zero shared-reference-source input discrepancy within the arrival window and reproduce all ten recorded fields bitwise. Seventeen have a nonzero input discrepancy but preserve every raw-slot spike count. Four change raw-slot spikes while preserving all 1 ms bin counts. The remaining 32 change the 1 ms counts. These categories are measured outcomes, not causal stages that every trajectory must pass through (Table 5).

Table 5. Four observed outcomes across 57 swap/input comparisons. Input equality uses common original-source activity and arrivals within the window; state equality covers the ten recorded fields.

| **Observed category** | **Comparisons** |
|---|---|
| Shared-source input equal; recorded states bitwise equal | 4 |
| Shared-source input changed; raw-slot spikes equal | 17 |
| Raw-slot spikes changed; 1 ms binned counts equal | 4 |
| 1 ms binned counts also changed | 32 |

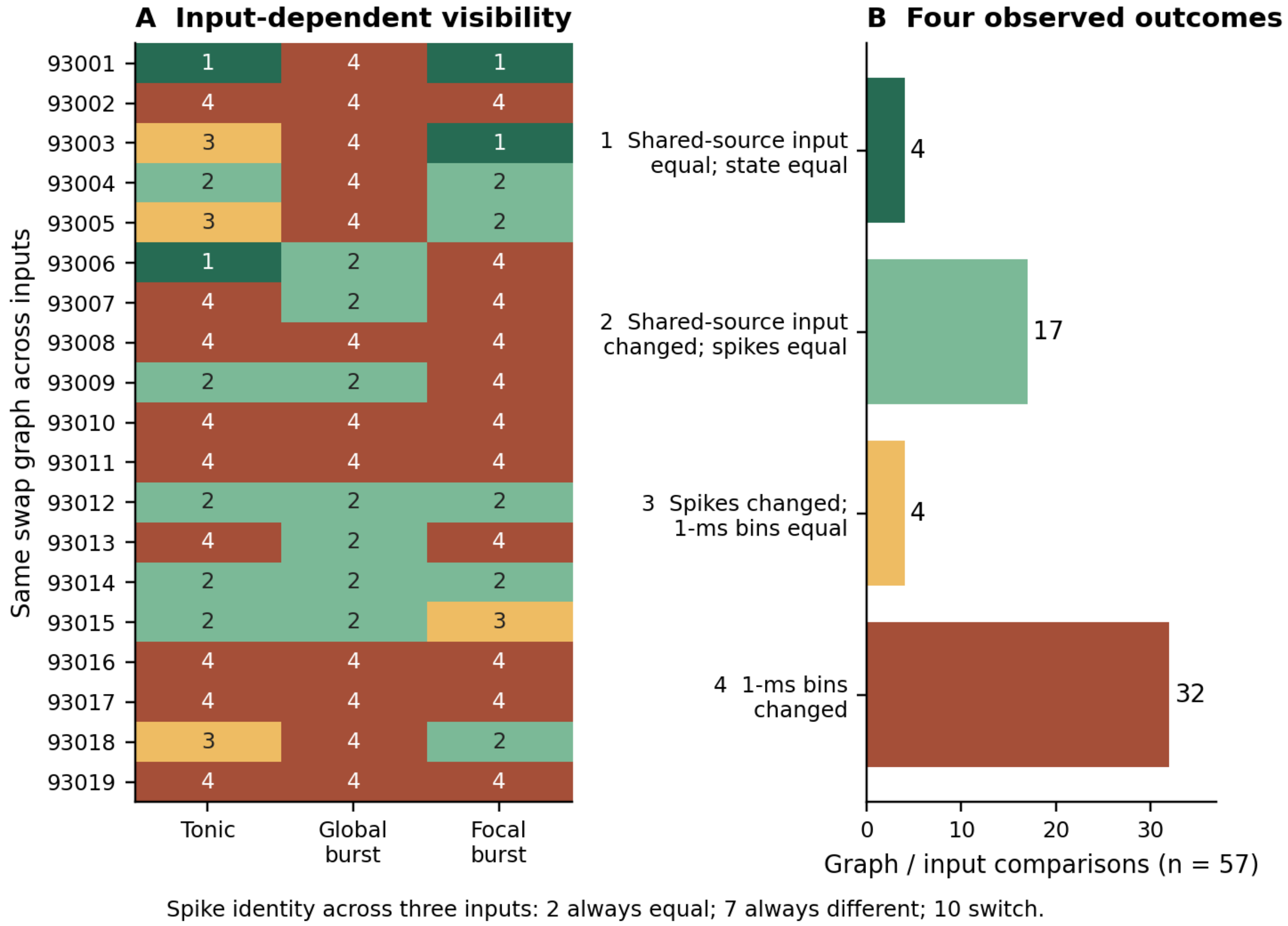

Figure 4. Input-conditioned outcomes of the same single-edge-pair swaps. Each row is one fixed swap graph; columns are the three inputs. The four mutually exclusive colors separate zero shared-input discrepancy with recorded state agreement, changed input with equal raw-slot spikes, changed spikes with equal 1 ms bins, and changed 1 ms counts. Two graphs are spike-identical under every input, seven differ under every input and ten switch. Input-zero labels refer to the archived in-window-arrival diagnostic, not a proof of the complete finite-horizon hypotheses. All classifications are post hoc.

One graph, seed 93006, illustrates all three levels across inputs. Under tonic drive, the shared-source input diagnostic is zero and the recorded states agree. Under global burst, the input diagnostic is nonzero but spikes agree. Under focal burst, 1,268 chunk/slot count cells differ. The paired input contrast thus reaches beyond a change in numerical error magnitude: the same graph modification can be input-invisible, input-visible but spike-invisible, or spike-visible in the tested protocols.

Corollary 1 provides a sufficient explanation for unchanged trajectories when the complete exact-arithmetic hypotheses hold; the source-pair identity makes the role of activity explicit. The empirical diagnostic uses the observed floating-point source sequence and only in-window arrivals. Recorded state equality is checked separately. Conversely, the 17 cases with changed inputs and unchanged spikes show why a nonzero input-map discrepancy is insufficient to predict spike divergence.

### 4.5 Visibility of fixed swaps under matched inputs

Matching global and focal external drive at every slot leaves 9/19 visibility switches under the original burst and 6/19 under the common rearranged schedule. Preserving each node's amplitude multiset while rearranging time gives 4/19 switches globally and 1/19 under matched focal drive. The matched-focal comparison retains the same binary visibility label for 18/19 swaps. Its one counterexample rules out exact determination by the amplitude multiset, but does not establish frequent switching; unchanged labels do not imply unchanged trajectories. Increasing focal amplitude at fixed support gives 3/19 switches (Table 6; Figure 5, panel A). Each comparison uses the same fixed swap and its own input's original reference. Visibility switching therefore does not require a difference in total drive, and per-node amplitude multisets do not fully determine it. These controls do not quantify how much total drive contributed to the original comparison. Output activity is not matched: original total spike counts for G_B, F_H, G_T and F_HT are 362,200; 280,500; 336,900; and 372,400, respectively.

Table 6. Paired spike visibility under controlled inputs. Same and different refer to each swap versus its own input reference. All rows include nineteen fixed swaps.

| Inputs | Same/same | Diff/diff | Same/diff | Diff/same | Switch |
|---|---|---|---|---|---|
| G_B / F_H | 2 | 8 | 5 | 4 | 9 |
| G_T / F_HT | 2 | 11 | 3 | 3 | 6 |
| G_B / G_T | 4 | 11 | 3 | 1 | 4 |
| F_H / F_HT | 5 | 13 | 1 | 0 | 1 |
| F_B / F_H | 5 | 11 | 2 | 1 | 3 |

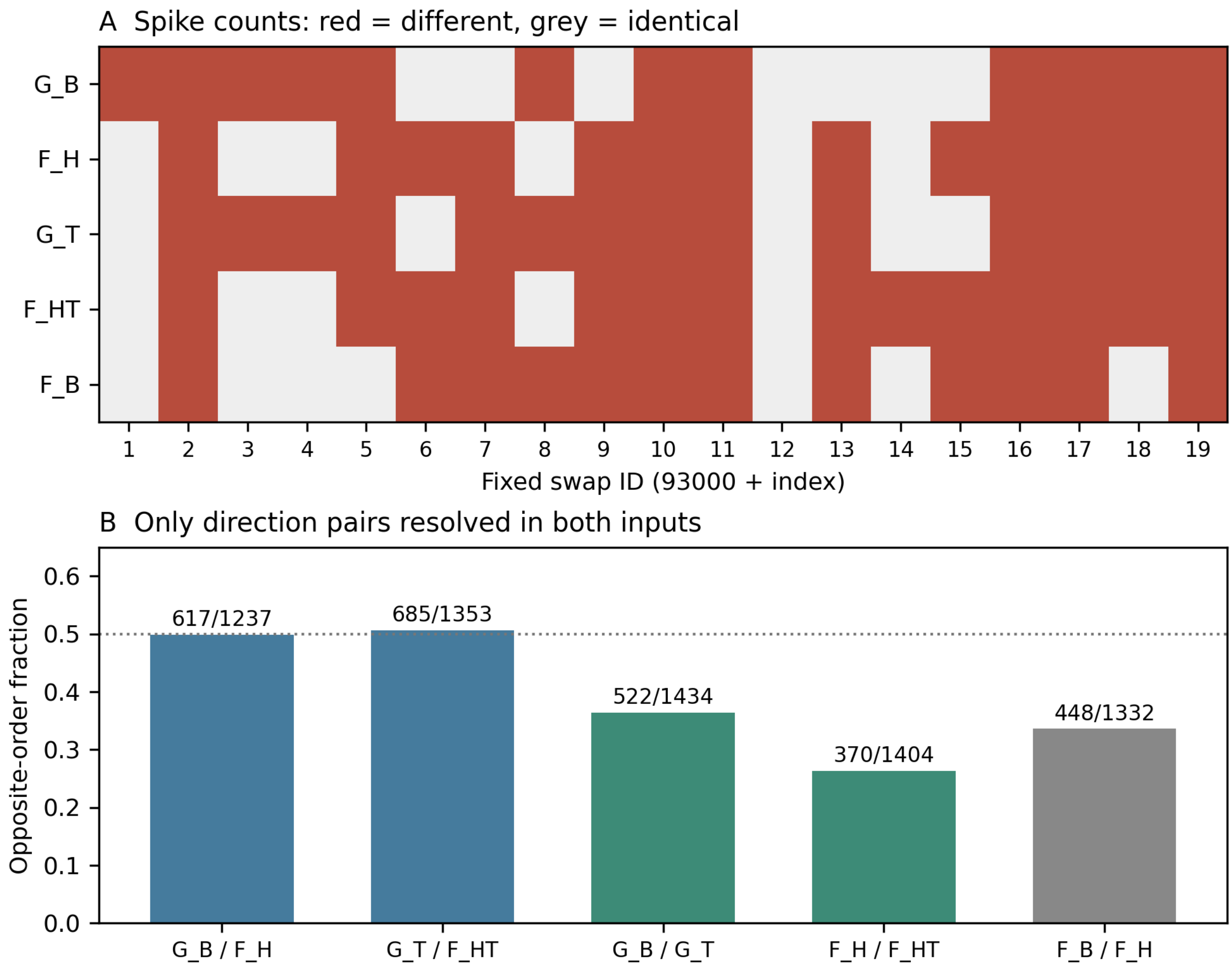

Figure 5. Locked input controls on the same original graph and nineteen fixed swaps. A: spike visibility under each input, compared with its own original reference; labels follow Section 3.7. B: fractions of opposite radius orders among pairs resolved in both inputs, with denominators shown. Global/focal comparisons match summed external amplitude at every slot; temporal comparisons preserve per-node amplitude multisets. Neither comparison isolates a single spatial or timing parameter. All observations share one graph and 30 ms; direction comparisons are dependent.

## 4.6 Visibility across originals and connection rules

Across the ten legacy graphs, G_B/F_H switches occur in 4–13 of 19 fixed swaps per graph (median 9), despite total-drive equality at every slot. After temporal rearrangement the matched-drive contrast G_T/F_HT gives 6–11 switches. Global temporal rearrangement gives 2–8; matched focal rearrangement gives 0–4. Ring seed 303 has 0/19 focal switches and therefore fails the preset all-graphs switching criterion for that contrast. Seed 99's 1/19 focal result is a counterexample to exact sufficiency, not evidence of frequent switching. Table 7 and Figure 6 (left) preserve graph-level counts rather than pooling the swaps into an independent sample.

Table 7. Visibility switches out of nineteen fixed swaps on each original. The first row is discovery; the other nine originals were prospectively fixed. A zero is retained, not replaced by resampling. Counts depend on marginal visibility rates and do not rank input dependence across columns or test excess switching above independence.

| Seed | Rule | GB/FH | GT/FHT | GB/GT | FH/FHT | FB/FH |
|---|---|---|---|---|---|---|
| 99 | Independent | 9 | 6 | 4 | 1 | 3 |
| 100 | Independent | 4 | 8 | 3 | 3 | 4 |
| 101 | Independent | 10 | 9 | 5 | 2 | 4 |
| 102 | Independent | 7 | 6 | 3 | 4 | 6 |
| 201 | Module | 8 | 8 | 4 | 4 | 2 |
| 202 | Module | 13 | 6 | 8 | 1 | 5 |
| 203 | Module | 8 | 10 | 7 | 1 | 2 |
| 301 | Ring | 11 | 11 | 2 | 2 | 6 |
| 302 | Ring | 9 | 8 | 4 | 1 | 7 |
| 303 | Ring | 11 | 9 | 4 | 0 | 4 |

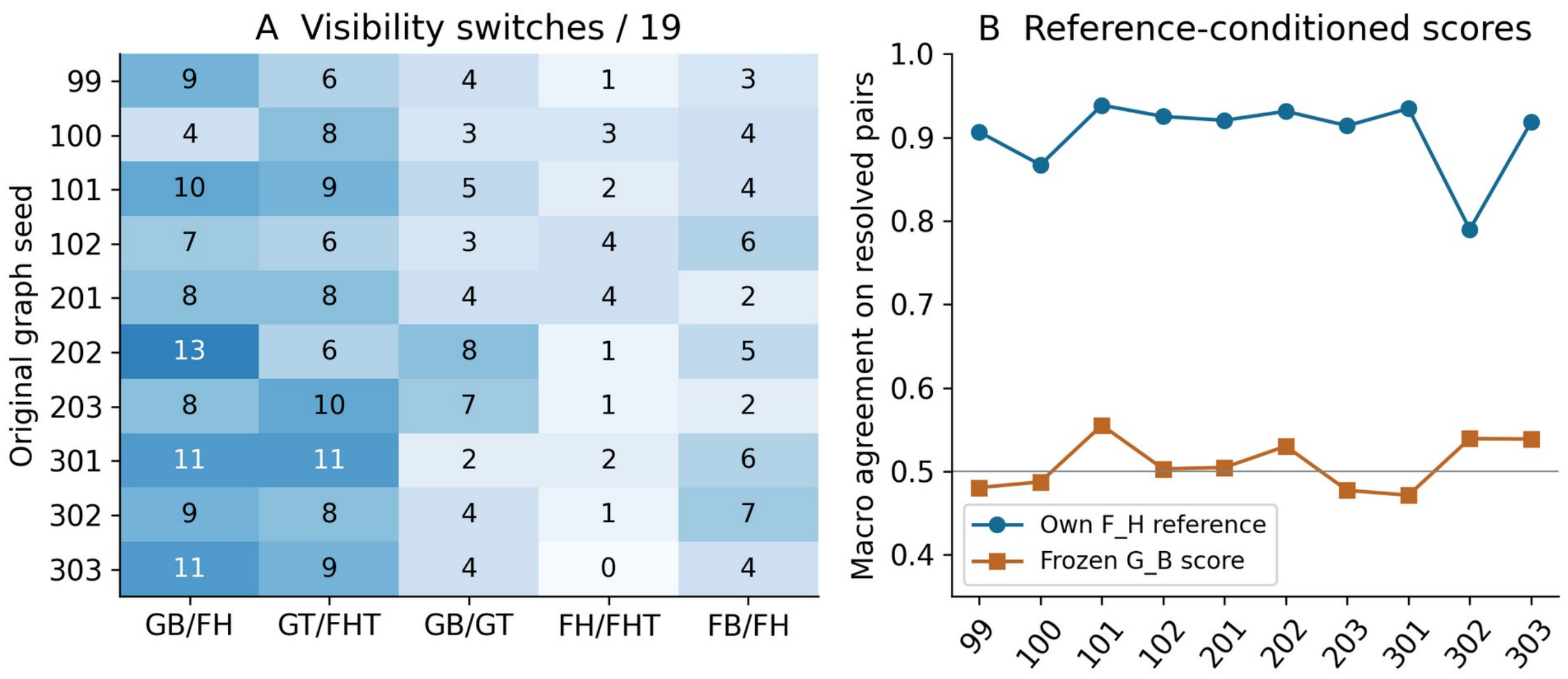

Figure 6. Graph-level replication in the legacy kernel. Left: switch counts for five controlled-input contrasts. Right: F_H direction-order agreement from its own reference versus frozen G_B scores. Points are graph summaries, not independent direction observations; missing or overlapping brackets remain in the reported coverage denominator.

### 4.7 Heterogeneous LIF replication and saturation of binary visibility

The paired LIF validation contains 600 scientific 300 ms network trajectories: three reused graphs, two neuronal parameter variants, five inputs and 20 graph conditions. In the heterogeneous model, 30 ms G_B/F_H switches are 7/19, 8/19 and 9/19 for seeds 100/201/301. At 300 ms, 116–125 of 125 chunks contain unequal unit spike trains, 401–500 of 500 neurons have fired, and mean rates are 21.59–40.36 Hz. The last 100 ms still contains 1069–1934 spikes. Thus the short-window counterexamples survive genuine within-chunk desynchronization and do not arise from a silent tail. Focal temporal rearrangement remains conditional: seed 201 already has 0/19 switches at 30 ms.

For heterogeneous LIF at 300 ms, every tested swap under every input differs from its own reference. Consequently all five input contrasts have zero visibility switches. This is saturation of the equality label:

both sides have become different, not mutually equal. The homogeneous controls still show 2/1/1 G_B/F_H switches across the three graphs. Neither switching nor its disappearance alone describes the error magnitude. Figure 7 (left) shows the complete equality-label composition; the right panel is discussed in Section 4.12.

The complete five-contrast homogeneous/heterogeneous counts are reported in Appendix N. At 30 ms the heterogeneous focal-time contrast is already zero on seed201, while the homogeneous G_B/F_H comparison retains 2/1/1 switches at 300 ms across seeds100/201/301. Thus parameter heterogeneity does not simply increase every input-switch count. Per-unit and chunk-count readouts are both retained; this batch has no case in which unit spikes differ but all chunk slot counts remain equal.

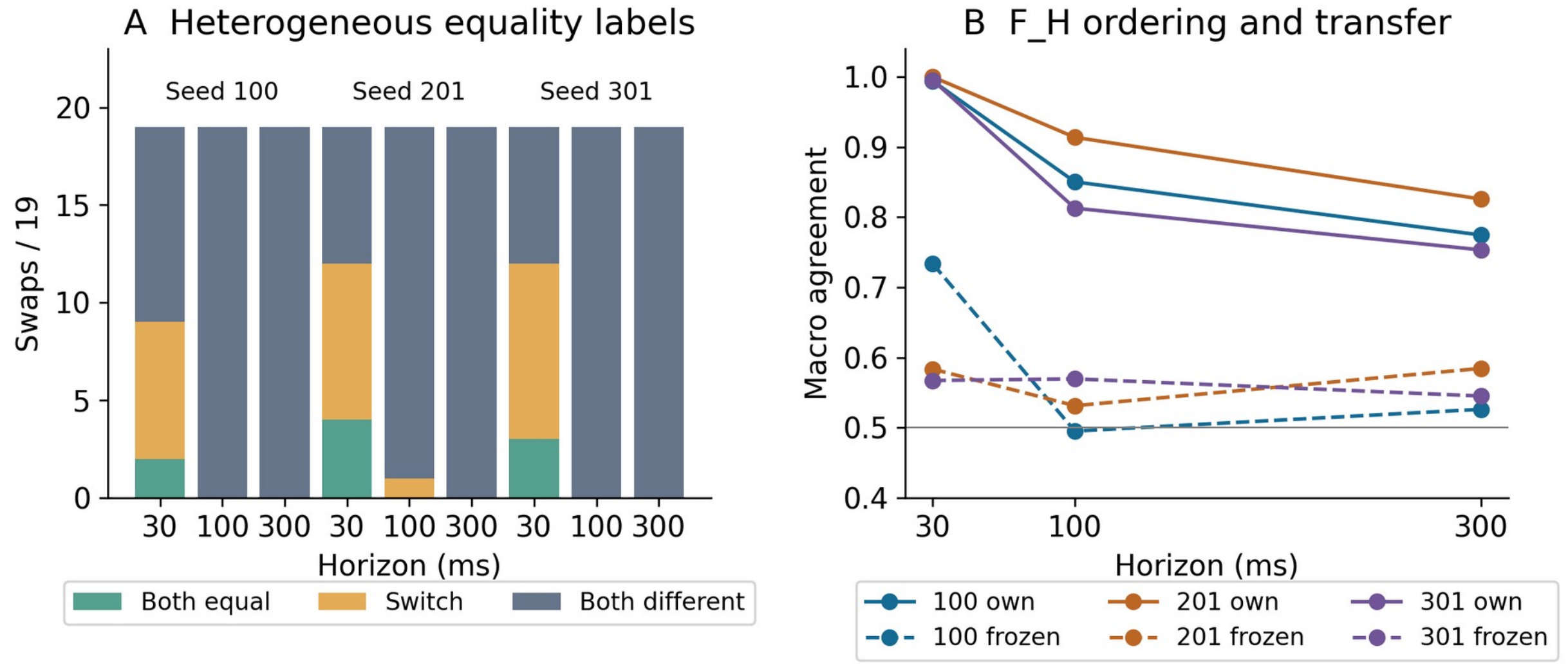

Figure 7. Heterogeneous LIF under nested horizons. Left: G_B/F_H swaps classified as equal under both inputs, switching equality label, or different under both. All 300 ms cases occupy the last category. Right: own F_H and frozen G_B score agreement; the same reference trajectory supplies each horizon prefix, and Section 4.12 gives coverage. Connecting lines summarize nested readouts, not independent time samples.

### 4.8 Nonzero perturbations can be certified without state equality

Certification first concerns singleton swap cases in the legacy kernel and then separately defined fixed-support weight families; these interventions do not share all structural constraints.

The singleton audit certifies 16 of the 17 previously observed input-changed, spike-equal cases, all four input-equal controls and none of the 36 spike-different cases. The remaining target case, swap93005 under focal input, wakes target 24 at slot 4 rather than reference slot 7. Its unchanged spikes do not satisfy the stronger execution-branch contract. Fourteen of the sixteen certified target cases have affected-target firing; continuous voltage differences may nevertheless reach about 662.84 uncalibrated units. This is spike and branch preservation, not state closeness.

The fixed-support family prototype gives all sixteen target cases nonzero one- and two-parameter certified regions. The fixed five-point and 25-point grids witness respectively five and 25 distinct binary32 weight vectors and event streams in each passing nonzero region; the proof covers the whole region rather than those samples alone. Four target cases per dimension reach the 25% cap, two of which have silent affected targets under focal input. The optimized singleton certificate remains slower than native local replay. In two dimensions the support certificate is cheaper than the implemented rigorous corner envelope, but 25-point replay remains cheaper with its weaker coverage. Appendix J retains the measured costs and failure semantics.

## 4.9 Regional coverage and actual failures over longer horizons

All 57 original-centred reference conditions match their archived states bitwise. The 912 direction/input conditions require 9,133 radius queries: 652 reach the width tolerance, 98 retain proof gaps and 162 certify to the 25% cap. Every condition has a positive certified lower bound; all 750 finite upper bounds have actual local-replay threshold mismatches. Every lower bound is rechecked. Final-witness prefixes agree bitwise with independent Python replay over 1,787,790 state endpoints; 25 prescribed lower-bound conditions pass 90,000 exact-rational endpoint checks. A separate-process repeat reproduces all nontiming results.

All 1,520 conditions have positive certified lower bounds: 1,049 reach the width tolerance, 230 stop at proof gaps, 239 certify through the 25% cap, and two exhaust the eight refinement steps. No extra refinement is added. All 100 network conditions reproduce 37,500,000 recorded state words exactly in a second run; 40 old network conditions and 608 old radius conditions bridge exactly. Independent Python replay checks 3,169,794 final-witness state endpoints; nine prescribed exact-rational checks cover 5,400 state rows, each with three two-sided coordinate intervals. Appendix K gives stopping and exclusion details. The control protocol was fixed before these new input outcomes; it remains a conditional test on the existing graph, not independent model validation.

Across the ten originals, the tested fixed-support conditions have 14,848 positive lower bounds out of 14,880. The 32 missing positive bounds are zero-radius enclosure failures in two sixteen-direction groups; baseline point replay still agrees. They are not evidence that the unperturbed dynamics fail. The six affinity graphs include 800 silent-target conditions among 8,960 direction/input conditions. Silence and target exclusions are retained, and the original seed-99 16/17 singleton result is not reinterpreted as a multi-graph swap-certification rate. Appendix L gives stage-specific denominators and preserves the failed all-graph focal-time criterion.

Positive certified radii are obtained in 4308/4320 LIF conditions and each covers at least three actually distinct floating weight/event vectors and endpoint state paths. Table 8 retains cap and refinement censoring. The twelve zero lower bounds all occur in heterogeneous seed 301, F_H, target-pair 18, 300 ms; their brackets are $[0, 2^{-20}]$, stopped at the preset absolute width tolerance with an actual failure upper witness. Their zero-radius certificates pass. These are resolution-limited lower bounds, not proof that no positive safe interval exists.

Table 8. LIF certified interval coverage. Medians include zeros and the 25% cap among 720 dependent conditions per row; Limit denotes the eight-refinement limit.

| Model | ms | Positive | Total | Median % | Cap | Limit |
|---|---|---|---|---|---|---|
| hom | 30 | 720 | 720 | 25.00000 | 518 | 1 |
| hom | 100 | 720 | 720 | 1.85547 | 63 | 2 |
| hom | 300 | 720 | 720 | 0.34485 | 57 | 3 |
| hetero | 30 | 720 | 720 | 4.19922 | 194 | 0 |
| hetero | 100 | 720 | 720 | 0.35629 | 0 | 1 |
| hetero | 300 | 708 | 720 | 0.09880 | 0 | 4 |

The median heterogeneous lower radius decreases from 4.19922% at 30 ms to 0.09880% at 300 ms. Lower bounds alone could reflect proof conservatism. The post hoc paired witness test is stronger: $U_{300}<L_{30}$ holds in 710/720 heterogeneous direction conditions and 630/720 homogeneous conditions. Thus a long-prefix failure lies inside a previously certified short-prefix range in those cases. This proves strict range loss for the corresponding executable family, without fitting a decay law, estimating a universal timescale or attributing the loss to chaos. Figure 8 (left) displays every matched bound, including the conditions that do not satisfy the strict test.

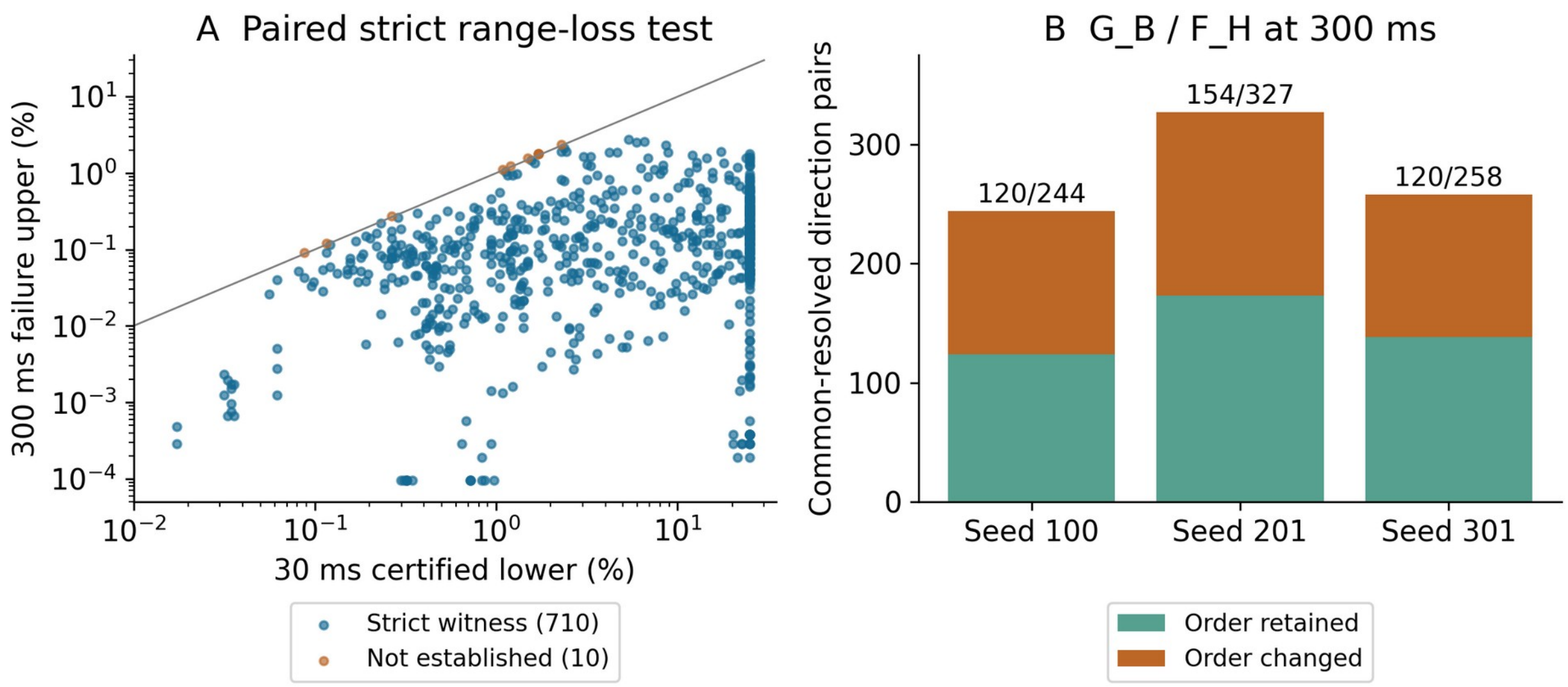

Figure 8. Post hoc tests on archived LIF brackets. Left: short certified lower $L_{30}$ versus long replay-failure upper $U_{300}$ for all 720 heterogeneous direction conditions; 710 lie below the equality line. Right: G_B/F_H direction pairs at 300 ms whose orders are resolved under both inputs, split by retained or changed order. Denominators are 244/327/258 rather than all 360 pairs per graph. A failure upper is a point witness, not an interval-certified boundary.

### 4.10 Equal budget direction order changes across inputs

Table 9 reports macro ordering concordance on the resolved radius pairs.

Table 9. Macro ordering concordance on resolved radius pairs. Comparisons are dependent; full sensitivity is an algebraic ceiling.

| Score | Tonic | Global | Focal |
|---|---|---|---|
| Random order | 0.479 | 0.528 | 0.489 |
| Raw projection | 0.511 | 0.571 | 0.571 |
| Mean activity | 0.672 | 0.654 | 0.724 |
| Independent shifts | 0.557 | 0.457 | 0.579 |
| Temporal margin | 0.908 | 0.881 | 0.911 |
| Full sensitivity | 1.000 | 1.000 | 1.000 |

The simplified temporal-margin score orders resolved pairs more consistently than mean activity, time-shifted activity or raw projection magnitude. Resolved coverage is 1,353/2,280, 1,811/2,280 and 1,624/2,280 direction pairs for tonic, global and focal inputs, with 16, 19 and 16 target pairs contributing to the respective macro averages. Unresolved pairs are not counted as correct. Restriction to firing, nonzero-projection conditions leaves the primary averages unchanged. Full sensitivity reaches 1.000 in each input, an algebraic consistency result rather than independent predictive evidence. The primary score locates the failed target and time within one slot in 577/750 witnesses (76.9%); no matched location-frequency baseline has yet been tested.

Across inputs, resolved orders disagree in 371/1,076 jointly resolved direction pairs for tonic/global, 498/1,008 for tonic/focal and 657/1,325 for global/focal (34.5%, 49.4% and 49.6%). At least one reversal occurs in 16, 14 and 14 of the nineteen target pairs, respectively. The near 50% focal comparisons indicate near decorrelation of ordering, not systematic inversion. Input does not merely rescale every direction equally in these comparisons. A lexicographically selected example, rather than the largest effect, is shown in Figure 9.

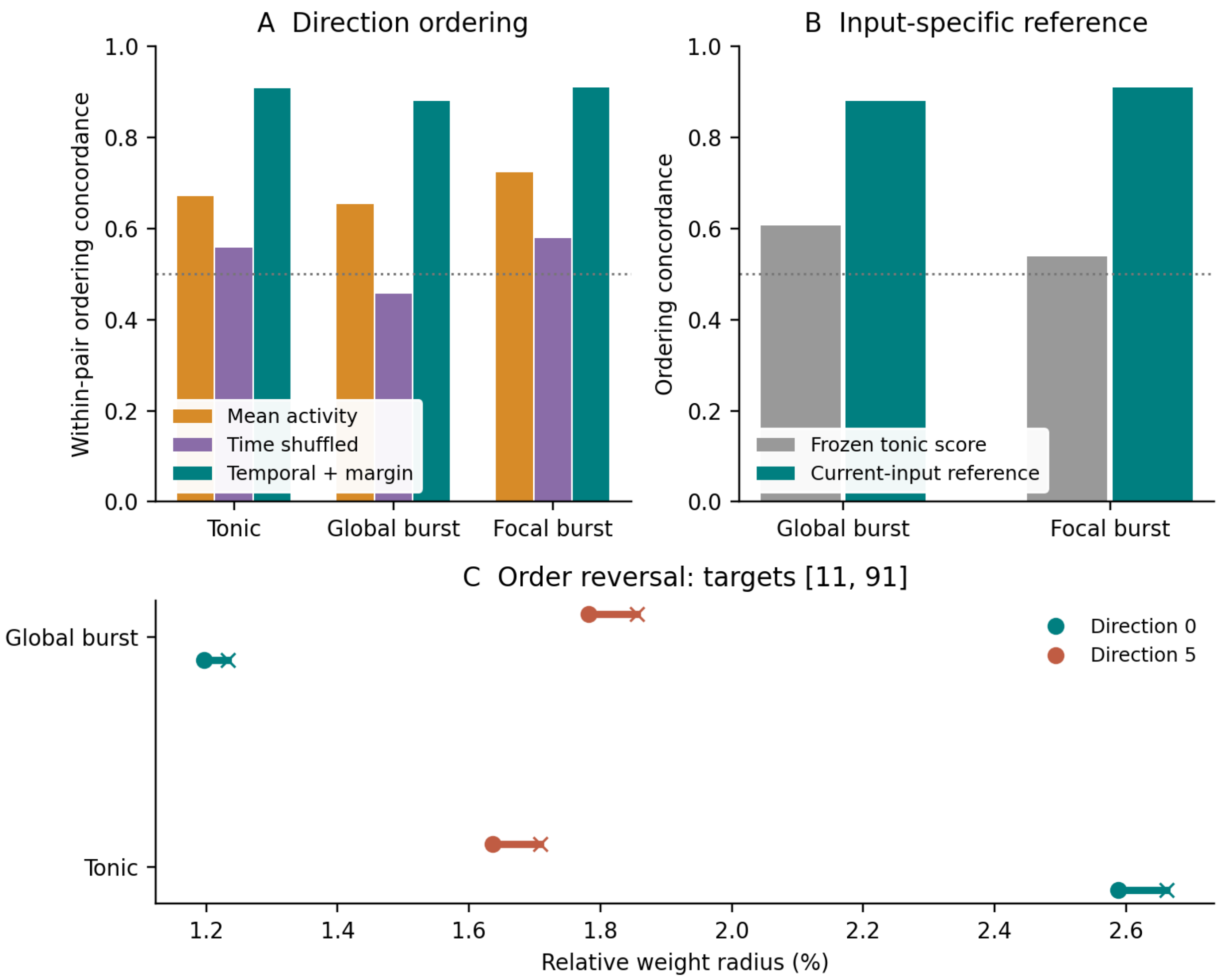

Figure 9. Paired direction safety and input dependence. A: macro ordering concordance on strictly separated radius brackets, not certification success or a confidence interval. B: frozen tonic scores versus the same rule using each input's own complete reference. C: the first lexicographic tonic/global reversal at targets 11 and 91, directions 0 and 5; bars are certified-lower/point-witness-upper bounds in relative weight percent. All panels use dependent comparisons on one original graph. Mean and shuffle in A are the original locked controls; the delivered-window correction is post hoc.

Freezing tonic scores yields macro concordance 0.607 under global input and 0.539 under focal input; recomputing the unchanged scoring rule from each input's own reference yields 0.881 and 0.911. This is reference-conditioned ranking, not input-blind transfer. Actual tonic-selected directions show the negative transfer results in Table 10. In particular, relative to mean selection, focal input gives three wins, ten losses, four operational ties and two unresolved comparisons. This supports input-specific direction choice, not a universal superior direction.

Table 10. Radius comparisons for the fixed tonic temporal choice against three controls. Ties denote identical directions or joint cap certification; overlap is unresolved.

| Input | Control | Win | Loss | Tie | Unresolved |
|---|---|---|---|---|---|
| Tonic | Random | 12 | 2 | 5 | 0 |
| Tonic | Mean | 10 | 2 | 6 | 1 |
| Tonic | Shuffle | 10 | 1 | 6 | 2 |
| Global | Random | 9 | 3 | 3 | 4 |
| Global | Mean | 7 | 6 | 3 | 3 |
| Global | Shuffle | 9 | 4 | 3 | 3 |
| Focal | Random | 5 | 8 | 4 | 2 |

| Input | Control | Win | Loss | Tie | Unresolved |
|---|---|---|---|---|---|
| Focal | Mean | 3 | 10 | 4 | 2 |
| Focal | Shuffle | 7 | 6 | 4 | 2 |

A post hoc score-only check addresses a two-slot boundary in the original controls. Full-window circular shifts preserve source counts over 300 slots but can change the counts arriving within the observation window; 10, 28 and 8 sources are affected, by at most one chunk firing each. Restricting mean and shift controls to the first 298 source slots preserves delivered counts. Corrected mean concordance is 0.673/0.654/0.725 and corrected shuffle concordance is 0.546/0.467/0.586. The primary ranking advantage and focal transfer failure against mean selection remain. No new perturbation run is added, and these corrected controls do not replace the original locked results.

A post hoc 2×2 factor audit uses the same eligible-slot mask and resolved direction set for every method (Table 11). Joint filtering and margin normalization gives 0.908/0.881/0.911, whereas filtering alone gives 0.587/0.588/0.643 and margin normalization alone 0.527/0.625/0.576. The gain supports complementarity within this score family, not a universal effect of threshold margins alone or a causal mechanism intervention. Mean and shuffled activity are not margin-free controls.

Table 11. Post hoc score-factor ablation on identical eligible slots and resolved pairs. Macro concordance; mean and shuffle rows retain the original locked full-window definitions.

| Score | Tonic | Global | Focal |
|---|---|---|---|
| Neither factor | 0.529 | 0.574 | 0.576 |
| Filter only | 0.587 | 0.588 | 0.643 |
| Margin only | 0.527 | 0.625 | 0.576 |
| Filter + margin | 0.908 | 0.881 | 0.911 |
| Mean + both | 0.672 | 0.654 | 0.724 |
| Shuffled + both | 0.557 | 0.457 | 0.579 |

Ordering concordance also depends on guaranteed separation. For g in (1,1.1], joint-score macro concordance is 0.735/0.717/0.709; for $g > 2$ it is 0.970/0.935/0.989. Small g gives only a weak lower bound on separation, not evidence of truly adjacent radii. Unknown orders include both capped and finite-overlap cases. Loose all-comparison score-completion ranges are [0.575, 0.913], [0.699, 0.886] and [0.664, 0.933]; these are not confidence intervals or accuracy estimates and need not be jointly realizable rankings. Appendix K supplies the coverage categories, denominators and excluded targets.

### 4.11 Direction ordering under controlled inputs and across originals

For weight directions, global/focal matched-drive comparisons disagree in 617/1,237 and 685/1,353 jointly resolved orders (49.88% and 50.63%). Time rearrangement changes 522/1,434 global and 370/1,404 matched-focal orders (36.40% and 26.35%; Figure 5, panel B). These are dependent direction comparisons; near 50% indicates near decorrelation rather than a universal reversal. Spike visibility of endpoint swaps and safety order of fixed-support weight changes remain different interventions.

Using each new input's reference gives joint-score macro concordance 0.906/0.923/0.940 for F_H/G_T/F_HT; frozen G_B scores give 0.480/0.601/0.517 on the same resolved sets. Coverage is 67.06%, 76.32% and 71.67%. Single-factor scores and the matched mean/shuffle controls remain weaker on average (Table 12). For the small guaranteed-separation bin, joint concordance is 0.662/0.726/0.714, so the overall result does not extend to all difficult comparisons. G_B-selected directions also fail to dominate all controls after transfer: against the frozen shuffled choices, F_H and F_HT each yield five wins and eight losses, with six other outcomes. The complete transfer table retains same-direction ties, joint caps and unresolved cases separately in the deposited CSV.

Table 12. Reference-conditioned factor scores under the controlled inputs. Target-level macro concordance on the same resolved sets; controls use the delivered source window.

| Score | G_B | F_B | F_H | G_T | F_HT |
|---|---|---|---|---|---|
| Neither | 0.574 | 0.576 | 0.570 | 0.537 | 0.590 |
| Filter only | 0.588 | 0.643 | 0.666 | 0.569 | 0.677 |
| Margin only | 0.625 | 0.576 | 0.606 | 0.567 | 0.636 |
| Joint score | 0.881 | 0.911 | 0.906 | 0.923 | 0.940 |
| Mean control | 0.654 | 0.725 | 0.688 | 0.671 | 0.721 |
| Shuffle control | 0.467 | 0.586 | 0.533 | 0.512 | 0.673 |
| Full sensitivity | 1.000 | 1.000 | 1.000 | 1.000 | 1.000 |
| Coverage % | 79.43 | 71.23 | 67.06 | 76.32 | 71.67 |
| Target pairs | 19 | 16 | 16 | 18 | 16 |

Legacy F_H scoring from its own reference yields graph-level macro agreement 0.790–0.939 across the ten graphs, versus 0.471–0.555 for frozen G_B scores (Figure 6, right). The result concerns ordering on resolvable brackets.

### 4.12 Horizon limits of LIF direction ordering and scores

A direct post hoc comparison of the certified brackets distinguishes order changes from predictor failure. At 300 ms the heterogeneous G_B/F_H comparison resolves both orders for 829 of 1080 possible direction pairs; 394 of those 829 pairs change order. The graph-specific counts are 120/244, 154/327 and 120/258 (Figure 8, right). These are shared-target, shared-direction descriptive counts, not 829 independent replications. They establish concrete changes of direction order under matched total drive, without interpreting a proportion near one half as systematic reversal or as a statistical equivalence to random order.

Input-conditioned F_H joint-score macro agreement is 0.994–1.000 at 30 ms and 0.753–0.826 at 300 ms (Figure 7, right); G_B-frozen scores give 0.567–0.734 and 0.526–0.585, respectively, on the same resolved sets (Table 13). Short-window frozen agreement of 0.734 on seed100 prevents an assertion that all transfer is random. Agreement near one is also conditional on resolved coverage: at 30 ms it is 41.7–76.7%, and two graphs have only two contributing target pairs. These results describe ranking with a known unperturbed reference, not prediction without observing the new input.

The joint score does not uniformly outperform activity controls. For heterogeneous F_H at 300 ms, joint/mean/shuffled agreement is 0.775/0.740/0.676 on seed 100, 0.826/0.774/0.748 on seed 201, and 0.753/0.753/0.759 on seed 301. Close radius pairs are harder: pooling resolved pairs descriptively by the guaranteed separation ratio gives joint agreement 0.676, 0.729 and 0.863 in ratios [1, 1.25), [1.25, 2) and [2, ∞), with 136, 321 and 417 pairs. The primary summaries remain target-pair macro averages. Selecting a direction is a distinct task: frozen G_B joint selection versus random at 300 ms F_H gives 4 wins, 3 losses and 2 unresolved outcomes over nine target pairs, not uniform transfer failure.

Longer windows also expose a scoring implementation boundary. In heterogeneous LIF, at least one eligible reference margin falls below the fixed $2^{-20}$ V floor in 15/45 target-pair/input groups at 30 ms, 34/45 at 100 ms and 45/45 at 300 ms. These group counts mean at least one clipped eligible time in each counted group; they do not measure the fraction of clipped times or show that clipping dominates the score. This post hoc diagnostic was not used to retune the score. It prevents attributing all ranking degradation to a physiological mechanism or claiming that the full frozen-branch score must order radii exactly. The certifier uses the actual thresholds, not that scoring floor.

Table 13. Heterogeneous LIF G_B/F_H switches and F_H score agreement. Own and frozen scores use the same resolved pairs; coverage has 360 possible pairs per graph/horizon. Macro n counts contributing target pairs.

| Seed | ms | Switch /19 | Own | Frozen | Cover % | Macro n |
|---|---|---|---|---|---|---|
| 100 | 30 | 7 | 0.994 | 0.734 | 41.7 | 2 |
| 100 | 100 | 0 | 0.850 | 0.495 | 85.6 | 3 |
| 100 | 300 | 0 | 0.775 | 0.526 | 71.9 | 3 |
| 201 | 30 | 8 | 1.000 | 0.584 | 76.7 | 3 |
| 201 | 100 | 1 | 0.914 | 0.531 | 97.2 | 3 |
| 201 | 300 | 0 | 0.826 | 0.585 | 95.6 | 3 |
| 301 | 30 | 9 | 0.995 | 0.567 | 60.3 | 2 |
| 301 | 100 | 0 | 0.813 | 0.570 | 95.6 | 3 |
| 301 | 300 | 0 | 0.753 | 0.545 | 75.3 | 3 |

## 4.13 Local and extensive changes have different temporal error composition

The following temporal-error, dose and selection results concern the seed99 legacy kernel and serve as interpretive controls for the two research questions; they are not new LIF tests.

For the 12 single-swap cases per input with nonzero raw-slot error, the median fractions $D_{count}/D_{slot}$ are 4.34%, 5.93% and 12.66% for tonic, global-burst and focal-burst inputs. Across all 19 extensive controls per input, the corresponding medians are 56.77%, 58.36% and 52.55% (Figure 10). Thus, in each input, most of the typical single-swap discrepancy cancels during time accumulation within the same identified node, whereas more than half of the extensive discrepancy remains in identified-node cumulative counts. The difference in these measured fractions is not imposed by the triangle inequality, which only bounds each fraction between zero and one (Table 14).

Table 14. Raw-slot error retained in per-node 30 ms counts, $R_{count}$ (%), median (minimum–maximum). Ratios are undefined for zero raw-slot error: 12/19 swap and 19/19 extensive comparisons contribute under each input.

| Input | One swap, n = 12 | Extensive, n = 19 |
|---|---|---|
| Tonic | 4.34 (0.00–46.42) | 56.77 (44.92–67.47) |
| Global burst | 5.93 (1.94–34.99) | 58.36 (48.77–63.43) |
| Focal burst | 12.66 (0.00–21.22) | 52.55 (44.10–60.52) |

The temporal-first decomposition also distinguishes cancellation within nodes from the subsequent cancellation between nodes. It identifies what the readout removes, not the dynamical mechanism that generated the discrepancies. In particular, time cancellation can pair excesses and deficits separated by much of the window; it does not establish small temporal jitter or one-to-one displaced spikes. A lower cumulative-count discrepancy does not imply a close fine trajectory. The comparison is conditional on spike-different cases; the seven spike-identical swaps under each input have undefined composition fractions and remain reported separately.

These are different intervention ensembles. Extensive controls remove 76.3–78.6% of original adjacency pairs, compared with 0.07030% for a single swap, approximately 1,100 times less. Their discrepancy composition therefore does not identify an effect of scale independent of endpoint selection, nor a universal relation between edit budget and error composition.

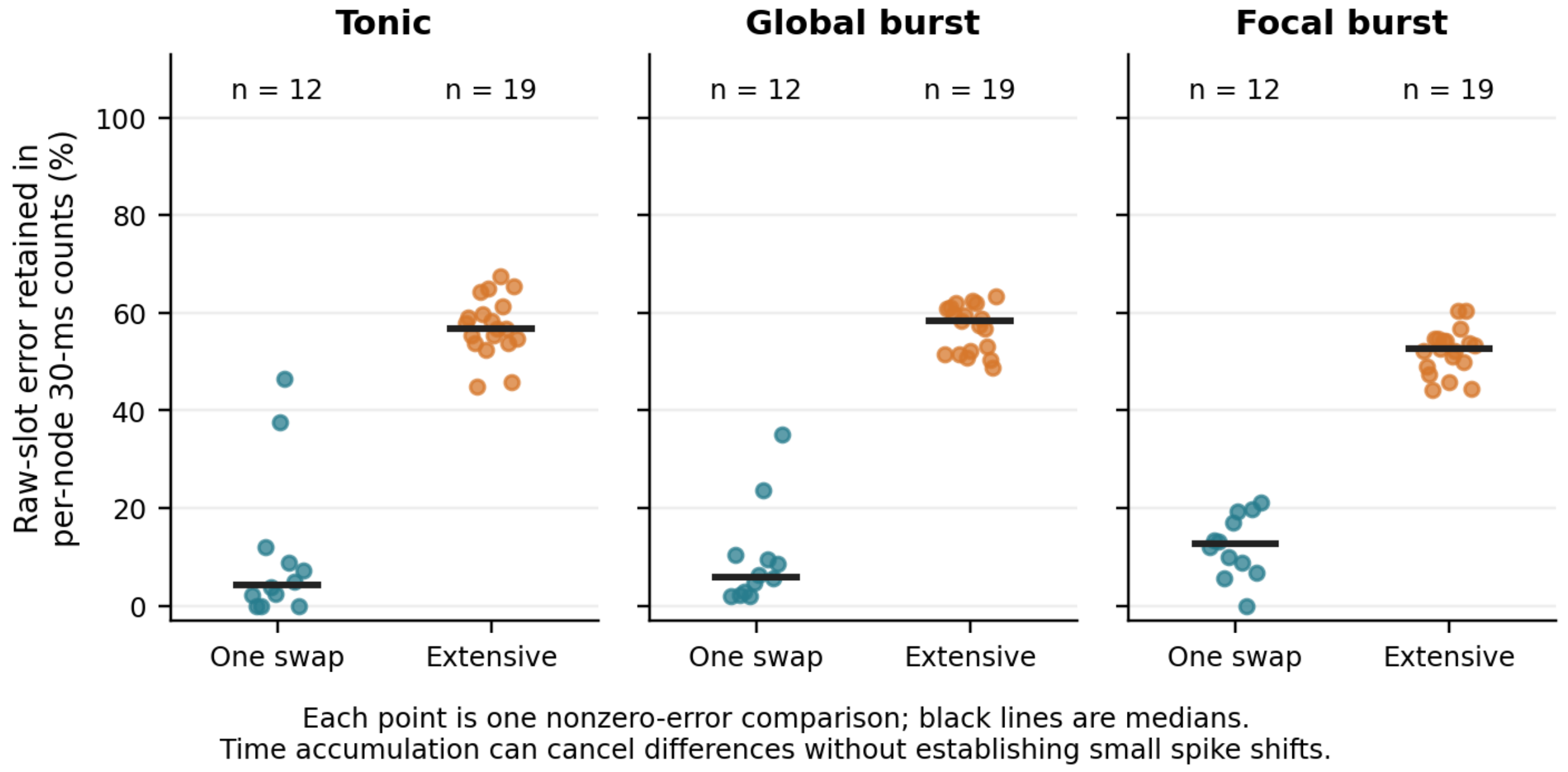

Figure 10. Raw-slot discrepancy retained in each identified node's 30 ms count. Each point is $D_{count}/D_{slot}$ for one nonzero-error comparison; black bars are medians. There are 12 defined single-swap and 19 extensive-control ratios per input. Seven spike-identical swaps per input have undefined fractions and are excluded. This is a temporal-first, finite-window count decomposition: the complement is canceled within nodes across time, not necessarily by short spike-time shifts. Individual temporal, spatial and global fractions are retained in the evidence tables.

A paired post hoc comparison of ten complete random chains supplies intermediate edit budgets without linking unrelated interventions (Table 15). At 0.984%, 1.968% and 9.912% adjacency edits, mean normalized raw-slot discrepancy X is 1.0659, 1.0746 and 1.1463; mean cumulative-count discrepancy Y is 0.2783, 0.2884 and 0.4380. Mean retained fractions are 25.47%, 26.60% and 38.33%. All thirty trajectories have nonzero slot and count error. The lowest tested budget already gives a large slot discrepancy; increasing the budget mainly enlarges cumulative-count discrepancy relative to the modest change in mean slot discrepancy. Normalized L1 is not a correlation coefficient, and one is not a decorrelation or saturation threshold.

Table 15. Ten paired random chains at three nested edit budgets. Values are means; X and Y use the same original spike total. R is the mean of individual ratios, not a ratio of means.

| **Edit %** | **Adjacencies** | **Exchanges** | **X** | **Y** | **R / %** |
|---|---|---|---|---|---|
| 0.984 | 28 | 14 | 1.0659 | 0.2783 | 25.47 |
| 1.968 | 56 | 28 | 1.0746 | 0.2884 | 26.60 |
| 9.912 | 282 | 141 | 1.1463 | 0.4380 | 38.33 |

From the lowest to highest budget, mean Y rises 15.97 percentage points. The exact per-chain product decomposition assigns 2.29 points to the magnitude term and 13.68 to the retained-fraction term. This is not a causal attribution. From 1% to 2%, X decreases in seven chains and increases in three, while Y increases in five and decreases in five. Across all three budgets, X, Y and R are nonmonotonic in seven, six and five chains, respectively. Even at 9.912%, the mean fraction canceled within nodes is 61.67%; cumulative-count bias does not dominate by default. The budgets have overlapping composition ranges and do not support a universal monotonic dose law. Appendix K retains paired changes and all bin-offset readouts.

### 4.14 Readout robustness preserves the identified-node range separation

At the raw 0.1 ms resolution, median single-swap count errors are 0.50%, 9.69% and 40.38%; with the original 1 ms bins, they are 0%, 1.35% and 7.05%. The largest focal-burst single-swap raw-slot error is 98.58%. Minimal edits therefore need not produce a small fine-time trajectory error, even when binning substantially lowers the score.

For every common width and offset in the five-width scan, the smallest extensive identified-chunk error exceeds the largest single-swap error: 1,089 of 1,089 input/width/offset readouts (Figure 11). At 1 ms, the smallest gaps over the ten common offsets are 6.40, 13.14 and 50.00 percentage points for the three inputs. This is an exhaustive check over the specified offsets of the existing trajectories, not 1,089 independent confirmations or evidence for untested inputs. E/I-pooled and globally pooled ranges overlap in most settings; the uniform separation applies to the identified-node readout.

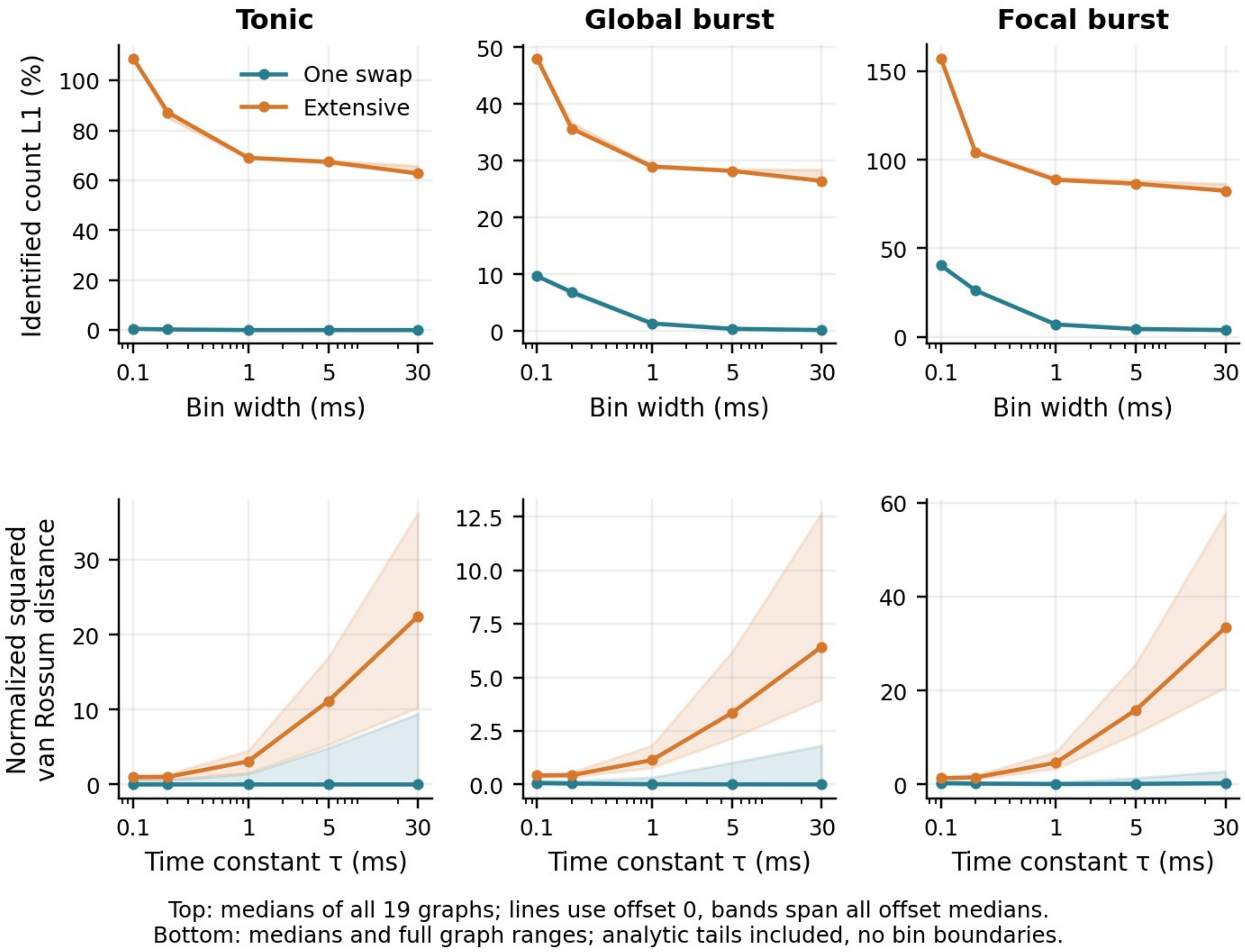

Figure 11. Readout robustness on the same archived trajectories. Top: identified-node count L1 at five bin widths. Lines show the median across all 19 controls at offset zero; shading spans the medians over every slot offset, with partial edge bins retained. It is not a confidence interval or the full graph range. Bottom: median and full graph range of the normalized squared van Rossum score at five time constants, including the analytic filter tails. These values are not L1 percentages. All identified-node bin settings and filter scales preserve the extensive-versus-single-swap range separation; E/I and global readouts, supplied in the evidence tables, usually do not. Repeated readouts are dependent analyses of the same controls.

The bin-free exponential-filter score also preserves strict range separation in all 15 input/time-constant combinations. This makes the observed ordering more than an artifact of the original 1 ms bin origin. It does not make error magnitudes readout-independent, guarantee the ordering under all possible spike distances, or show that temporal cancellation consists of short displacements. The robustness claim concerns the separation of the two sampled intervention ensembles while node identity is retained.

### 4.15 Activity scores and closed loop fidelity at fixed edit budgets

At 9.9121% adjacency replacement, minimizing the coactivity score lowers it relative to the random choice in all ten banks. The paired relative reductions have mean 19.33%, median 19.82% and range 6.00–30.35%. This improvement follows selection from the shared candidate bank; it does not establish a dynamical benefit. Mean primary held-out raw-slot L1 is 1.1161 for random selection and 1.2440 for coactivity selection. The paired difference is +0.1279, or +12.79 percentage points, with eight losses and two wins for coactivity. The corresponding differences against mean-activity and shuffled selection are +0.0324 and +0.0112. Training-input raw-slot error also increases by 0.09255 on average, with eight losses and two wins, so the mismatch is already present under the input used to compute the scores.

The predefined secondary readouts vary with the input. For the split-burst condition on the original target mask, coactivity-minus-random mean differences are −6.23 percentage points for 1 ms L1, −6.68 points for cumulative-count L1, −0.4380 for the normalized squared exponential-filter distance and −0.0776 for standardized state RMS. All four improve in the same seven banks, while raw-slot L1 worsens in eight. Their correlated improvements do not replace the primary outcome. The interpretation of this readout-dependent reversal was formed after the results were observed.

At 0.9842% and 1.9684% edits, paired relative reductions of the coactivity score against random have means 54.87% and 46.90%, respectively, with nine strict reductions and one shared choice at each budget. Training raw-slot L1 improves by mean differences −0.08591 and −0.11382, or −8.59 and −11.38 percentage points; each budget has seven wins, one tie and two losses. Mean coactivity errors are nevertheless 0.9799 and 0.9608, so these are relative improvements at substantial absolute discrepancy (Table 16; Figure 12).

Table 16. Training-input outcomes of activity-guided graph selection. Each budget and rule has ten candidate-bank memberships; overlapping selections remain as ties. J_co is the frozen-reference coactivity score; spike error is normalized identified raw-slot L1. The 9.912% condition is the previously completed result.

| Replaced / % | Selector | Mean J_co | Mean spike error |
|---|---|---|---|
| 0.984 | Random | 808 | 1.0659 |
| 0.984 | Mean activity | 592 | 0.9386 |
| 0.984 | Coactivity | 312 | 0.9799 |
| 0.984 | Shuffled | 314 | 0.9755 |
| 1.968 | Random | 1.81e+03 | 1.0746 |
| 1.968 | Mean activity | 1.25e+03 | 1.0011 |
| 1.968 | Coactivity | 889 | 0.9608 |
| 1.968 | Shuffled | 890 | 0.9897 |
| 9.912 | Random | 9.62e+03 | 1.1463 |
| 9.912 | Mean activity | 8.57e+03 | 1.2714 |
| 9.912 | Coactivity | 7.7e+03 | 1.2388 |
| 9.912 | Shuffled | 7.95e+03 | 1.2534 |

In every row, including the prior 9.912% condition, zero coactivity scores and zero raw-slot spike discrepancies each occur in 0/10 candidate-bank memberships.

The 0.9842% budget has mean paired differences +0.04133 against mean-activity selection and +0.004404 against shuffled selection. It fails the comparator safeguards. The 1.9684% budget passes its individual rule, but the protocol requires both budgets to pass. The joint decision is therefore to stop, and no lower-budget changed-input condition is simulated. Coactivity and shuffled selection choose the same graph in nine of ten banks at each budget; their remaining mean differences are driven by one bank per budget. These comparisons do not isolate a benefit specific to cross-source coactivity. Table 17 reports the paired comparisons against random, mean-activity and shuffled selection, together with the stopping criteria.

Table 17. Paired small-budget differences and stopping decision. Differences are coactivity minus comparator in normalized raw-slot L1, expressed in percentage points (pp); negative favors coactivity. Each row contains ten candidate-bank pairs. Same-graph counts are reported separately; overlapping selections remain among ties.

| Replaced / % | Comparator | Mean / pp | Median / pp | W/T/L | Same graph | Criterion |
|---|---|---|---|---|---|---|
| 0.984 | Random | -8.59 | -11.25 | 7/1/2 | 1/10 | Pass |
| 0.984 | Mean activity | +4.13 | 0.00 | 4/2/4 | 2/10 | Fail |
| 0.984 | Shuffled | +0.44 | 0.00 | 0/9/1 | 9/10 | Fail |
| 1.968 | Random | -11.38 | -13.11 | 7/1/2 | 1/10 | Pass |
| 1.968 | Mean activity | -4.03 | -2.81 | 6/1/3 | 1/10 | Pass |
| 1.968 | Shuffled | -2.89 | 0.00 | 1/9/0 | 9/10 | Pass |
| 0.984% budget: all three comparator conditions required | | | | | | Fail |
| 1.968% budget: all three comparator conditions required | | | | | | Pass |
| Joint continuation: both budgets must pass | | | | | | Stop |

W/T/L uses unrounded normalized-L1 differences Δ before conversion to percentage points and frozen tolerance $\varepsilon = 10^{-12}$: $\Delta < -\varepsilon$, $|\Delta| \leq \varepsilon$, $\Delta > \varepsilon$. Random requires mean and median $< -\varepsilon$ and at least seven wins. Mean activity and shuffled require mean $\leq \varepsilon$ and wins ≥ losses, with no median threshold. Every condition requires ten complete pairs. These are operational continuation criteria, not significance or noninferiority tests; no new migration runs were performed.

None of the 1,280 lower-budget candidates has a zero value for any of the four recorded scores, including the analytic shuffle diagnostic. Across the 80 selector–budget training records, none has zero raw-slot, 1 ms or cumulative-count discrepancy, or exact agreement of all ten recorded fields. The records correspond to 57 unique selected graphs plus one original reference, giving 58 scientific runs and one separate technical repetition. Thus the observed selection gains are not a collection of exact-preservation cases hidden by averaging, and lower-budget transfer remains untested. The three budgets, nested candidate construction and re-selection at each budget do not establish a critical edit fraction or a monotonic error law.

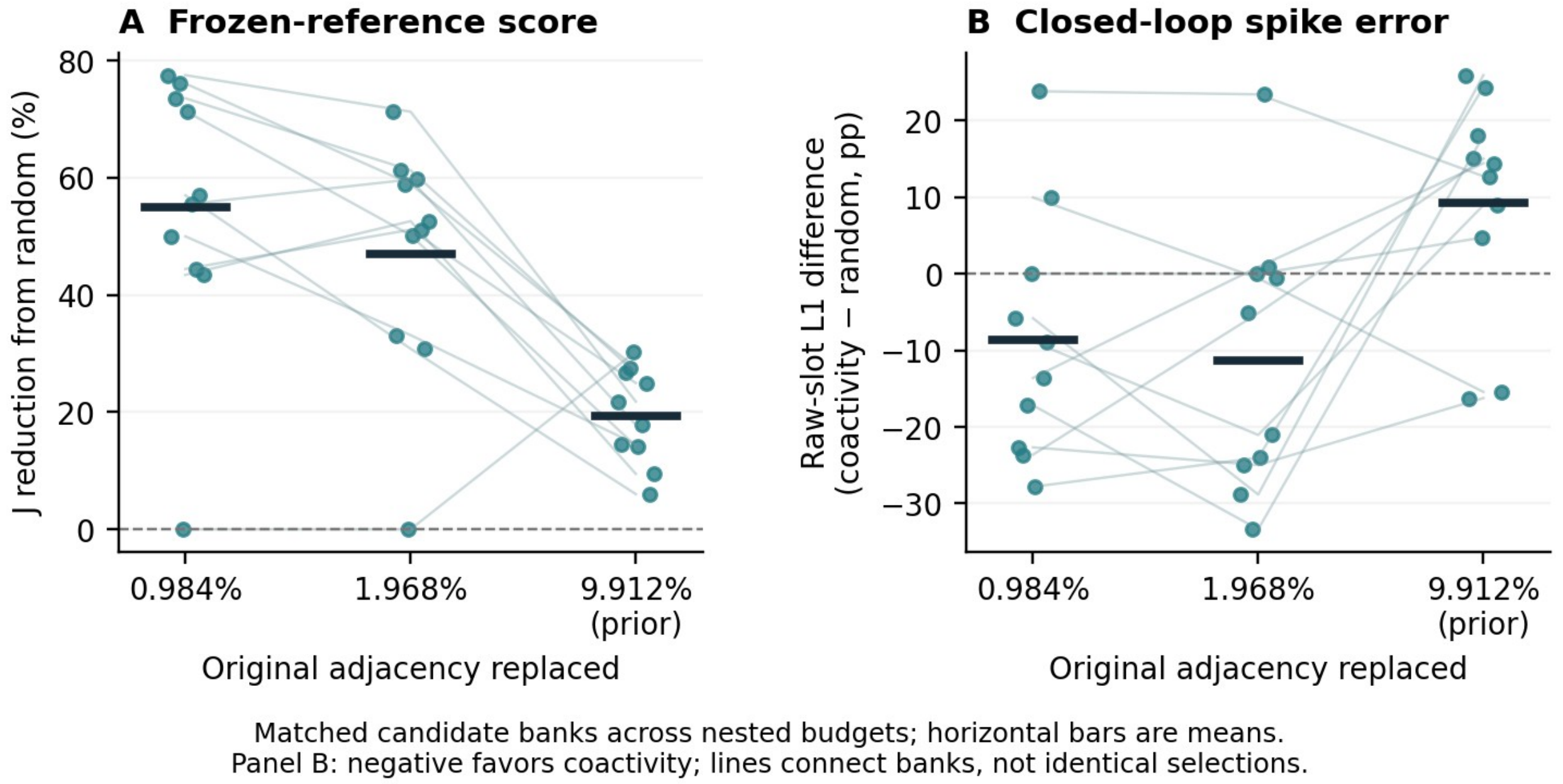

Figure 12. Training input scores and trajectory differences across three adjacency-edit budgets. The left panel gives the paired relative reduction of the coactivity score from random to coactivity selection; the right gives coactivity-minus-random raw-slot identified L1 on the training input. Positive score reduction favors coactivity, whereas positive L1 difference indicates larger trajectory error. Each point is one of ten matched candidate banks on the same original network. Lower budgets use nested prefixes of the archived full-budget exchanges; points across budgets are dependent. The historical full-budget primary endpoint averages three held-out inputs, while this figure uses training input consistently for the budget comparison. Table 17 supplies paired comparisons against all three controls and the stopping criteria.

## 5 Discussion

The controlled-input counterexamples and direction brackets answer different parts of the same question. Matched connection and external-input summaries do not determine the effect of every structural intervention. Conditional on a specified reference, a sound enclosure can nevertheless identify nonzero perturbation families that preserve the finite spike sequence. Input transfer changes some certified direction orders, and longer observation can turn a previously safe perturbation into a witnessed failure. These findings are complementary, but fixed-support weight certification does not fully explain the effects of endpoint swaps.

The original readout and dose results remain important for interpreting this distinction. Within-node time accumulation can cancel much of a local edit's fine-slot discrepancy, and pooled statistics can hide changes at identified nodes. The paired random chains show that increasing edits from roughly 1% to 10% adds relatively little to normalized fine-slot L1 while increasing cumulative-count discrepancy. Because chains are nonmonotonic, the single-swap-to-smallest-dose gap is unsampled, and L1 near one is not a calibrated decorrelation measure, these findings support readout-dependent interpretation rather than a critical edit threshold. The failed selection gate likewise constrains the inference from a low open-loop input score to closed-loop fidelity.

The LIF comparison removes complete within-chunk synchrony and the legacy unbounded numerical voltage scale as necessary conditions for the short-window phenomenon. It also changes several aspects of the kernel; only the homogeneous/heterogeneous LIF pair isolates the specified parameter heterogeneity. Three reused graphs and one fixed LIF configuration do not establish generality over neuronal models, arbitrary heterogeneous connectivity, noise, time discretization or sustained nonperiodic input. The 300 ms prefix is ten times the original observation window, not an asymptotic result. The ten legacy graphs and three LIF graph selections are reported separately because the latter are a subset.

The LIF background is sufficient for isolated-cell tonic firing. Accumulated phase shifts in this operating regime are a possible contributor to long-window equality-label saturation and shrinking certified ranges, but the present comparisons do not isolate that mechanism or establish that background drive dominates the network firing rate. They also do not explain the earlier temporal-error results in the different legacy kernel. The evidence does not cover a fluctuation-driven LIF regime with subthreshold background, and finite-horizon certification does not imply asymptotic stability.

The score is a reference-dependent diagnostic whose filter and margin reflect the same dynamics used by the certificate. Factor comparisons quantify its restricted ordering performance, but neither prove a unique causal mechanism nor establish a runtime advantage over direct certification. Likewise, unsuccessful open-loop graph selection and successful ordering of fixed-support weight directions are different tasks; one does not causally explain the other. Approximation and dose results remain informative controls (Section 4.13 and Appendix H), not evidence that memory proxies compress graphs or accelerate simulation. The score-floor audit motivates a future unit-consistent choice proportional to each unit's threshold-to-reset distance. Such a relative floor would require a new fixed protocol and validation; it is not used here, and neither improved ordering nor elimination of clipping is established by this post hoc proposal.

The certificate contribution is coverage of an explicitly defined parameter family under an auditable execution contract. Numerical validation is necessary for that implementation but does not replace the enclosure proof; its many checked state values are not many independent scientific discoveries. Zero lower bounds, proof gaps, capped radii, unresolved orders and failed selection gates remain separate outcomes. These distinctions make a finite fidelity claim reviewable without assigning every failure the same dynamical interpretation.

The empirical unit of replication must follow the design. There are ten original legacy graphs, with one LIF original per connection rule drawn from that same set. Module and ring generators share a weighted sampler and are not matched in per-node degree; their comparison is not a causal test of geometry. Directions, inputs, graph versions and nested windows are dependent. The LIF validation uses one parameter regime, one timestep, deterministic broadcast projections and a repeated 30 ms drive. Three hundred milliseconds extends the original window tenfold but supplies no stationary or infinite-horizon conclusion. The 710/720 and 394/829 bracket comparisons are post hoc witnesses within these fixed conditions, not population rates or universal laws.

## 6 Conclusion

In the tested graphs, matching total external drive does not determine a fixed swap's finite spike effect. This counterexample survives a paired heterogeneous LIF implementation. Fixed-support weight certificates and replay failures then locate input- and horizon-dependent fidelity ranges: direction order can change with input, and longer prefixes can invalidate perturbations certified for shorter ones. Exact-equality switching and heuristic score agreement both have observable limits. The resulting contribution is a controlled, reproducible account of conditional structural fidelity, with explicit sufficient guarantees and retained negative results.

## Data and code availability

The accompanying reproducibility package contains the bilingual manuscripts, all current source files, numerical tables and figure scripts, the unchanged original experimental release, and the additional graph and brian2-atlas studies. All methods, proofs, principal results and necessary controls are described in this article and its Appendices A–N; consulting a previous manuscript is not required to follow the scientific argument. Archived experiment paths in Appendices A–K are relative to evidence/legacy-release unless a study-specific path is given. Additional studies are under evidence/studies. The current analysis/ directory includes the post hoc horizon witnesses and input-order comparisons with unresolved cases. The default rebuild regenerates analysis tables, figures and documents from archived data; it does not execute a new scientific experiment. Immutable source manifests and earlier frozen manuscript hashes are checked. The frozen complete source-and-evidence package and checksum manifests are archived at https://doi.org/10.5281/zenodo.24089361 under CC BY 4.0. The public repository, current manuscripts and availability-update sources are available at https://github.com/RockLi/structural-fidelity. The archived package preserves the pre-DOI manuscripts; this update changes only availability information.

## Appendix A Reproducible model and workload parameters

All graph probabilities and weights below are ordered (EE, EI, IE, II). Each recurrent workload has 80% excitatory chunks and 20% inhibitory chunks, 100 units per chunk, a stored projection delay of one slot and dt = 0.1 ms. For heterogeneous weights, spread 50% means multiplication by Uniform(0.5, 1.5). Self-chunk EE and II edges are excluded; the excitatory and inhibitory chunk sets are disjoint.

Table S1 lists the graph parameters.

Table S1. Graph parameters. Probability and weight tuples use EE, EI, IE, II order; spread is the fractional variation around the base weight.

| Setting | Connection probabilities p | Base weights w | Seed / spread |
|---|---|---|---|
| E1 / E1 gain scan | (.10, .20, .20, .15) | (.8, .8, 1.2, .6) | 42 / 0% |
| E3 heterogeneous | (.15, .25, .25, .20) | (.6, .8, 1.4, .5) | 200 / 50% |
| E3 homogeneous | (.15, .25, .25, .20) | (.6, .8, 1.4, .5) | 99 / 0% |
| E4 stable | (.10, .20, .20, .15) | (.3, .5, 1.6, .3) | 42 / 0% |
| E4 balanced | (.10, .20, .20, .15) | (.8, .8, 1.2, .6) | 42 / 0% |
| E4 excitable | (.10, .20, .20, .15) | (1.2, 1.0, .8, .4) | 42 / 0% |
| E5 | (.10, .30, .30, .15) | (.3, 1.0, 1.8, .4) | 55 / 0% |

Table S2 specifies the explicit-edge selection rules.

Table S2. Explicit-edge selection. I-source denotes an inhibitory source; an edge in both union components is retained once.

| Setting | Retention predicate |
|---|---|
| E1 / E1 gain scan / E3 homogeneous | I-source w ≥ 1.0; pure E1: none |
| E3 heterogeneous main | I-source w ≥ 1.4 |
| E3 budget sweep | None (0%); I-source w ≥ θ; θ ∈ {2.0, 1.75, 1.4, 1.1, 0.8} |
| E3 edge classes | I-source w ≥ 1.0; I→I w ≥ 0.25; E→I w ≥ 0.70; union of first two |
| E4 stable / balanced / excitable | I-source $w \geq 0.8w_{IE}$ = 1.28 / 0.96 / 0.64 |
| E5 | I-source w ≥ 1.5 |

The E1 medium gain grid is $g_{fE} \in \{1.0, 2.5, 5.0\}$ and $g_{fI} \in \{0.1, 0.4, 0.8\}$. The local grid is $g_{fE} \in \{2.0, 2.5, 3.0\}$ and $g_{fI} \in \{0.30, 0.40, 0.50\}$. Both methods retain their default slow gains and time constants. The center pair (2.5, 0.4) occurs twice. The history output ignores slow gains as specified in Appendix H.1.

All sustained workloads inject amplitude 10 per slot into every excitatory chunk. E5 injects amplitude 150 in slots 0–79 and zero in slots 80–299. Local fields start at zero, receive the post-step update $F_E \leftarrow \exp(-dt/10)F_E + S_k$ and $F_I \leftarrow \exp(-dt/10)F_I$, and are injected before the next active unit step with coefficient 0.01. Initial voltage, threshold, reset and refractory rules are given in Section 2.1.

## Appendix B Additional comparisons

The edge-class policies below use the same heterogeneous E3 realization. The retained fractions are 17.4% for inhibitory-source selection, 3.7% for I-to-I, 13.1% for E-to-I and 21.1% for their specified union. Figures S1, S2 display deterministic parameter comparisons without error bars; the gain points are not independent repetitions.

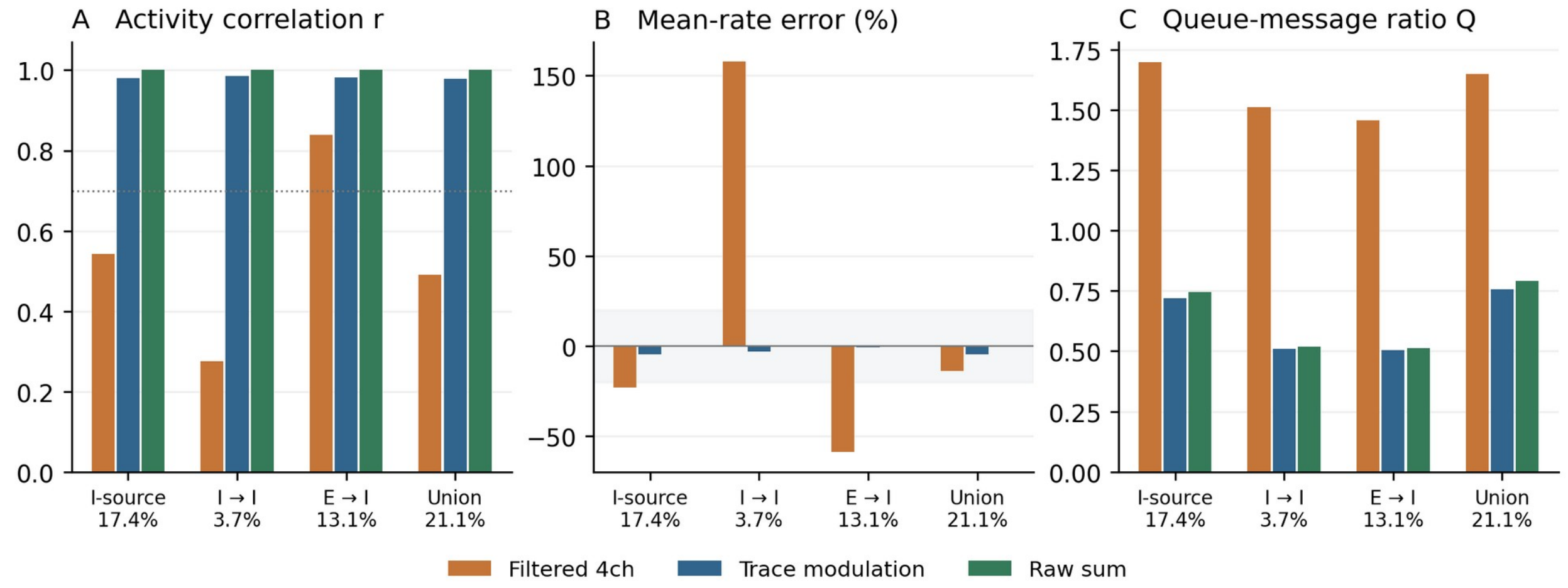

Figure S1. Edge-class selection on heterogeneous E3. Labels give each policy's retained-edge fraction. The union combines inhibitory-source and I-to-I predicates with separate cutoffs. Budgets are unequal, and the predicates do not detect graph cycles. Zero raw-sum rate errors lie on the horizontal axis.

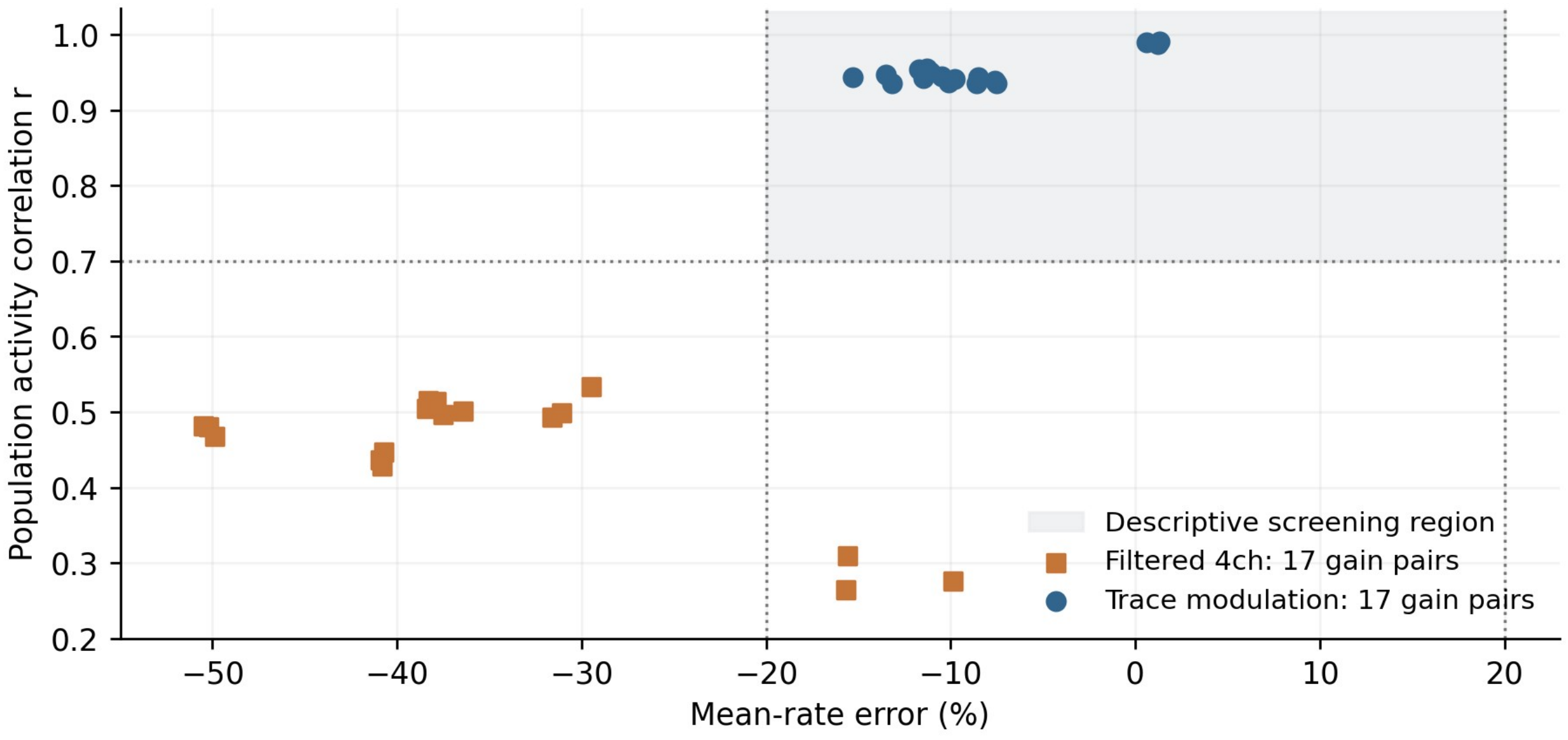

Figure S2. Gain sensitivity in the 12,500-unit E1 workload over 20 ms. Each method has 17 distinct gain pairs from 18 evaluations; the repeated center pair is plotted once. In the history rule, these gains map to $\kappa_E$ = 0.970–1.030 and $\kappa_I$ = 0.9895–1.021; the scans do not match effective input perturbations across methods. The shaded region is the descriptive screening interval r ≥ 0.70 and absolute rate error ≤ 20%. It is not a confidence region.

Table S3 reports the static-gain and history-modulated outputs for each principal setting. Raw-sum output has r = 1, zero mean-rate error and zero population-count mismatches in all eight settings. Ratios are computed from integer queue counts where available.

Table S3. Output-rule controls. Each r / δFR entry reports correlation and percent rate error. Raw-sum population counts agree in every setting.

| Setting | History r / δFR% | κR r / δFR% | Raw Q | Raw mismatches / slots |
|---|---|---|---|---|
| E1 pure | 0.992 / -8.4 | 0.996 / -5.9 | 0.157 | 0 / 500 |
| E1 hybrid | 0.980 / -16.0 | 0.998 / -4.1 | 0.494 | 0 / 500 |
| E3 heterogeneous | 0.983 / -3.5 | 0.991 / -2.3 | 0.672 | 0 / 200 |
| E3 homogeneous | 0.972 / -0.3 | 0.978 / +1.8 | 0.827 | 0 / 300 |
| E4 stable | 0.775 / +3.5 | 0.844 / +3.0 | 0.917 | 0 / 200 |
| E4 balanced | 0.904 / -1.3 | 0.895 / -1.4 | 0.904 | 0 / 200 |
| E4 excitable | 0.559 / -1.0 | 0.625 / -0.4 | 0.642 | 0 / 200 |
| E5 pulse | 0.973 / -0.3 | 0.987 / -0.1 | 0.891 | 0 / 300 |

## Appendix C Source identity and execution record

The frozen source is identified by the 55-file manifest evidence/original/source-sha256.json in the accompanying archive. Its SHA-256 is 9a266203cf86bbe43b201e7af53fab3131c9d89deddbb64d6181f30199a6e448. The source manifest, rather than the enclosing repository revision, identifies the Rust files used here because that directory was untracked. Reproduction left their hashes unchanged.

Validation explicitly selected rustc 1.98.1 (LLVM 22.1.8) on Apple arm64 Darwin. The library and test sources were compiled directly with edition 2024 and optimization level 3, and tests ran serially. The unmodified library and six integration targets yielded 86 passing tests and one ignored 500,000-unit test. An initial run used the host default rustc 1.86.0; both environments and their logs are archived. Rebuilding with 1.98.1 reproduced all 111 diagnostic records, including integer counts and metrics printed to nine decimal places. This is a build and numerical consistency record, not a hardware performance comparison.

The original implementation received measurement instrumentation, while isolated copies implemented the post-hoc raw-sum and static-gain outputs. Delivery remained at n + 2 for spikes emitted in n; current-input emission conditions and internal state updates remained present. The archived reproduction driver verifies the source manifest and builds each variant separately. It records recurrent queue insertions, emitted spikes and differing total-population-count slots.

The E3 zero-retention extension uses a separate additive test and the same frozen source, compiled with rustc 1.98.1. It verifies that all 2,840 edges are aggregated and no recurrent messages use the explicit subpath. Three rerun 2.0% controls reproduce the archived metrics to their recorded precision, including exact available integer counts. Six new records, per-slot counts, output-only patches and validation logs are archived separately from the original 111 records.

The initial static-gain run stopped at an inherited spectral assertion whose analysis passed a timestep expressed in milliseconds to functions expecting seconds. After preserving the failure log, the isolated diagnostic copy replaced that assertion with a logged warning so the remaining time-domain probes could execute. Continuation does not indicate that the spectral assertion passed. Those spectra are excluded from this study; all reported temporal metrics use the per-slot spike counts and the physical timestep stated above.

## Appendix D Delivery fixture specification

All fixtures start with voltage −65, zero input and local-field states, and dormant chunks, except B2, which assigns voltages (−65, −32, 0, 29) within each chunk. Local-field decay uses 10 ms. Event slots are zero-based and all events are enqueued before the first step. B1 and B2 send positive amplitudes (256, 512) to chunk 0 and (128, 256) to chunk 1 at slot 2; correct grouping sums within each target. B3 and B4 send one zero-amplitude event to chunk 0 at slot 2; the candidate drops or preserves it, respectively. B5 and B6 send (256, 512, −256, −512) to chunk 0 at slot 2; the candidate keeps separate polarity sums or replaces them with a zero-amplitude wake-up, respectively. B7 and B8 send (256, 512) at slot 2 and (−128, −128) at slot 4 to chunk 0. B7 preserves those two arrival slots; B8 delivers both channel sums at slot 2 and retains a zero-amplitude wake-up at slot 4.

The fixture runner compiles the unchanged frozen library and the additive test with rustc 1.98.1, edition 2024 and optimization level 3. Its event counts include all prescribed messages and wake-up markers; they are not the recurrent-message ratio Q used in E1–E5. Two executions produced byte-identical trajectory CSV files. The archive contains the test, reproduction script, exact event lists, compiler provenance, commands, per-slot f32 bit patterns and validation records.

## Appendix E Rewiring diagnostics and reproducibility

The complete rewiring package is in evidence/rewiring. Its protocol and primary analysis were fixed before the 360 scientific runs; the E/I rate-distribution analysis was added afterward and is explicitly post hoc. The original example trace uses rewiring seed 91001, chosen without selecting an outcome. Figures S3, S4 provide the message/error comparison and that fixed example.

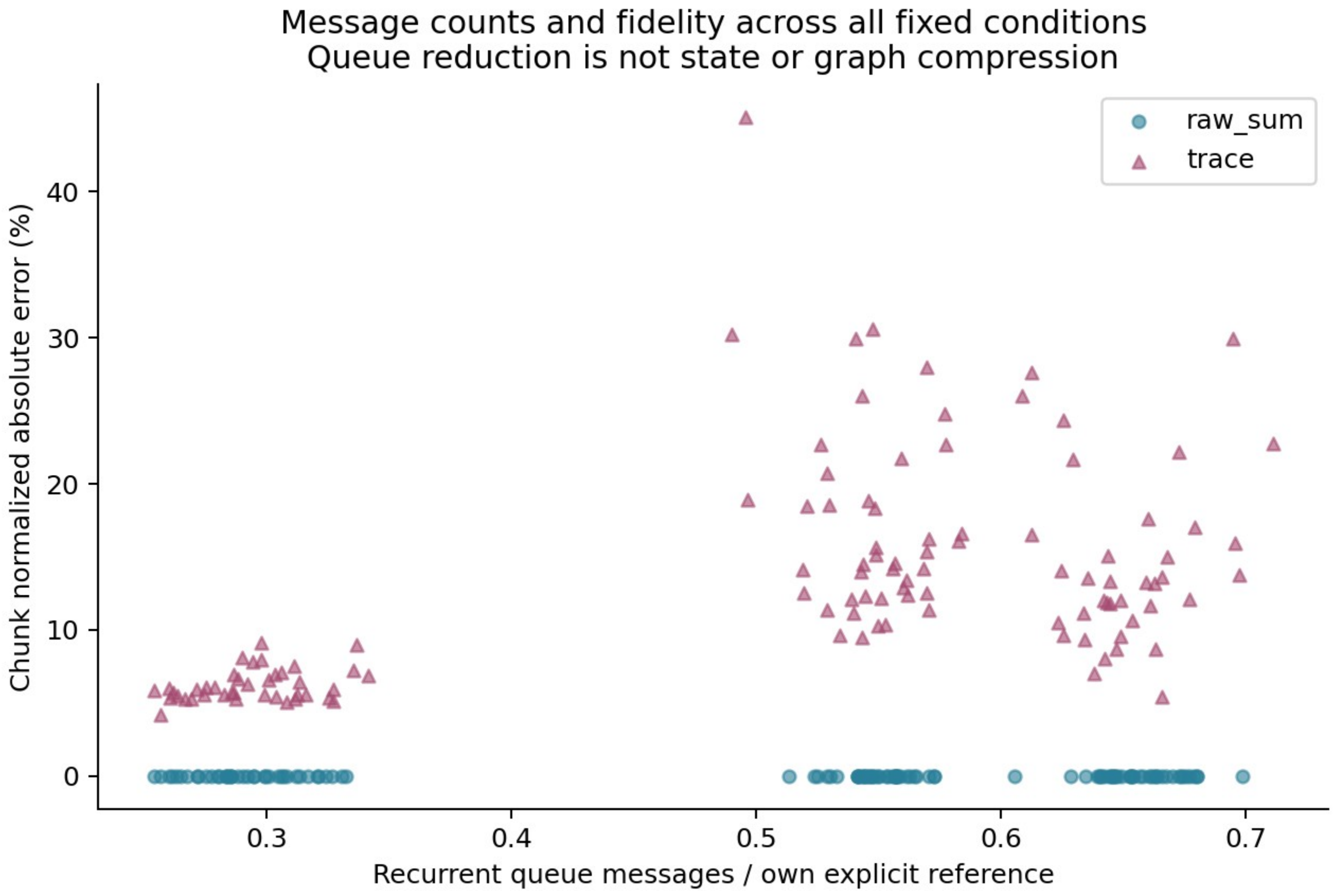

Figure S3. Recurrent message ratios versus chunk L1 errors on original and rewired graphs. Ratios use each graph's explicit reference. raw_sum retains all active-source edge visits despite reducing messages. The plot does not report runtime or graph compression.

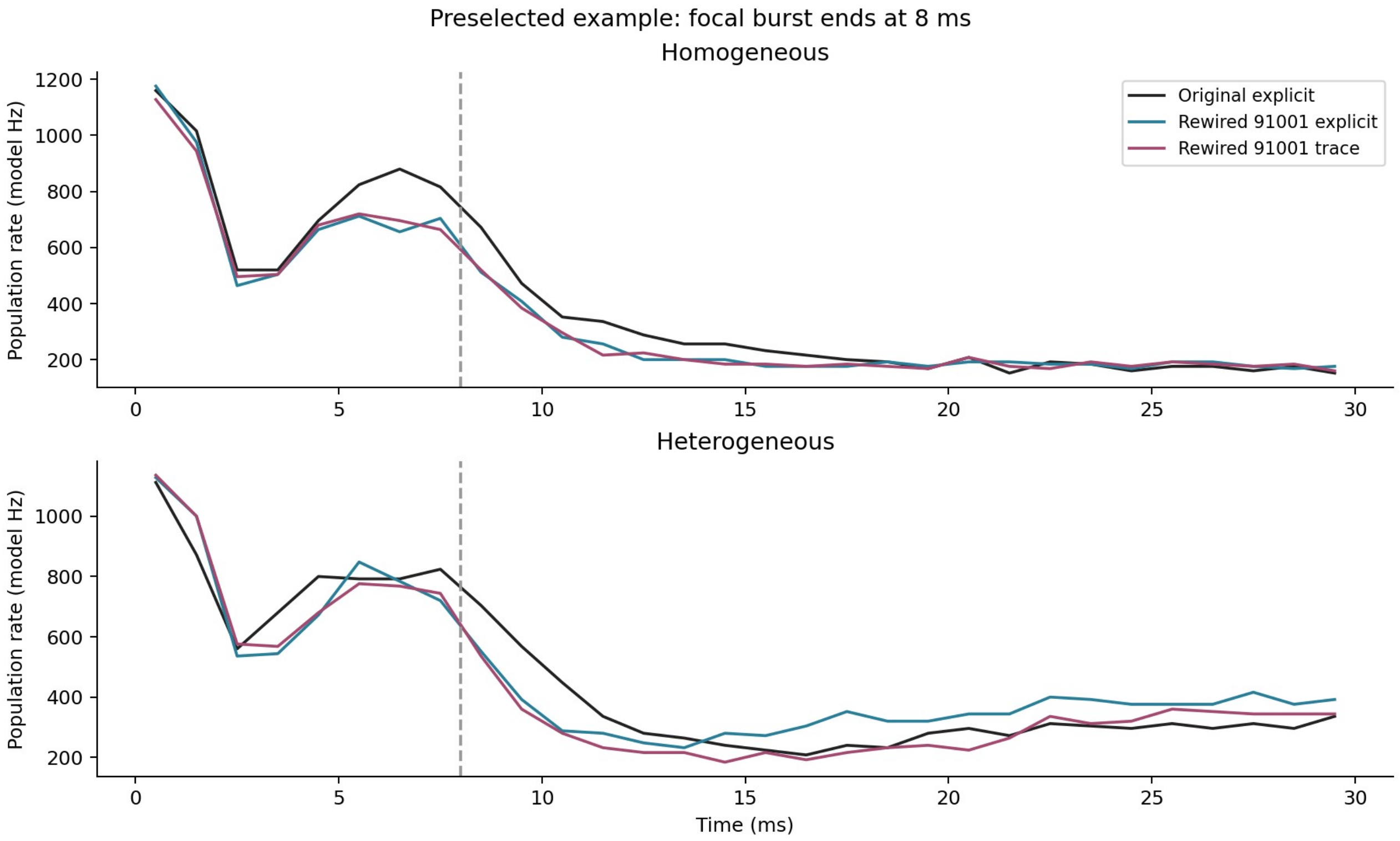

Figure S4. Fixed example of focal-burst activity for rewiring seed 91001 in both families. Original explicit, rewired explicit and rewired trace trajectories show distinct structural and approximation comparisons. The seed was fixed before inspecting outcomes.

The post-hoc diagnostic compares sorted per-chunk 30 ms mean rates within E and I groups using Wasserstein-1 distance, normalized by the original group's mean. It discards within-group identities. For homogeneous global-burst inputs, the median normalized E/I distribution distances are 6.71%/14.41%, versus 22.42%/30.54% for identity-matched mean-rate absolute differences. These different observables can differ without contradiction; no formal distribution-equivalence claim is made. The CSV field post_stimulus_rate_delta_pct denotes slots 80–299; for tonic input it is a late window with continuing drive, not a post-stimulus interval.

The two source changes are explicit source sorting and a raw-output control copied from the frozen stateful path; build.patch records both. All 55 original source hashes are retained and verified. Each run saves [300,125,10] uint32 state-bit arrays, including spike counts, representative unit variables, local context, refractory deadline and chunk status. Equality of all 100 units within each chunk is checked before recording. Pending queues are not fully saved. Original spike/message CSVs, compressed state payloads, graph audits, independent metric reductions, commands and compiler versions accompany the analysis. A new-directory rebuild reproduces graphs, per-slot CSVs, state payloads and primary error CSVs exactly; timing equality is neither required nor interpreted scientifically.

The theory check independently verifies the partition constraints for all 19 homogeneous controls and four projection arms, exercises the residual and group-total identities, and verifies the four-node counterexample with exact rational arithmetic. These checks support implementation correspondence; they do not replace the proofs or claim the bound predicts the nonlinear trajectory errors.

## Appendix F Phase and perturbation protocol

The follow-up is archived in evidence/perturbation. Its 16 graphs retain the homogeneous seed-99 original and five controls per treatment with seeds 92001–92005. The inherited first-hit sampler stops when each projection arm lacks exactly floor(0.35 $M_{arm}$) of its original endpoint pairs. Source rows, weight bits and delays remain fixed; source and target degrees and strengths are preserved per arm. B preserves node-level input/output strengths for the ten declared subgroups. C uses an equally sized E/I-respecting random relabeling fixed by seed 73109. The feasible sets differ across treatments despite their common changed-adjacency budget.

The archived phase analysis uses 48 explicit trajectories, each with 300 slots and 125 chunks. Phase RMS errors use channel scales from the corresponding original input's full trajectory. Weighted squared phase errors reconstruct all 45 rewired full-window scores to within $10^{-12}$. The rank-five state PCA basis and its channel scales are fitted only to the original E0–19 trajectory and held fixed across inputs and phases. Inside and outside error energies share the same denominator and sum to total error energy. These post-hoc diagnostics are distinct from the subsequently frozen source-activity prediction test.

The perturbation runs retain the original library and add observation fields to an explicit driver. Inputs are applied before each update. Row n is recorded at completed time (n + 1) × 0.1 ms, and slot ranges are half-open. The primary response window is [200,1000); secondary windows are conditioning [0,80), pre-perturbation [80,200), pulse [200,210), early [210,300), middle [300,500), late [500,1000), and terminal [900,1000). Each graph's three runs share identical stored fields through slot 199.

For source-activity prediction, the uncentered training SVD retains mean activity as part of the basis. In addition to rank five and the training-only 90%-energy basis, complete replay distortion is the Frobenius norm of $\Delta G_g\, S_{train}$ divided by that of $G_0\, S_{train}$, where $S_{train}$ is the archived source-count matrix. Structural descriptors are relative Frobenius change, absolute changes in reciprocal-pair and directed-three-cycle counts, leading-source singular-subspace distance and singular-value distance. The original E0–19 trajectory error is an empirical comparator. The near-constant correlation guard is range $\leq 10^{-12}$ times the larger of one and the maximum absolute value. Analyses report all 15 controls, each treatment separately (n = 5), and each omitted-treatment subset (n = 10), separately by amplitude, without p-values or fitted prediction coefficients. Multiple windows and amplitudes remain dependent observations.

At each slot, own-response magnitude is the RMS of Eq. (23) in Appendix I.2 across the 375 coordinates. Recovery is searched after pulse end and no earlier than the maximum response over slots 200–999. The first eligible sample must remain at or below 5% of that peak through the recorded horizon, with at least 50 samples remaining; otherwise recovery is right-censored. Zero responses are classified separately. The short protocol description omits the post-pulse restriction present in the locked implementation; this omission changes none of the 32 measured recovery classifications or times. Peak, response area, terminal magnitude, spike-response differences and deviation from amplitude doubling are also saved.

Eight independent kernel and scheduling checks precede the new simulations. A 300-slot bridge matches the earlier original E0–19 trajectory's ten fields and count records exactly. The follow-up stores 14 fields, adding pre-update refractory deadline, actual execution, refractory skip and due-event observations. A spike in slot n sets the deadline to n + 2, skips only n + 1, and permits input accumulation during that skip. In the 48 archived phase trajectories, the reconstructed pre-update gate fraction is 4.7049%; testing post-update deadlines instead would incorrectly report 9.4214%. Status after a step does not identify whether that step was executed.

The model has no lower voltage bound or calibrated membrane conductance units. Minimum voltages are −5500.2495 in the archived phase inputs and −4692.4697 in the new conditions. During the final 10 ms, 90.4–100% of chunk/slot updates still execute, with 16,100–57,300 allocated-unit spikes per run. A finite or small residual is not a test of convergence or asymptotic stability. All 48 new conditions pass finite-state, within-chunk equality, pre-perturbation identity and execution/gating checks; all 32 response scores are independently recomputed. A fresh-directory rebuild reproduces all saved state payloads and count CSVs exactly. Derived floating-point CSV fields use relative and absolute tolerances of $10^{-12}$ for last-bit reduction variation; labels, integer fields and missingness remain exact, and all differences are logged. Pending queue contents are not completely serialized, and instrumented timings are excluded from performance claims.

## Appendix G Sensitivity calibration and interpretation

The calibration package is evidence/sensitivity. Nineteen extensive controls with seeds 91001–91019 and three original trajectories are inherited from the homogeneous explicit arm of the earlier study. The new 19 local controls use seeds 93001–93019, with exactly one accepted swap per graph; the 19 initial-voltage directions use seeds 94001–94019. Each input uses the same swap graphs and initial directions. The three doses are repeated measurements along the same direction, not independent samples. All inputs and code are hashed before new scientific runs, with zero and inverse-swap technical shams listed separately.

The actual nonzero f32 initial offsets are 0.00009918212890625, 0.00099945068359375 and 0.01000213623046875, corresponding to 13, 131 and 1,311 representable steps at −65. Population-wise positive and negative offsets cancel exactly; one inhibitory chunk is unchanged. First-execution observations confirm that dormant blocks retain the supplied voltage until their first update, without additional wake-up events. All three new original trajectories and all six shams reproduce the archived originals' recorded states and counts exactly.

The calibration uses per-input original channel standard deviations over all 300 slots and 125 chunks, with population variance (ddof = 0). These scales are fixed for extensive rewiring, local swaps and initial perturbations, and for the 0–8, 8–15, 15–30 and complete 0–30 ms windows. They differ from the old E0–19 training scales used for the separate pulse-response analysis. CSV files retain absolute voltage RMS and maximum differences alongside standardized state errors. Independent integer reductions check every population and chunk L1 numerator and denominator.

Figure S5 reports one minus the population/chunk L1 ratio within each comparison. Its median values among the single-swap cases with nonzero chunk error are 62.07%, 55.28% and 47.16%, with denominators of nine, 12 and 11 controls, respectively. All 19 extensive controls contribute for each input. All initial-voltage cases have zero chunk error and therefore undefined cancellation fractions, rather than zero cancellation. A cancellation fraction describes aggregation of existing errors; it does not test equivalence or assign biological relevance to their magnitude.

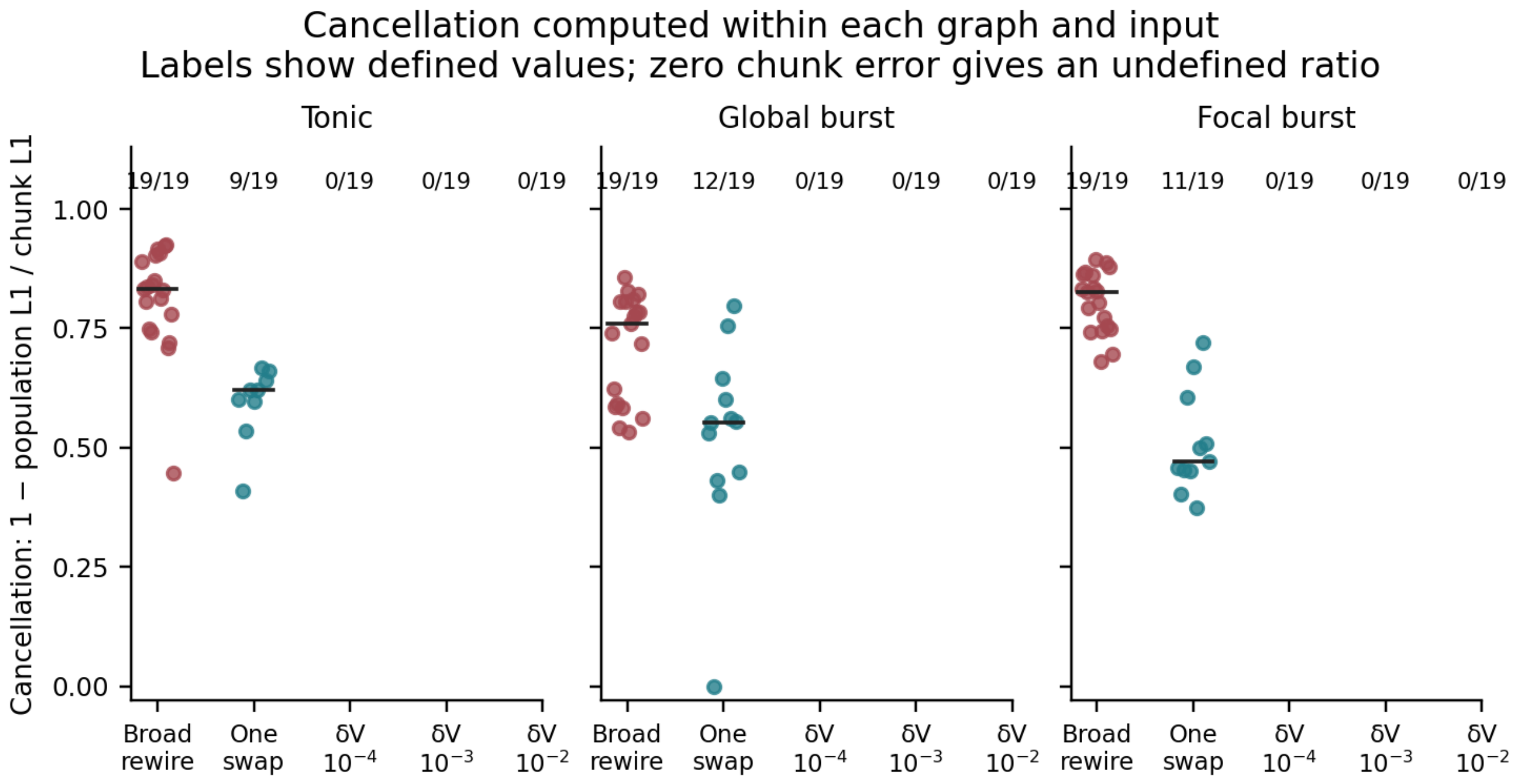

Figure S5. Error canceled by population aggregation. Each defined point is one minus population L1 divided by identified-chunk L1 for the same comparison. Labels give the number of defined ratios out of 19. A zero chunk error gives an undefined ratio and is not plotted as zero cancellation; all initial-voltage ratios are therefore undefined. Bars mark medians of the defined individual ratios.

For a swap a→b, c→d to a→d, c→b, the shared-source discrepancy at its two targets is determined by the difference between source counts $S_a$ and $S_c$ times the common weight. The diagnostic replays those counts from the original only; feedback divergence is assessed separately in the new simulation. Source slot n arrives at n + 2, so only source slots 0–297 can contribute within the 300-slot horizon. Four swap/input cases have zero discrepancy throughout that eligible window: three involve equally active source sequences and one involves two silent sources. All four reproduce the ten recorded fields exactly. Seventeen further spike-identical cases have nonzero shared-source input discrepancy. No controls are excluded because their exchange is unactivated or their output is unchanged.

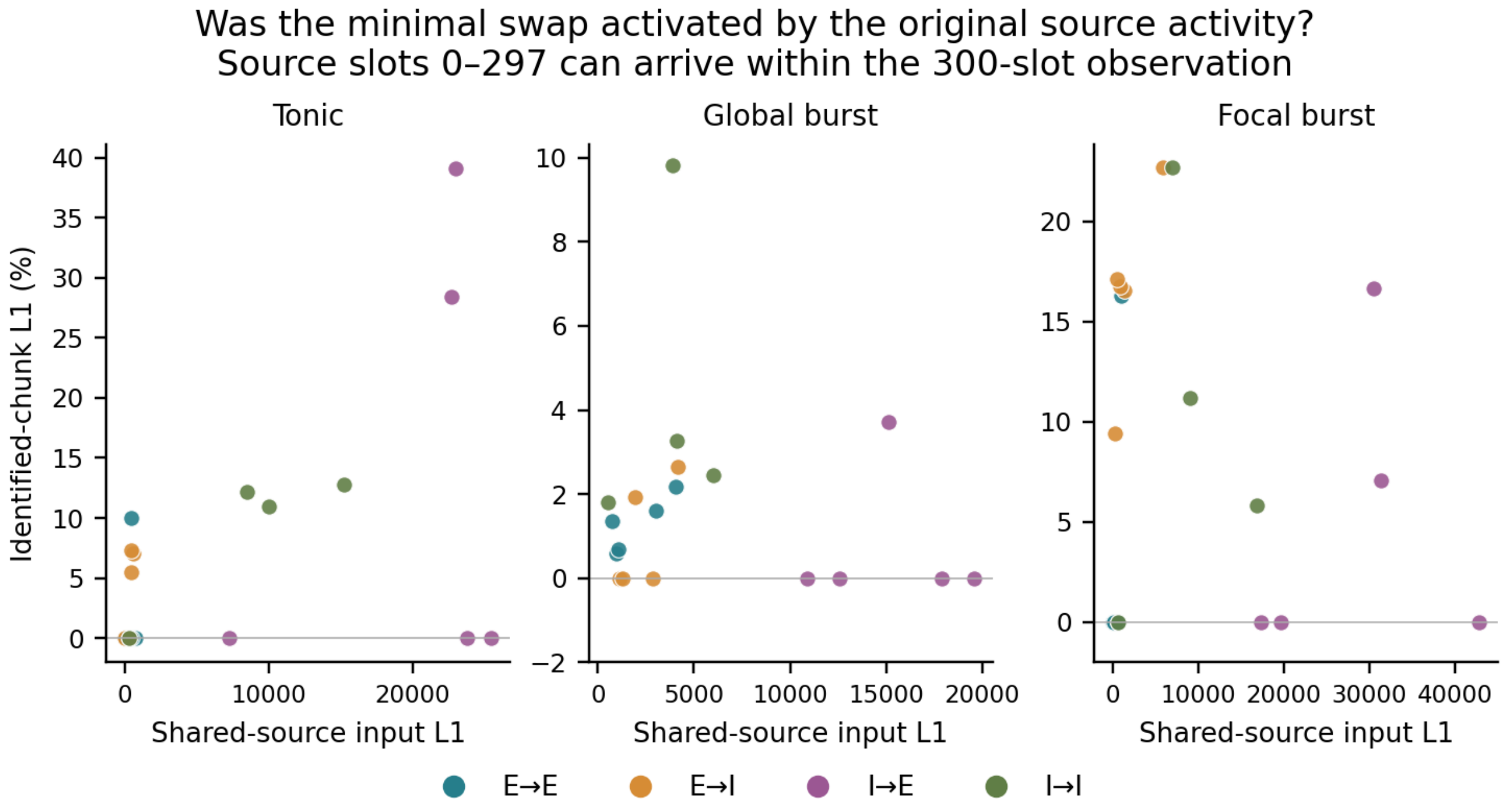

Figure S6. Shared-source activation of the single swaps. The horizontal axis is the L1 input discrepancy obtained by replaying the original source counts through the changed input map; the vertical axis is the resulting closed-loop chunk L1. Only source slots 0–297 can arrive within the observation. Colors identify the exchanged projection arm, whose fixed allocation is 5/5/5/4. Four zero-discrepancy cases also reproduce the recorded states bitwise, but nonzero discrepancy does not necessarily produce a spike difference.

The complete time courses distinguish cumulative 1 ms count error, normalized by the full-window original spike total, from state RMS in each 1 ms interval (Figure S7). Shaded ranges span the 19 controls and are not confidence intervals. For initial-voltage perturbations, all final recorded states coincide, although no complete trajectory is bitwise identical. All chunks have a common reset under tonic and global-burst inputs; only 71 do so under the focal input. The remaining focal chunks recoincide without firing, consistent with contraction and finite-precision rounding. Accordingly, final agreement cannot be attributed entirely to spike reset, and initial agreement of spikes cannot be interpreted as equality of the intervening continuous states.

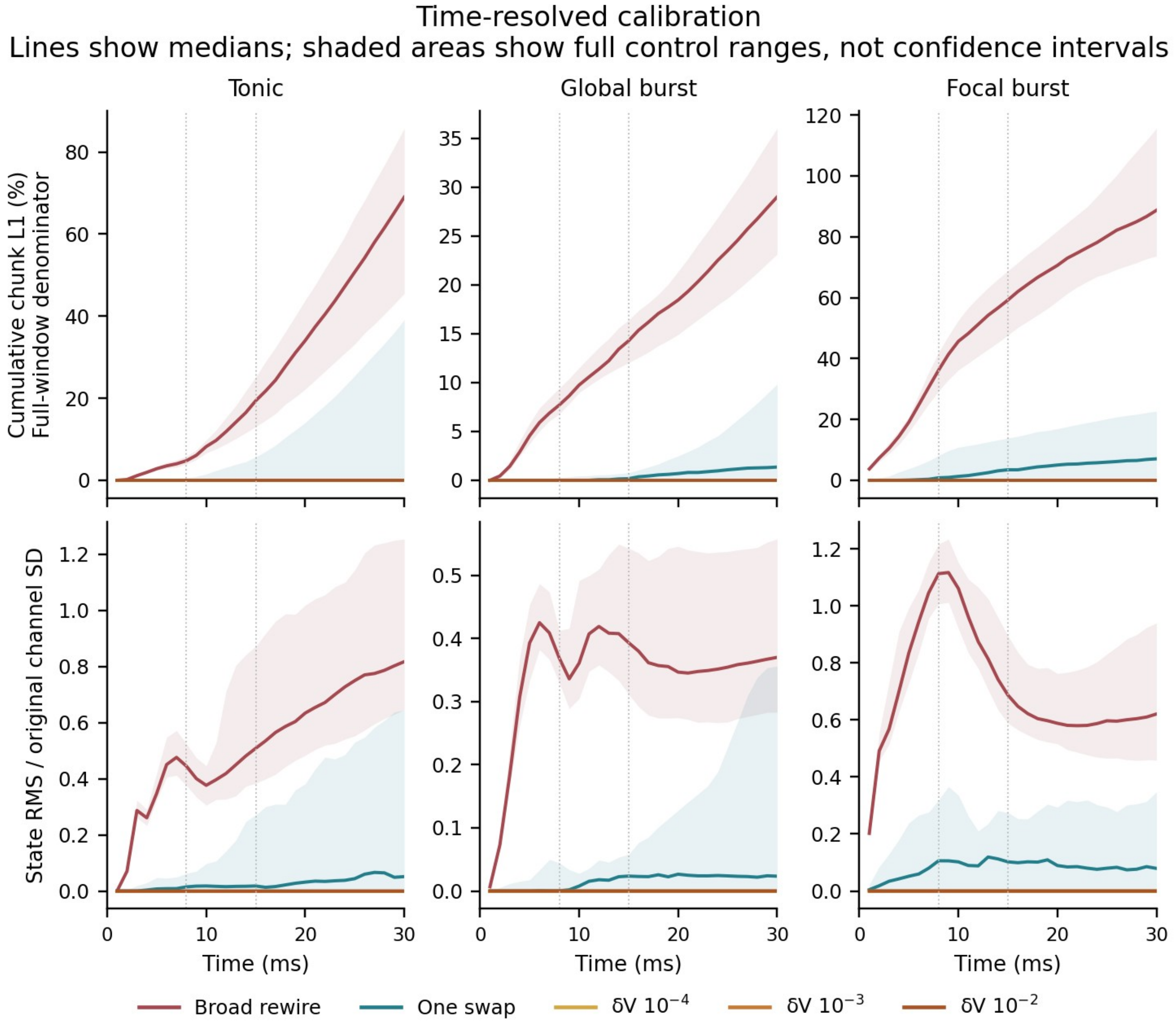

Figure S7. Time-resolved sensitivity calibration. Top: cumulative binned chunk L1 divided by the complete original spike total. Bottom: standardized state RMS in each 1 ms interval, using fixed per-input original V, E-input and I-input scales. Lines are medians of 19 controls and shading spans their full ranges. Vertical dotted lines mark 8 and 15 ms; intervals are not confidence bounds.

This calibration compares chosen interventions on one original over 30 ms. Minimal and extensive rewiring have different changed-edge budgets and may engage different source patterns; initial-voltage changes do not alter connectivity. Their separated error ranges therefore provide an empirical scale, not a derivative with respect to graph distance, a matched effect-size intervention, a test of chaos, or evidence that microscopic initial variations are harmless in general.

## Appendix H Earlier execution and approximation comparisons

### H.1 Approximation rules

The filtered and history-modulated rules are prototype alternatives defined by the heuristics specified in this appendix. They provide concrete comparison cases, without representing optimized proxies or the full range of possible local-circuit approximations.

The history-modulated rule transforms the same raw contribution $R_{kq}$. For each destination and polarity it maintains fast and slow filters $x_{kqa}$, where a ∈ {f, s}, and a first-order trace $h_{kq}$, all initialized to zero:

$$x_{kqa}[n] = d_{qa} x_{kqa}[n - 1] + (1 - d_{qa}) R_{kq}[n] \quad (20a)$$

$$d_{qa} = \exp[-dt / max(\tau_{qa}, 0.1)] \quad (20b)$$

$$h_{kq}[n] = (1 - \alpha) h_{kq}[n - 1] + \alpha R_{kq}[n] \quad (20c)$$

The four filter time constants are 3 and 80 ms for excitation, and 7 and 80 ms for inhibition. The trace coefficient is α = clip(dt/20, 0.005, 0.25), giving α = 0.005 at the experimental timestep. With ε = 0.001, the output is expressed below. In Eq. (21a), $x_f = x_{kqf}$[n], $x_s = x_{kqs}$[n] and h = $h_{kq}$[n]; c and s refer to the same destination, polarity and slot. Equation (21b) suppresses the common slot index [n].

$$c = \frac{x_f - h}{x_f + h + \varepsilon}, \qquad s = clip\left(\frac{x_s}{h + \varepsilon}, 0, 1.5\right) \quad (21a)$$

$$D_{kq} = \kappa_q [1 - 0.03c + 0.01(s - 0.5)] R_{kq} \quad (21b)$$

The factor in brackets is bounded by 0.965 and 1.040 for nonnegative states. Static gains are $\kappa_E$ = clip[0.955 + 0.015 max($g_{fE}$, 0.05), 0.9, 1.08] and $\kappa_I$ = clip[0.985 + 0.045 max($g_{fI}$, 0.05), 0.9, 1.08]. Principal runs use $g_{fE}$ = 2.5 and $g_{fI}$ = 0.4, yielding $\kappa_E$ = 0.9925 and $\kappa_I$ = 1.003. The history path does not use the slow-gain or population-normalization parameters. It emits only when the current $R_{kq}$ is positive, so residual filter states alone do not emit messages.

A static-gain control uses $D_{kq} = \kappa_q R_{kq}$, removing the history factor. The raw-sum control removes both factors. These exploratory output ablations preserve the history path's delivery time and retain its internal state-update code, although those states no longer affect output. They isolate delivered magnitudes without constituting optimized runtime implementations.

The two-channel rule sums weighted source spike fractions $S_j$/U, with unit gains and one output channel per polarity. The filtered four-channel rule uses the same population normalization and fast and slow states updated as $x_{qa}[n] = \exp(-dt/\tau_{qa}) x_{qa}[n - 1] + g_{qa} \Sigma_e w_e S_j[n]/U$. Its output is the sum of the fast and slow states. Unlike the history filters, its innovation term lacks the factor $1 - \exp(-dt/\tau_{qa})$, and its residual state can continue to generate output after source activity stops. Default gains, ordered as excitatory fast/slow and inhibitory fast/slow, are (1.0, 0.5, 1.0, 0.3). Both normalized methods use the earlier arrival slot specified in Section 2.2.

H.2 Raw sums preserve measured population counts with fewer messages

In the E1 workload of 500 chunks comprising 50,000 allocated units, explicit routing and pure raw-sum aggregation each emit 1,606,400 spikes over 500 slots. Their total population spike counts agree in every slot. Recurrent queue insertions fall from 1,196,827 to 187,953, giving Q = 0.157043, or an 84.30% reduction. Retaining 24.0% of edges also preserves the count series, with 591,580 messages and Q = 0.494290. Thus, on this workload, retaining explicit messages increases queue traffic without improving the already identical measured counts (Table S4; Figure S8).

Table S4. E1 comparison. Retention is the fraction of chunk edges routed individually; Q is the recurrent queue-insertion ratio.

| **Method** | **Retained / %** | **r** | **δFR / %** | **Q** |
|---|---|---|---|---|
| Raw sum | 0 | 1.000 | 0.0 | 0.157 |
| Raw-sum hybrid | 24.0 | 1.000 | 0.0 | 0.494 |
| Static gain | 0 | 0.996 | −5.9 | 0.159 |
| Static-gain hybrid | 24.0 | 0.998 | −4.1 | 0.488 |
| History modulation | 0 | 0.992 | −8.4 | 0.164 |
| History-modulated hybrid | 24.0 | 0.980 | −16.0 | 0.451 |
| Filtered four-channel | 0 | 0.435 | −9.7 | 0.405 |
| Filtered hybrid | 24.0 | 0.183 | −52.6 | 0.636 |
| Two-channel | 0 | −0.217 | +323.6 | 0.388 |

Both raw-sum settings match the reference population count in all 500 slots. The reference emits 1,606,400 spikes.

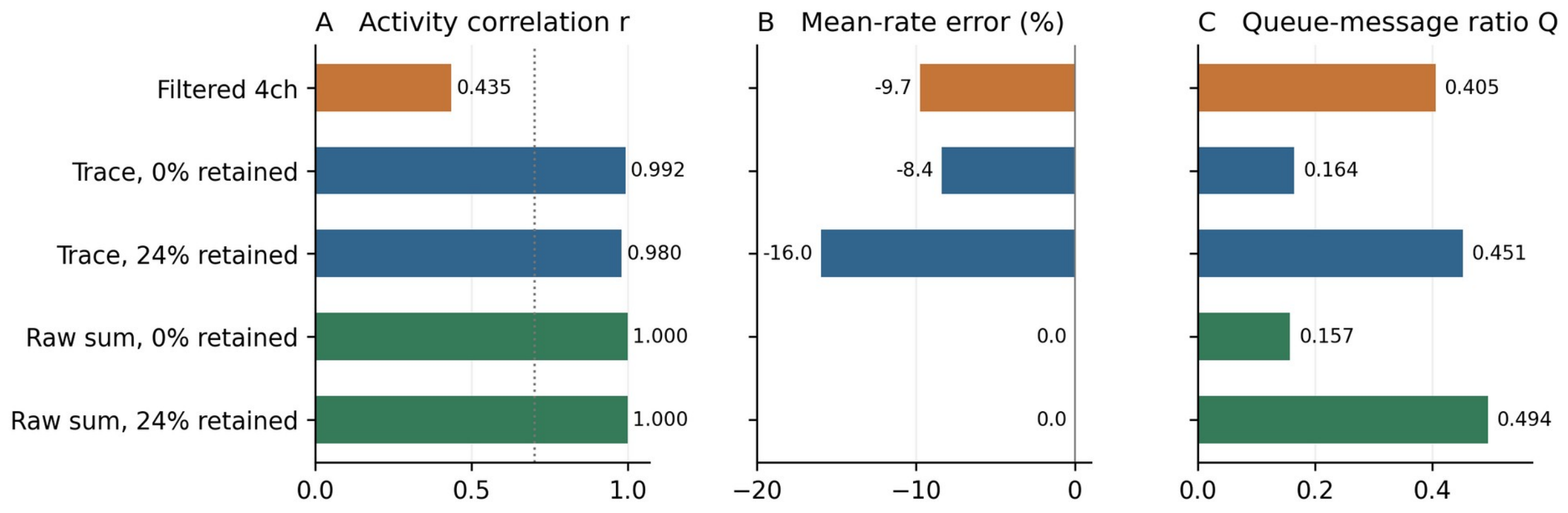

Figure S8. Population fidelity and recurrent queue traffic in E1. The workload contains 50,000 allocated units and runs for 50 ms. Trace denotes history modulation. Raw-sum variants have zero mismatching total-population-count slots. Q is the ratio of enqueued recurrent messages. The dotted line marks the descriptive correlation threshold r = 0.70.

The result extends to the heterogeneous E3 main setting at 11.2% retention: both implementations emit 124,500 spikes, with zero mismatching slots, while messages decrease from 32,247 to 21,663 (Q = 0.671783). Raw sums also give zero count mismatches in homogeneous E3, all three E4 weight settings and E5 (Table S5). This agreement is an empirical result for the total population sequence, not a measurement of every unit state.

### H.3 Output modulation introduces measurable differences

The history-modulated pure proxy in E1 achieves r = 0.992 and $\delta_{FR}$ = −8.4%, with Q = 0.164. It differs from the reference count in 426 of 500 slots. Keeping only the static gain improves correlation to 0.996 and reduces the magnitude of the rate error to 5.9%, but still changes 439 slots. A larger correlation can therefore coexist with more differing slots: correlation measures the shape of the activity series, while mismatch count records any difference. Removing both output factors yields zero mismatches.

The filtered four-channel method gives r = 0.435 and $\delta_{FR}$ = −9.7% in the pure E1 setting. Its 24.0%-retention hybrid gives r = 0.183 and $\delta_{FR}$ = −52.6%. History modulation provides much higher agreement than these filtered variants. However, the methods simultaneously change arrival time, population normalization, filter scaling and dynamics, residual output, and the way gains affect delivered input. The aligned controls establish that history modulation is unnecessary for the observed exact population-count agreement; they do not identify or rank the individual causes of the filtered method's error.

H.4 Retention effects depend on the aggregation rule

At zero E3 retention, filtered four-channel delivery gives r = 0.435, $\delta_{FR}$ = +69.2% and Q = 1.421. History modulation gives r = 0.986, $\delta_{FR}$ = −2.5% and Q = 0.472, differing from the reference in 146 of 200 slots. Raw sums preserve all 200 population counts and the reference total of 124,500 spikes, with 15,319 messages versus 32,247 for explicit routing (Q = 0.475; Figure S9).

Across the five nonzero inhibitory-source budgets, filtered four-channel correlation ranges from 0.492 to 0.544, and mean-rate error ranges from −28.4% to +20.1%. History modulation gives r = 0.980–0.986 and errors from −4.7% to −2.5%. Its zero-retention point already meets both screening criteria and has the lowest observed message ratio for that rule. At 2.0% retention, its population-count sequence is unchanged from zero retention, while Q rises to 0.507; at 11.2%, Q is 0.658.

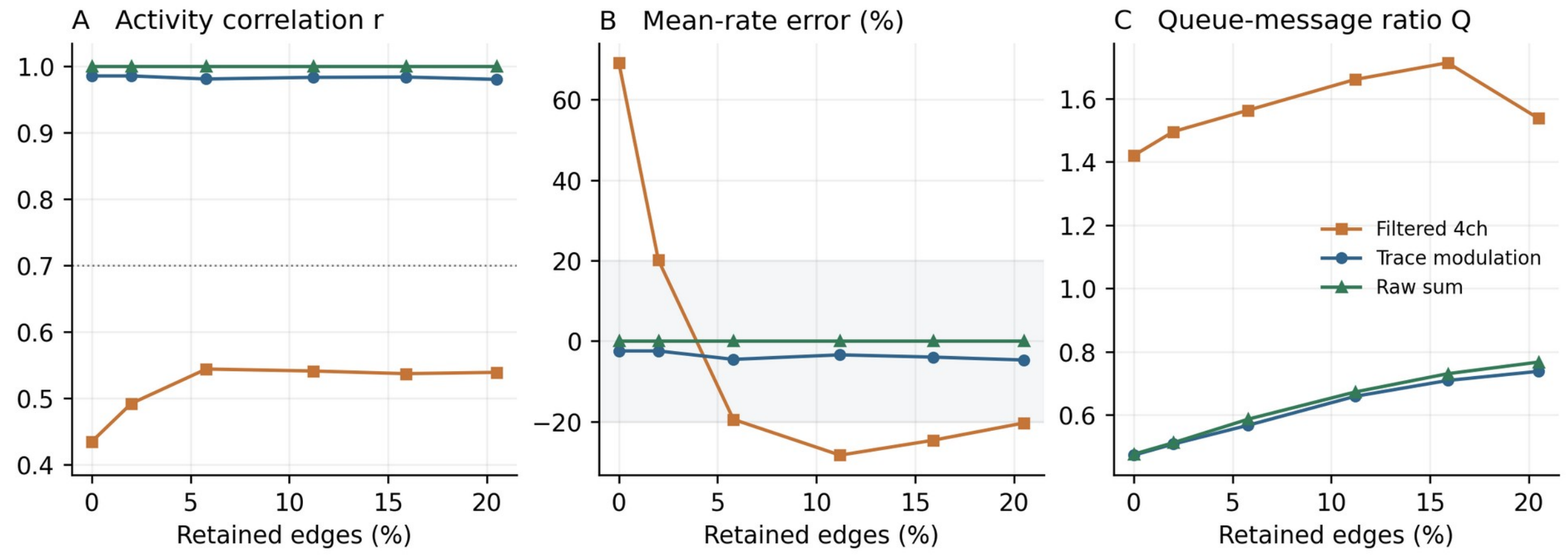

Figure S9. Inhibitory-source retention sweep on the heterogeneous E3 graph. Each point uses 12,500 units and 20 ms of activity. Lines connect the tested budgets. Zero retention uses an empty edge-selection policy. Raw-sum count sequences agree with the reference at all six budgets. Shading marks ±20% rate error; the dotted line marks r = 0.70.

Raw sums preserve the count sequence at all six tested budgets, while Q increases from 0.475 at zero retention to 0.511 at 2.0% and 0.767 at 20.5%. Within this tested range, additional explicit routing increases message cost without improving raw-sum count fidelity. Figure S10 displays the archived zero-retention trajectories and their slotwise differences without smoothing: the high correlation of history modulation coexists with 146 differing population-count slots.

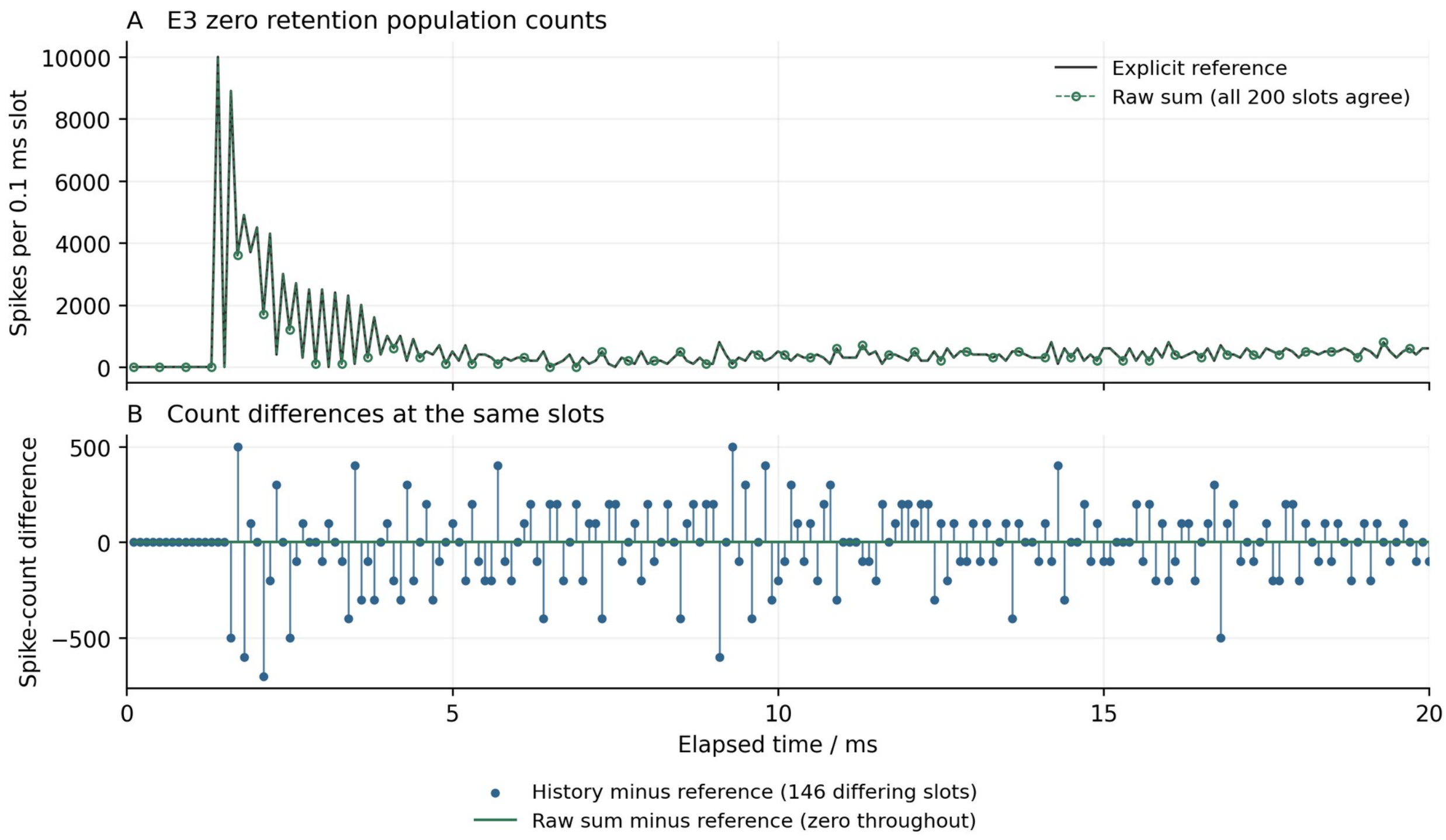

Figure S10. Archived E3 zero-retention trajectories. A: explicit and raw-sum population counts coincide at all 200 slots; open markers show every fourth raw-sum point for legibility, while lines include all points. B: history-minus-reference and raw-minus-reference count differences at the same slots. History modulation differs in 146 slots despite r = 0.986. No smoothing, binning or lag adjustment is applied. Elapsed time uses the archived one-based completed-slot index multiplied by 0.1 ms.

The edge-class comparison shows the same interaction (Figure S1). Under filtered aggregation, I-to-I retention produces a +157.8% rate error, while E-to-I retention produces −59.0%. History modulation reduces those errors to −3.1% and −1.0%, respectively; raw sums yield zero in both cases and in the other two policies. Since the policies retain 3.7% and 13.1% of edges, respectively, this comparison also combines edge identity with a change in budget. It characterizes policy–approximation interactions rather than an independent functional role for either class.

### H.5 Gain scans and pulse responses delimit the comparison

In the 12,500-unit E1 gain scan, the best filtered four-channel correlation is 0.534, accompanied by a −29.5% rate error. None of its 17 distinct gain pairs meets both screening criteria. History modulation gives r = 0.936–0.991 and absolute rate errors of 0.6–15.3%; all evaluated pairs meet both criteria (Figure S2). These values describe the sampled grid on one graph. The nominal gains act directly on filtered inputs but map to $\kappa_E$ = 0.970–1.030 and $\kappa_I$ = 0.9895–1.021 in the history rule. The scans therefore do not match effective input perturbations across methods; they also leave the history time constants and modulation coefficients fixed.

Agreement in mean rate does not ensure agreement in timing. Across the stable, balanced and excitable E4 weight settings, the history rule has rate errors of +3.5%, −1.3% and −1.0%, but correlations of 0.775, 0.904 and 0.559. In the excitable setting the average rate remains close while temporal agreement falls below the descriptive threshold. During the E5 pulse and recovery window, history modulation gives r = 0.973 and $\delta_{FR}$ = −0.3%. Raw sums preserve the reference count series in all four probes (Table S5).

Table S5. Fidelity and message counts across workloads. Every raw-sum setting has r = 1, δFR = 0 and zero mismatching population-count slots.

| Setting | Retained / % | History r | History δFR / % | History Q | Raw Q |
|---|---|---|---|---|---|
| E3 heterogeneous | 11.2 | 0.983 | -3.5 | 0.658 | 0.672 |
| E3 homogeneous | 22.8 | 0.972 | -0.3 | 0.822 | 0.827 |
| E4 stable | 26.4 | 0.775 | +3.5 | 0.940 | 0.917 |
| E4 balanced | 26.4 | 0.904 | -1.3 | 0.900 | 0.904 |
| E4 excitable | 26.4 | 0.559 | -1.0 | 0.641 | 0.642 |
| E5 pulse | 27.7 | 0.973 | -0.3 | 0.888 | 0.891 |

## H.6 Proxy errors across original and rewired graphs

Across all 120 graph/input pairs, raw_sum matches its own explicit reference at every slot and chunk. None of the 120 pairs has all recorded state bits identical: maximum absolute differences over all pairs are 0.01275635 in voltage, 0.03515625 in excitatory conductance and 0.01171875 in inhibitory conductance, in the code's state units. This is finite-window spike-count agreement, not floating-point state equivalence or a guarantee beyond 30 ms.

Trace has positive chunk error in all 120 comparisons. In the homogeneous family its original-graph errors are 13.43%, 7.98% and 10.50%, while the rewired medians are 13.96%, 6.42% and 14.99% for tonic, global and focal inputs. In all six family/input combinations, the original error lies within the corresponding control range (Figure S11). This does not establish statistical equivalence or exceptional performance of the original graph. This extension does not repeat the static-gain ablation, so its trace errors cannot be assigned solely to memory rather than the combined output transformation.

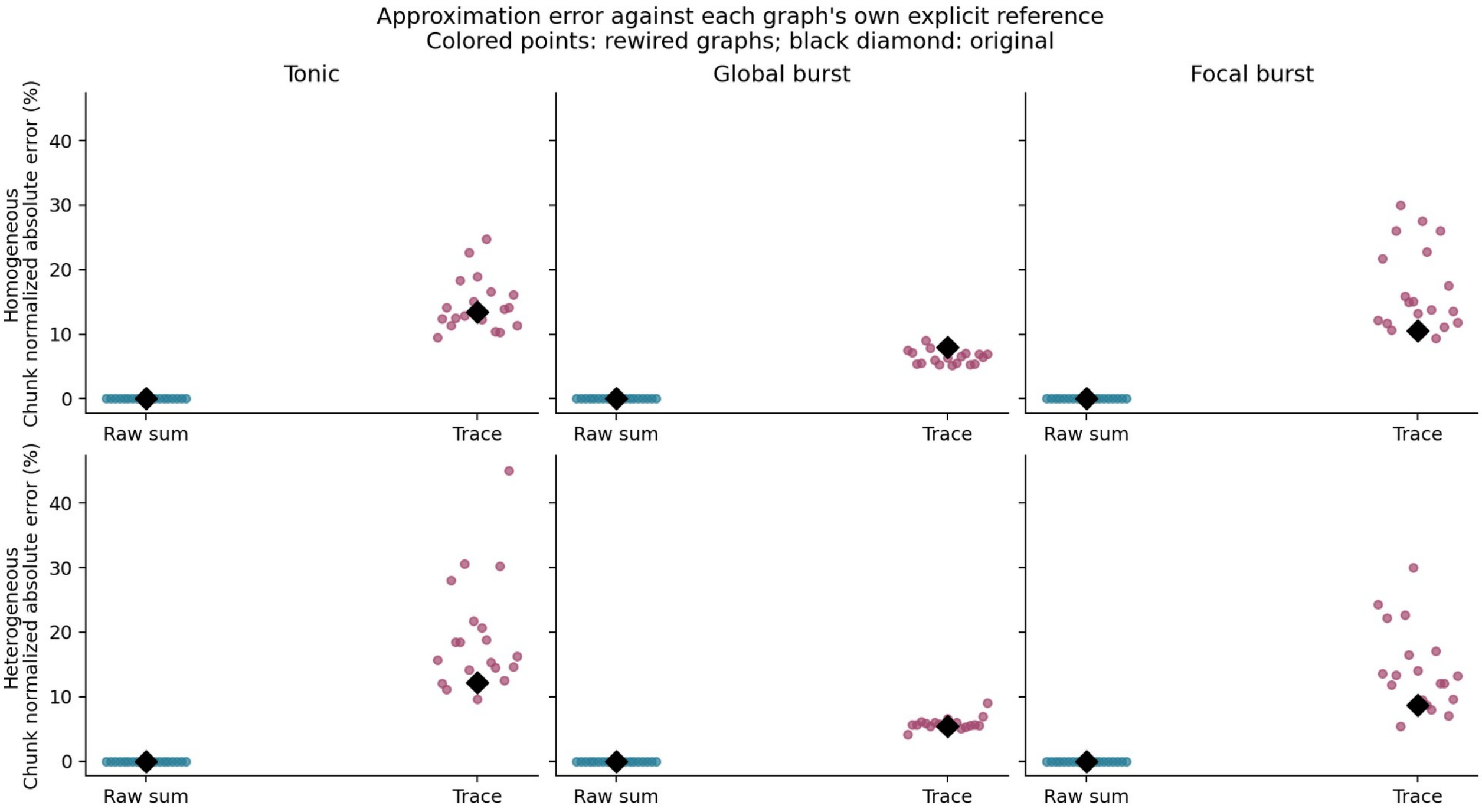

Figure S11. Proxy error against each graph's own explicit reference. Points show 19 rewired controls; black diamonds mark original graphs. raw_sum has zero chunk-count error in every case. Trace includes its gain and history transformation. Original errors lie within the corresponding control ranges; this is not an equivalence test.

Raw_sum reduces recurrent messages by 30.1–74.6% across these conditions but retains exactly the explicit reference's active-source edge visits. All adjacency and unit states remain stored, and unused filter/trace updates remain in the raw ablation. Instrumented timings include observation and counting overhead; they are not speedup measurements. A complete fresh rebuild and rerun reproduces all 40 graph files, 360 per-slot count/message CSVs and 360 recorded state payloads exactly on the specified environment. This is technical replication, not 360 additional independent samples.

## Appendix I Secondary phase and pulse response studies

I.1 Group-constrained controls and phase diagnostics

A separate structural control starts from the same homogeneous E3 original (seed 99) and uses five rewired graphs in each of three treatments. A preserves every node's per-arm E/I in/out degrees and strengths. B additionally preserves incoming and outgoing strengths between each node and ten declared subgroups: five groups of 20 E chunks and five groups of five I chunks. C imposes the same constraints on an equally sized, E/I-respecting random partition. These groups are declared labels, not inferred communities. Each treatment removes exactly 994 of the 2,845 original adjacency pairs (34.9385%), with arm-specific missing-pair counts matched across treatments. The first-hit constrained sampler gives no uniform null-sampling guarantee.

The original and 15 controls have archived explicit trajectories for three 30 ms inputs: a burst to E0–19, a shifted burst to E60–79, and a dispersed burst to E0, E5, …, E95. All inject amplitude 150 in slots 0–79. These 16 × 3 = 48 archived trajectories are reanalyzed, rather than counted as new simulations. For each input, state differences in voltage and E/I input states are divided by that input's original full-trajectory channel standard deviations. RMS errors use fixed scales across the 0–8, 8–15 and 15–30 ms phases. A separate rank-five state PCA basis, fitted to the original E0–19 trajectory, decomposes error energy inside and outside a fixed reference subspace. This linear projection is distinct from the source-activity basis used below.

I.2 Perturbation response and activity-based prediction

We retain these 16 graphs, the numerical kernel and explicit canonical-order routing for 48 new conditions. Each run lasts 100 ms. A conditioning burst of amplitude 150 targets E40–59 during slots 0–79, a location unused for the previous contiguous-group bursts. During slots 200–209, the same targets receive an additional amplitude a = 0, 1.5 or 3.0 per slot. The two nonzero doses are 1% and 2% of the conditioning amplitude per slot; this choice does not assume a linear response regime. Graphs, training inputs, predictor definitions and analysis code were hashed before these runs. The design is an exploratory follow-up with a new stimulus on existing graphs.

The predictor uses only the original archived E0–19 source spike-count matrix, with 125 rows and 300 slots. An uncentered SVD defines $U_5$ as the 125 × 5 matrix of leading left singular vectors, a rank-five source basis. Let $G_g$ vertically stack the E and I input matrices of graph g, let $G_0$ denote the original, and let $\Delta G_g = G_g - G_0$. The primary structural distortion is

$$F_5(g) = \frac{\|\Delta G_g U_5\|_F}{\|G_0 U_5\|_F} \quad (22)$$

A secondary basis retains at least 90% of the training activity's squared energy; the training data select rank 19. Complete training-activity replay and fixed structural descriptors provide additional comparisons. No basis is refitted to the new trajectories.

At slot n, $X_g{}^a[n]$ is the 375-component vector of voltage and E/I input states for the 125 chunks. D is diagonal, repeating the three old E0–19 reference standard deviations (705.80399884, 520.79332709, 752.89656332) for each chunk. The standardized response relative to that graph's own unperturbed trajectory is

$$r_g{}^a[n] = D^{-1}\left(X_g{}^a[n] - X_g{}^0[n]\right) \quad (23)$$

Its fidelity to the original response is measured by

$$Y_g(a) = RMS(r_g^{\ a} - r_0^{\ a}) \quad (24)$$

The RMS in Eq. (24) covers slots 200–999 and all 375 coordinates. Dividing this error by the original response RMS gives a relative error whose value is one for a zero-response prediction. Response magnitude and empirical recovery to each graph's own baseline are recorded separately. Each amplitude contributes 15 graph-level values to the predictor evaluation; the original is excluded. Spearman associations, within-treatment results and leave-one-treatment-out checks are descriptive. Appendix F specifies the secondary features, recovery criterion, time windows and technical validation.

### I.3 Structural fidelity depends on the observation phase

The matched-budget controls exhibit a phase-dependent ordering (Table S6; Figure S12). For the E0–19 burst, B has lower median standardized state error than A during stimulation (0.6198 versus 0.7223) and during the following 7 ms (0.8305 versus 0.8772), but greater error in the 15–30 ms phase (1.0037 versus 0.5597). The shifted input gives a different late ordering: B remains below A and close to C. The median ordering therefore depends on the stimulus and phase in these five-graph groups; their overlapping ranges do not establish a population-level advantage of one constraint.

Table S6. State error across three windows of module_burst. Medians over five conditional rewires per group; V, gE and gI use fixed channel scales from the original 30 ms trajectory. Lower values indicate closer trajectories.

| Rewiring constraint | 0–8 ms | 8–15 ms | 15–30 ms |
|---|---|---|---|
| A E/I | 0.7223 | 0.8772 | 0.5597 |
| B aligned | 0.6198 | 0.8305 | 1.0037 |
| C random | 0.6477 | 0.8975 | 0.8404 |

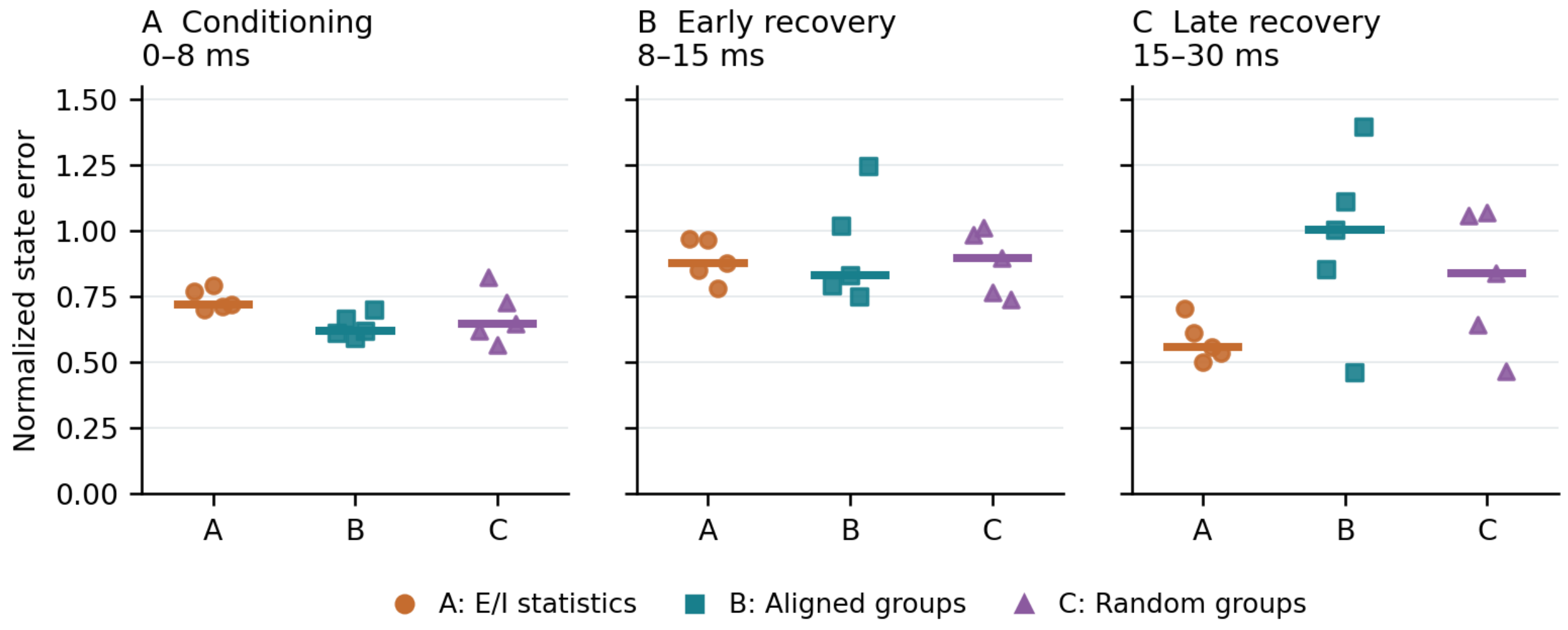

Figure S12. Phase-specific state fidelity for module_burst (E0–19) in matched-budget controls. A, B and C respectively preserve E/I statistics, declared-group profiles, and randomly relabeled-group profiles. Points show the five graphs in each treatment; horizontal bars mark medians. All remove 34.9385% of original adjacency pairs. State errors use this input's original full-trajectory channel scales, fixed across phases. The displayed phase ordering is conditional on the input; it is not a statistical equivalence or population-inference test.

Across the 45 graph/input comparisons, median fractions of error energy outside the fixed state PCA subspace are 88.80%, 84.23% and 78.08% in the three phases. These fractions describe error relative to the reference linear subspace; neither visual overlap in a low-dimensional projection nor preservation of static group profiles establishes a shared invariant manifold.

I.4 Recovery and response fidelity separate

The original graph's standardized response RMS is 0.046824 at amplitude 1.5 and 0.060187 at amplitude 3.0. Its response peaks occur at 80.2 and 87.1 ms, and neither response satisfies the recovery criterion by 100 ms. In contrast, 9 of 15 rewired graphs at the smaller amplitude and 7 of 15 at the larger amplitude return below the criterion 4.1–4.3 ms after the pulse ends. These are precisely the graphs whose spike sequences remain unchanged from their own baselines, despite small nonzero state responses. Recovery therefore measures a different property from preservation of the original response.

The smallest relative response errors among controls are 0.999948 and 0.999872, essentially the zero-response comparator; the largest are 10.896 and 9.022. The weak-response graphs' relative errors differ from one another by only about $1.12 \times 10^{-6}$ and $3.14 \times 10^{-6}$. Their fine ordering does not identify a reliably best-preserved graph. Only 0–2.5% of target chunk/slot updates during the pulse are refractory-gated, and delivered input can accumulate while gated. The weak responses cannot be described as stimulation being mostly blocked by refractoriness.

The primary source-activity predictor has Spearman correlations 0.279 and 0.182 with response error (Table S7; Figure S13). Within-treatment correlations are positive in A and B but −0.3 in C at both amplitudes, and leave-one-treatment-out values vary. The rank-19 and complete training-replay features also lack a consistent strong association. The rank-five basis captures 58.81% of training source energy but only 22.48% of new original-baseline activity in the response window; corresponding rank-19 values are 90.75% and 57.14%. The evaluation does not demonstrate a stable predictive association in this conditional sample. Reduced coverage is a candidate explanation, not an isolated cause or a proof that activity-based prediction is generally ineffective.

Among the prespecified comparisons, absolute changes in directed three-cycle counts and singular-value distances show larger associations. Their interpretation remains descriptive: selecting the strongest associations after observing several features, near-ties among weak responders and treatment-specific feasible graphs limit inference. The three-cycle statistic counts unsigned directed cycles and does not identify their nodes or E/I pattern. These results do not establish a causal cycle mechanism or an optimal graph geometry.

Table S7. Spearman associations between fixed structural predictors and perturbation-response error. Each amplitude uses 15 conditional rewires, excluding the original. Larger correlations among the controls remain exploratory.

| Predictor | Amplitude 1.5 | Amplitude 3.0 |
|---|---|---|
| Activity rank 5 (primary) | 0.279 | 0.182 |
| Activity rank 19 (secondary) | 0.000 | 0.225 |
| Full training replay | 0.211 | 0.057 |
| Directed three-cycle change | 0.503 | 0.735 |
| Singular-value distance | 0.539 | 0.575 |
| Old stimulus state error | -0.093 | 0.150 |

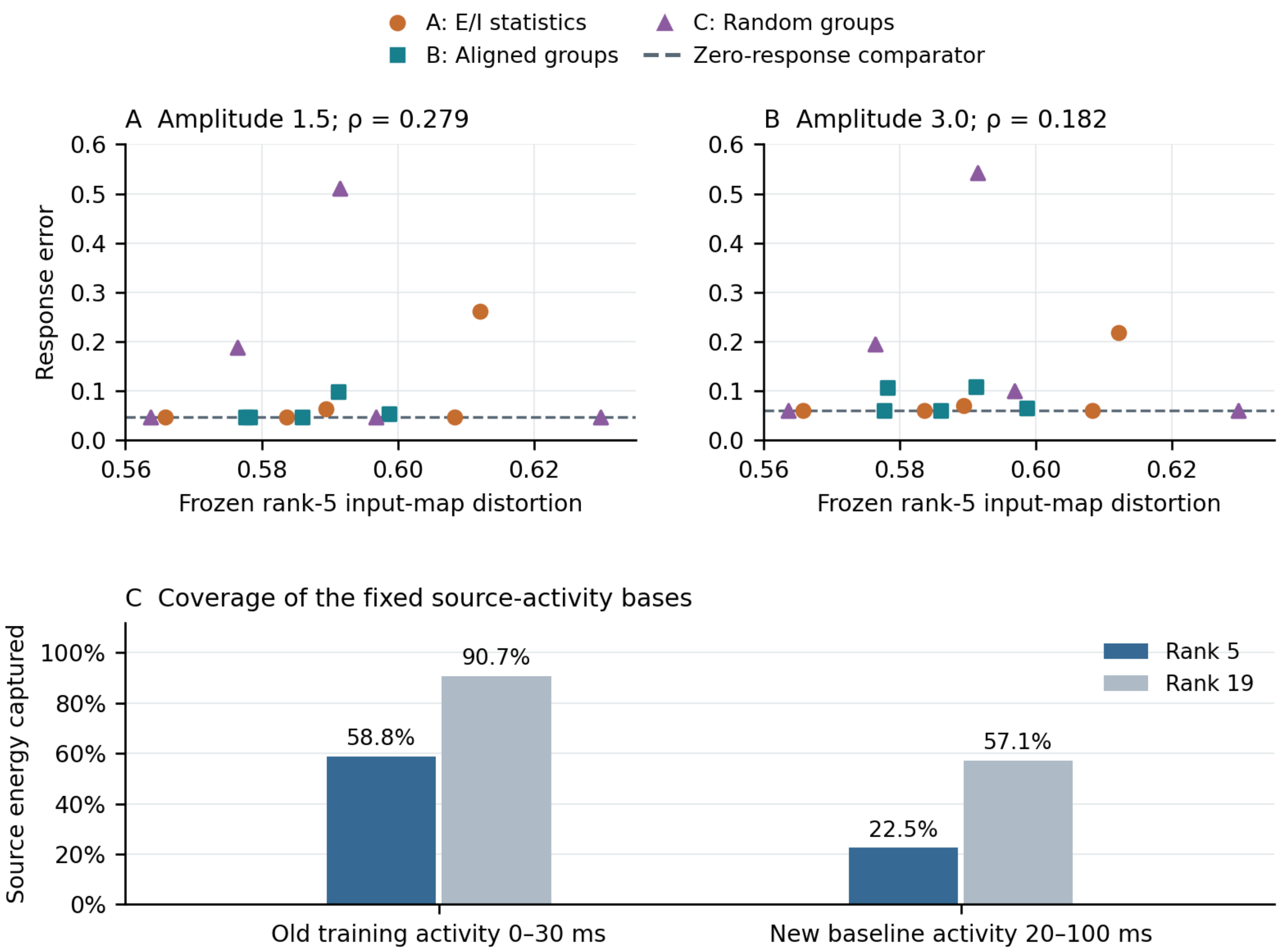

Figure S13. Fixed source-activity prediction at a new stimulus location. Rank-five structural distortion is fitted only to old original E0–19 source activity. Scatter plots relate this frozen feature to original-response fidelity error for 15 controls at each pulse amplitude, with no test refitting. The coverage comparison reports the squared source-activity energy retained by the training rank-five and 90%-energy bases in training and the new original-baseline response window. The basis is a linear source subspace, not an estimated invariant manifold; the two amplitudes are dependent observations on the same graphs.

# Appendix J Finite horizon certificates and parameter families

## J.1 Sufficient condition and proof

**Proposition 4 (common branch finite horizon enclosure).** Fix a deterministic discrete implementation, initial state, external input and finite window. Let a family of candidate graphs change incoming streams only at targets A. Construct each candidate stream from the reference emissions, including actual products, delivery order, event presence and pending-event effects. Suppose initial candidate states are enclosed, interval updates enclose every allowed parameter realization, and all execution, eligible-threshold, status and next-active guards resolve to the reference decisions at every affected target and slot. Then every family member has the reference spikes, execution decisions, status labels and refractory deadlines throughout the window; its continuous states lie inside the enclosures. Incoming streams and initial states of unaffected targets must be unchanged. Support-independent assertions about parameter families additionally require support and event-presence invariance or validated set-valued bookkeeping.

**Proof.** At initialization, the candidate states belong to their sets and the initial queues and statuses have their specified values. Assume jointly that all earlier emissions and relevant branch decisions agree. Routing these emissions through any fixed candidate gives exactly the conditional event streams used by the checker, in their real order, with the true due and pending flags. The scheduler guard therefore fixes the same execute/skip decision. Sound event and state updates enclose the true candidate pre-reset voltage. The one-sided threshold checks establish the same spike decision, hence the same reset and refractory deadline. Local fields depend only on their earlier values, common spikes and execution history, so their archived reference values remain valid. The residual-dynamics and pending-event guards establish the same status and next-active membership. At unaffected targets the incoming ordered streams and previous states agree, so their updates agree. This completes the induction simultaneously over all targets. A first disagreement cannot occur while the hypotheses hold. For a parameter set the reasoning applies to every member because every update and guard is enclosed uniformly. Failure of a sufficient guard leaves the property undecided unless a concrete counterexample is supplied.

The argument does not certify arbitrary candidate trajectories after the first branch mismatch. It permits nonzero input and state differences but no unaccounted execution differences. A common spike raster alone is insufficient to justify reading reference local fields in this scheduler. Positive retained weights in the tested families make due and pending flags parameter-independent under common emissions. The proof is an application of enclosure and branch preservation; it does not claim a new general stability theorem.

J.2 Arithmetic and affine representation

The frozen execution uses nearest-even binary32 operations, finite values, gradual underflow and unreassociated source order. Let $u = 2^{-24}$ and $\eta = 2^{-149}$. Each basic operation with exact result z is enclosed using error $u|z| + \eta$, assuming no overflow [27]. For m sequential additions with $m\,u < 1$, the checker uses $\gamma = m\,u/(1 - m\,u)$, bounding accumulation error by γ times the sum of operand magnitudes plus $m\eta/(1 - m\,u)$. The singleton implementation validates outward binary32 projection with exact rational comparisons; the optimized path encloses computations in binary64 and falls back to that oracle when necessary. The decay factor is the actual rounded binary32(1 − h), h = binary32(0.1).

For voltage, the exact affine expression $(1 - h)V + h(E - I)$ retains contraction, but its error budget follows the original operation order: E − I, addition of −V, multiplication by h, then addition to V. If the successive rounding bounds are $r_1$ through $r_4$, the added voltage error is $h(r_1 + r_2) + r_3 + r_4$. Refractory steps still accumulate inputs but hold voltage and conductance decay; skipped states freeze. A common spike resets voltage and its uncertainty to −65 while retaining conductance uncertainty. The deposited THEORY.txt specifies the interval formulas and every scheduling guard.

For families, each scalar state is represented as $c + \Sigma\ p_j\ \alpha_j + \varepsilon$ with $|\alpha_j| \le 1$ and $|\varepsilon| \le r$. The scalar box enclosure is $c \pm (\Sigma|p_j| + r)$. The rigorous corner alternative enumerates the same affine extrema and adds the same uniform residual radius, including uncertainty in calculated coefficients. It is therefore a same-guarantee baseline; ordinary corner simulation is not. Entry weights are rounded once to binary32 and emitted events are rounded again after multiplication by 100. Both errors, sequential addition and the original voltage/decay rounding are enclosed. The retained coefficients are validated quantities, not exact unvalidated derivatives. Full formulas, assumptions and exact-Fraction oracle code are included in THEORY-AND-METHODS.txt and oracle.py.

J.3 Coverage and cost of the preceding audits

The family prototype centres on the old swapped candidates, unlike the later original-centred matched-direction comparison. With d = 1 or 2, it uses ideal weight $w_0(1 + \rho\ \Sigma\ s_j\ \alpha_j/d)$, fixed source-sorted sign patterns and $\alpha_j$ in [−1,1]. The seven tested radii are 0, $2^{-16}$, $2^{-12}$, $2^{-8}$, $2^{-6}$, $2^{-4}$ and 1/4. The one-dimensional target counts passing at these levels are 16, 16, 16, 12, 11, 8, 4 out of 17; two-dimensional counts are 16, 16, 16, 12, 11, 9, 4. These are nested tests, not independent samples or maximal-radius estimates. All twenty passing centres across the 57 cases obtain nonzero regions in both dimensions. Distinct actual weight, event and state witnesses rule out coverage that is only a single floating-point plateau.

Table S8 reports the matched-work costs from the preceding audits.

Table S8. Matched-work costs from preceding audits, in ms. First two rows concern singletons; remaining rows concern two-parameter families and are a different workload.

| Method | ms | Coverage |
|---|---|---|
| Singleton certificate kernel | 0.13115 | One candidate |
| Singleton local replay kernel | 0.003430 | One candidate |
| Preparation + support certificate | 0.15839 | Whole box |
| Preparation + rigorous corners | 0.20909 | Whole box |
| Four-point replay + generation | 0.01706 | Four points |
| 25-point replay + generation | 0.10372 | 25 points |

The cost table reports first-process medians over sixteen full-horizon passing target cases, with no traced state writes in either timing path. Singleton kernels use eleven rounds of 100 native repetitions; family kernels use eleven rounds of twenty, symmetric warmup and alternating method order. Model preparation is shared within each comparison. Component sums are formed per case before medians, and exclude separately recorded reference setup and graph I/O. A second process gives the same qualitative ordering. The original Python/Fraction implementation is much slower, but changing language and trace collection is not an algorithmic speedup against replay. C kernels use Apple clang 21.0.0, -O3 -std=c11 -fno-fast-math -ffp-contract=off; the scientific environment records Python 3.13.5 and NumPy 2.5.4. This does not change the Rust 1.98.1 provenance of the original network trajectories.

The family support and rigorous corner methods agree on all 798 tested boxes, including failure target, slot and guard. Every failed box at these sampled levels has a failed corner, but that empirical fact does not make corner sampling a proof. Earlier prototype failure reporting is target-major and need not be globally earliest in time; the paired-direction study separately checks both targets and takes the earliest witness. Neither report is an unrestricted simulation of the divergent closed loop.

J.4 Scope of paired ranking and post hoc checks

The matched-direction radius certifies one joint weight direction at a time, over a symmetric parameter interval. Its sixteen sampled directions do not reconstruct the complete multidimensional safe set. A finite failed point bounds common-branch radius; enclosure failure alone does not. All 98 proof-gap conditions, cap censoring and unresolved comparisons remain in the evidence. Among 912 conditions, 32 focal conditions have silent target pairs and no condition has zero channel-input projection. Removing silent cases does not change the primary macro concordance because they contribute no resolved direction orders.

The original mean/shuffle controls operate on the complete 300-slot source record. The score-only boundary audit, explicitly post hoc, instead preserves every source total over slots 0–297, whose events arrive in the window, and leaves the final two source slots unchanged. Mean and shuffle concordances become 0.673/0.654/0.725 and 0.546/0.467/0.586. Relative to the original tonic temporal choice, corrected-mean win/loss/tie/unresolved counts are 9/3/6/1, 7/6/3/3 and 3/10/4/2. Corrected-shuffle counts are 10/1/6/2, 9/5/4/1 and 8/5/5/1. Changed shift ranges also change the random realization, so this is sensitivity analysis, not a causal isolation of the window effect. Existing references were already available; locking new direction rules is not a public preregistration or independent-data claim.

## Appendix K Controlled inputs and archived analysis details

K.1 Score coverage and factor definitions

F0/F1 denotes unfiltered/filtered absolute net projection; M0/M1 denotes no margin denominator/the original per-slot threshold margin. All four use the same eligibility mask. Legacy raw projection maximized over all slots and is retained separately. Independent-shift and mean controls use the original full source window in the old factorial audit; the locked controlled-input study uses the 298-slot delivered window. These are distinct reported definitions, not silently substituted baselines. The factorial audit additionally repeats all factors for mean and shifted activity, and retains strict-equality tie diagnostics. The high margin-only tie counts are actual equal floating-point scores, not merely the $1e-10$ ranking tolerance.

Table S9 separates the unresolved-order categories in the score audit.

Table S9. Unresolved-order categories in the post hoc score audit. Categories are mutually exclusive; proof-gap overlap takes precedence over other finite overlap. Each input has 2,280 comparisons.

| Input | Resolved | Both cap | Gap overlap | Finite overlap | Other cap |
|---|---|---|---|---|---|
| Tonic | 1353 | 506 | 123 | 298 | 0 |
| Global | 1811 | 82 | 102 | 285 | 0 |
| Focal | 1624 | 335 | 57 | 262 | 2 |

Score-completion ranges assign every unresolved non-tied prediction a possible credit from zero to one, while a predicted tie always receives one half. They ignore transitivity constraints and therefore are loose logical ranges, not attainable confidence bounds. Tonic excludes target pairs (37, 38), (52, 96), (82, 95), all active but cap-censored; focal excludes (37, 38), (37, 69) because of silent caps and (107, 114) because all finite brackets overlap. Global includes all nineteen. The controlled F_H/F_HT exclusions are pair IDs 3, 4, 16; G_T excludes ID 3; G_B has no excluded target pair. F_B, F_H, F_HT each have 32 silent conditions but no null channel-input projection.

K.2 Paired dose details

The common zero-edit reference has X = Y = 0 and undefined R, not ten independent zero-dose samples. The three nonzero budgets have replaced-adjacency fractions 0.9842%, 1.9684%, 9.9121%; the earlier nineteen single swaps are different chains and cannot be joined as their first dose point. All thirty trajectories have nonzero slot and cumulative-count discrepancy. From the lowest to middle budget the mean Y change is 1.01 percentage points, with −0.09 from magnitude and 1.10 from retained fraction; from middle to highest the change is 14.96, split 2.31 and 12.65. From lowest to highest, Y increases in nine chains and ties in one, while R increases in nine and decreases in one. Across the five widths 0.1, 0.2, 1, 5, 30 ms and every integer-slot offset, 10,890 dependent readouts are retained. The highest-budget mean exceeds the other two at each width and offset, but chain ranges overlap. Only offset zero at 30 ms equals complete-window count discrepancy; other offsets split the observation. No breakpoint or saturation model is fitted.

K.3 Controlled direction search and transfer

Table S10 reports the controlled-input direction-search outcomes.

Table S10. All controlled-input direction search outcomes, 304 per input. Refinement limit means the existing eight-step budget was exhausted.

| Input | Tolerance | Proof gap | Certified cap | Step limit |
|---|---|---|---|---|
| G_B | 238 | 41 | 25 | 0 |
| F_B | 218 | 32 | 54 | 0 |
| F_H | 194 | 33 | 76 | 1 |
| G_T | 198 | 83 | 22 | 1 |
| F_HT | 201 | 41 | 62 | 0 |

The two refinement-limit conditions retain their last valid bounds. Enclosure failure with passing endpoints is a proof gap; this record does not distinguish conservative wrapping from possible interior branch changes. Cap certification is a lower bound through 25%, with no inferred finite upper bound. All 1,520 positive lower intervals and 3,040 endpoints are checked. Nine exact-rational cases cover 5,400 state rows, or 32,400 coordinate endpoints; the historical field named exact_fraction_values counts rows. The kernel, compiler and operation order are unchanged. Source/event invariants include positive weights, group sign balance, equal direction budgets and matching group weight multisets at common parameter values.

Table S11 reports transfer of the frozen G_B selections to all inputs.

Table S11. Frozen G_B selections transferred to all inputs. Joint-score choice versus frozen controls; ties combine same direction and joint cap, with unresolved cases separate.

| Input | Control | Win | Loss | Tie | Unresolved |
|---|---|---|---|---|---|
| G_B | Random | 14 | 1 | 1 | 3 |
| G_B | Mean | 8 | 5 | 4 | 2 |
| G_B | Shuffle | 14 | 1 | 2 | 2 |
| F_B | Random | 8 | 6 | 2 | 3 |
| F_B | Mean | 6 | 6 | 5 | 2 |
| F_B | Shuffle | 8 | 6 | 3 | 2 |
| F_H | Random | 9 | 6 | 2 | 2 |
| F_H | Mean | 7 | 4 | 4 | 4 |
| F_H | Shuffle | 5 | 8 | 3 | 3 |
| G_T | Random | 10 | 6 | 1 | 2 |
| G_T | Mean | 8 | 3 | 4 | 4 |
| G_T | Shuffle | 12 | 4 | 1 | 2 |
| F_HT | Random | 7 | 5 | 2 | 5 |
| F_HT | Mean | 7 | 4 | 4 | 4 |
| F_HT | Shuffle | 5 | 8 | 3 | 3 |

Selections and control IDs are fixed on G_B before radius outcomes, then retained under every other input. Same-direction ties and joint-cap outcomes are combined only for this compact table; the underlying CSV distinguishes them. Unresolved intervals never count as wins. These outcomes do not claim significant superiority, and reference-conditioned ranking is not an unseen-input forecast. All controls, including the low-drive F_B bridge, remain in the archive.

## Appendix L Original graph validation design

L.1 Generator and sampling specifications

The independent-edge follow-up fixes seeds100–102 as the next three integers after discovery seed99 and uses the unchanged E/I generator. Module seeds201–203 and ring seeds301–303 sample exactly 1467/610/648/120 EE/EI/IE/II edges. For every legal directed pair, a uniform U in (0,1] gives key −log(U)/affinity; the required number of smallest keys is selected with source/target tie breaking. Self and parallel edges are excluded. PCG64(seed) supplies edge keys, with source-major, target-minor ordering within each arm.

Module labels are five groups of twenty E and five I nodes, assigned by separate E/I permutations with PCG64(seed+1000000). Affinity is four within a group and one otherwise. Ring positions are a permutation of the 125 node identities onto j/125, using the same seed offset; affinity is exp(−d/0.15) for circular distance d. No connected backbone, dynamical acceptance filter or output-based resampling is imposed. Equal arm totals across originals do not imply matched per-node degrees or strengths.

Swap i uses random.Random(100000×graph_seed+93001+i), accepting the first legal equal-weight/delay same-arm exchange within 200000 proposals. Allocation is 5/5/5/4, with source rows, weight bits and delays retained. Shared sources or targets, self-loops and parallel edges are rejected. Swaps may violate the generator's module or distance preference; preservation of geometric affinity is not an additional constraint. The sixteen direction pool signs use seed20261108+sorted_pair_index and at most10000 draws. Unavailable directions trigger a technical stop, not a replacement target.

L.2 Decision rules and sample accounting

For each input contrast, replication within a new three-graph family requires at least one switch in every graph; comparison across the two affinity rules requires both families to pass. The focal-time contrast fails this criterion because ring seed303 has 0/19. Score comparisons use per-target macro agreement, then equal input averaging per graph; consistent superiority requires a positive joint-minus-control difference above $10^{-12}$ on every graph in the declared family, and an undefined input prevents that claim. These are descriptive replication rules, not hypothesis tests. Cost timing is diagnostic and does not supply a speed claim.

The nine new legacy originals contribute 900 scientific network conditions and 13360 direction/input conditions, in addition to seed 99's controlled-input 100 networks and 1520 directions. Target-pair deduplication yields 4400 directions for the three independent-edge graphs and 8960 for the six affinity graphs. Positive lower bounds are 4384/4400 and 8944/8960; the missing 32 conditions are zero-radius enclosure failures on two groups of 16, not evidence of actual baseline failure. Seed 99 has 1520/1520 positive bounds. Combining these gives 14848/14880 covered legacy conditions, but does not make them independent samples. The archived multigraph endpoint count 8800 means point replays; valid enclosure inclusion endpoints are 8768, a distinction retained in this release.

Each positive lower interval is rechecked, together with −L, 0 and +L point realizations and state containment. Uncertified zero intervals receive baseline replay but are not credited with enclosure containment. Fixed first/middle/last pair indices, direction0 and all five inputs undergo exact-rational affine checks; final failure witnesses are independently replayed with exact parameter rounding. Scores and G_B selections are frozen across every new graph before any new radius result. The archived original, bridge and repetition counts are distinct from the scientific network conditions.

## Appendix M LIF execution and enclosure proof

M.1 Execution contract and parameter mapping

The published distribution is brian2-atlas 0.2.0 in an isolated environment; its frontend imports as brian2 and exports the model to its bundled native b2-runner reference-f64 backend. No upstream Brian2 distribution or fallback is used. Every block edge expands to sixteen ordered neuron pairs. The positive unit weight is float64(float32(w_block))×0.001/4 volts. This new unit normalization is part of LIF, not a reinterpretation of the original U=100 dynamics. Per-chunk τ values are 18, 19.333333333333332, 20.666666666666668 and 22 ms; thresholds are −51, −50.333333333333336, −49.666666666666664 and −49 mV. Independent within-chunk permutations use PCG64(20261109+graph_seed). Means match the homogeneous control and are fixed across all interventions.

The current-like states are updated by their ordered Euler expressions even during refractory voltage clamping. Thresholds use ≥, reset voltage is −0.060 V and resting voltage is −0.065 V. A reference spike determines the subsequent 2 ms eligibility mask. Affected-target replay includes all four neurons in each of the two target chunks. If all affected spikes agree, every unaffected neuron receives the reference event stream; until the first affected discrepancy, this local replay is therefore the exact full-network comparison. Local replay after its first discrepancy is not used to claim the later full-network trajectory. The same induction underlies the family certificate, but its arithmetic bounds must hold for every covered parameter.

M.2 Outward affine enclosure derivation

This appendix supplies the model-specific sufficiency premises used by Proposition 4 for LIF. Arithmetic is binary64 round-to-nearest ties-to-even, with fixed operation order, no fused multiply-add and no fast-math. Constants are the stored binary64 values. Finite, nonoverflowing arithmetic is required; nonfinite voltage bounds reject a certificate. Let $u=2^{-53}$, Γ=32u, and η=64 times the minimum positive subnormal. All radius operations use upward nextafter on nonnegative sums and products. The executable voltage expression has at most eight primitive operations; Γ exceeds γ8=8u/(1−8u). Absolute expression-condition sums bound rounding under cancellation; η covers gradual underflow, enlarged by 1+1/τ for division in voltage updates [27].

Represent a state as x=m+aα+δ with |δ|≤e for |α|≤ρ. Define M(x)=|m|+|a|ρ+e with upward arithmetic. Centers follow the actual rounded nominal expression. If an ideal affine expression has coefficients c, computed coefficients and slopes have certified defects dc and ds, and each actual or center evaluation has error at most R, then the updated residual is bounded by the coefficient-weighted old radii, ρds and 2R. The factor 2 accounts for subtracting a rounded center from a rounded actual update. The following recurrences instantiate this argument; all displayed nonnegative bounds are rounded outward in code.

For synaptic decay the executable expression is RN(RN(cx)+x), c=RN(−dt/τsyn). Let b=RN(1+c), db bound |b−(1+c)|, and set a′=RN(ba). The slope defect is bounded by db|a|+Γ|ba|. Use R=Γ(1+|c|)M(x)+η and assign the radius by Eq. (25). The implementation takes db=up(eps_machine(1+|c|)), a conservative coefficient bound for the fixed normal constants.

Here e always denotes the stored nonnegative enclosure radius, whereas δ denotes the actual residual. In the following equations, Eval↑ denotes outward evaluation of the bound through its nonnegative intermediate operations in the specified implementation order, not a single nextafter applied to an ordinarily rounded whole expression. Each update assigns e′ to that conservative computed value and establishes |δ′|≤e′; the stored radius is not required to be below the exact-real right-hand side.

$$e' := Eval\uparrow[(|b| + d_b)e + \rho\ d_s + 2R] \quad (25)$$

For ordered weight addition, the ideal event is w+σwα with σ∈{−1, 0, 1}. Its nested rounded construction has an error bounded by ew, obtained by outward evaluation of Γ|w|(1+ρ)+η. Set the new center to RN(m+w), slope to RN(a+σw), and a slope-defect bound ds obtained by outward evaluation of Γ(|a|+|σw|)+η. Choose the addition roundoff bound R as the upward evaluation of Γ(M(x)+|w|(1+ρ))+η. The new stored radius is the outward evaluation of e+ew+ρds+2R, using validated upper bounds ew and ds. The same bound is conservatively used for unchanged weights. This keeps one common α and its signed cancellation; it does not replace the direction by an independent-parameter box.

For eligible voltage updates let B=RN(−0.065+0.018) V. The actual operation order is RN(RN(RN(dt×RN(RN(B+RN(drive+$g_E$))−RN($g_I$+v)))/τ)+v). Write h=dt/τ in exact real arithmetic on the stored inputs, $h_0$=RN(dt/τ), $b_0$=RN(1−$h_0$). Let dh bound |$h_0$−h| and db bound |$b_0$−(1−$h_0$)|. The slope is

RN(b0av+$h_0$(aE−aI)) with the actual multiplication/addition ordering. Its defect includes (db+dh)|av|+dh(|aE|+|aI|), plus Γ times |b0av|+|$h_0$|(|aE|+|aI|). In particular dh affects the voltage coefficient as well as the current coefficients.

$$e' := Eval\uparrow[(|b_0| + d_b + d_h)e_v + h_u(e_E + e_I) + \rho\ d_s + 2R_v] \quad (26)$$

In Eq. (26), hu=|$h_0$|+dh and Rv is chosen by outward evaluation of Γ times M(v)+hu(|B|+|drive|+M($g_E$)+M($g_I$)+M(v)), plus η(1+1/τ). The fixed constants use dh=up(eps_machine|$h_0$|), db=up(eps_machine(1+|$h_0$|)). Remaining subnormal slope errors are covered by the conservative additive underflow allowance. These bounds and the triangle inequality propagate |δ|≤e through each expression. Every threshold uses the outward interval [m−(|a|ρ+e), m+(|a|ρ+e)]. Silent reference updates require the upper endpoint strictly below θ; firing reference updates require the lower endpoint at least θ. Resets assign (m, a, e)=(−0.060, 0, 0). Refractory voltage is unchanged while current enclosures continue updating. This proves the enclosure induction used in Proposition 4 for the fixed LIF arithmetic.

The executable source includes all rounding and rejection operations, and the package preserves its pre-query algebra audit. The proof concerns representable α under the stated parameter construction. It does not certify unrounded real-valued weights, other compiler contraction rules, a different delay implementation or arbitrary combinations of separately tested directions. The archived legacy binary32/affine implementation has a different state and guard system and remains a separate instance, with its own full source and derivation.

M.3 Sufficient branch proof and failure semantics

**Proof.** The initial states satisfy the enclosure premise. Assume enclosure containment and equal discrete histories up to an update. Sound propagation contains every candidate next state; guard inclusion forces the reference decision for all of them. The same reset and refractory transition then preserves the induction premise. Equal emitted spikes preserve subsequent source event streams, while fixed support and ordering preserve their routing structure. Unaffected neurons receive unchanged events; all affected neurons are enclosed. Induction over every state update and event delivery proves the finite-prefix claim. In the legacy scheduler, wake-up, freezing and field guards must also be enclosed. In the clock-driven LIF kernel, the same spike history determines refractory eligibility, and all four units in each affected chunk are checked. The guarantee is relative to this discrete execution, not to exact continuous-time integration.

This is an implementation-specific instance of Proposition4. All bounds in M.2 must be propagated before checking the reference threshold decision. A reset sets voltage slope and remainder to zero, while synaptic enclosures persist. Equal spike history fixes the refractory mask; voltage cannot be treated as freely evolving during the clamp, and currents cannot be frozen with it. An unresolved guard rejects the sufficient certificate. Only an actual replay discrepancy is an unsafe-point witness. Thus interval conservatism, iteration limits, zero radius resolution and genuine failure remain distinct outcomes.

## Appendix N LIF complete controls and verification

N.1 Complete visibility and score controls

Table S12 reports the complete per-unit visibility switch counts.

Table S12. Complete per-unit visibility switch counts, out of nineteen swaps in every cell. Each horizon is a prefix of the same trajectory. At heterogeneous 300 ms all zeros mean different under both inputs.

| Model | Seed | Contrast | 30 ms | 100 ms | 300 ms |
|---|---|---|---|---|---|
| hom | 100 | G_B/F_H | 4 | 7 | 2 |
| hom | 100 | G_T/F_HT | 5 | 8 | 2 |
| hom | 100 | G_B/G_T | 0 | 5 | 0 |

| Model | Seed | Contrast | 30 ms | 100 ms | 300 ms |
|---|---|---|---|---|---|
| hom | 100 | F_H/F_HT | 3 | 0 | 0 |
| hom | 100 | F_B/F_H | 1 | 2 | 1 |
| hom | 201 | G_B/F_H | 5 | 4 | 1 |
| hom | 201 | G_T/F_HT | 6 | 4 | 0 |
| hom | 201 | G_B/G_T | 0 | 3 | 0 |
| hom | 201 | F_H/F_HT | 1 | 1 | 1 |
| hom | 201 | F_B/F_H | 0 | 4 | 1 |
| hom | 301 | G_B/F_H | 6 | 4 | 1 |
| hom | 301 | G_T/F_HT | 6 | 2 | 0 |
| hom | 301 | G_B/G_T | 1 | 4 | 1 |
| hom | 301 | F_H/F_HT | 3 | 0 | 0 |
| hom | 301 | F_B/F_H | 0 | 1 | 0 |
| hetero | 100 | G_B/F_H | 7 | 0 | 0 |
| hetero | 100 | G_T/F_HT | 13 | 0 | 0 |
| hetero | 100 | G_B/G_T | 11 | 0 | 0 |
| hetero | 100 | F_H/F_HT | 5 | 0 | 0 |
| hetero | 100 | F_B/F_H | 2 | 1 | 0 |
| hetero | 201 | G_B/F_H | 8 | 1 | 0 |
| hetero | 201 | G_T/F_HT | 10 | 0 | 0 |
| hetero | 201 | G_B/G_T | 8 | 1 | 0 |
| hetero | 201 | F_H/F_HT | 0 | 0 | 0 |
| hetero | 201 | F_B/F_H | 3 | 0 | 0 |
| hetero | 301 | G_B/F_H | 9 | 0 | 0 |
| hetero | 301 | G_T/F_HT | 11 | 0 | 0 |
| hetero | 301 | G_B/G_T | 7 | 0 | 0 |
| hetero | 301 | F_H/F_HT | 3 | 0 | 0 |
| hetero | 301 | F_B/F_H | 3 | 0 | 0 |

Table S13 reports the F_H score controls and resolved coverage for both LIF variants.

Table S13. F_H score controls and resolved coverage across both LIF variants. Entries are target-level macro agreement; absent scores are undefined. The possible denominator is 360 per row.

| Model | Seed | ms | Own | Frozen | Mean | Shuffle | Cover % | n |
|---|---|---|---|---|---|---|---|---|
| hom | 100 | 30 | 0.975 | 0.415 | 0.771 | 0.576 | 32.8 | 1 |
| hom | 100 | 100 | 0.927 | 0.613 | 0.784 | 0.475 | 64.7 | 2 |
| hom | 100 | 300 | 0.832 | 0.602 | 0.718 | 0.587 | 63.9 | 2 |
| hom | 201 | 30 | 1.000 | 0.500 | 0.925 | 0.602 | 31.4 | 1 |
| hom | 201 | 100 | 0.957 | 0.446 | 0.780 | 0.662 | 93.9 | 3 |
| hom | 201 | 300 | 0.905 | 0.514 | 0.681 | 0.580 | 95.0 | 3 |

| Model | Seed | ms | Own | Frozen | Mean | Shuffle | Cover % | n |
|---|---|---|---|---|---|---|---|---|
| hom | 301 | 30 | 1.000 | 0.652 | 0.938 | 0.756 | 37.2 | 2 |
| hom | 301 | 100 | 0.852 | 0.586 | 0.710 | 0.699 | 91.1 | 3 |
| hom | 301 | 300 | 0.854 | 0.485 | 0.724 | 0.633 | 94.7 | 3 |
| hetero | 100 | 30 | 0.994 | 0.734 | 0.710 | 0.700 | 41.7 | 2 |
| hetero | 100 | 100 | 0.850 | 0.495 | 0.726 | 0.625 | 85.6 | 3 |
| hetero | 100 | 300 | 0.775 | 0.526 | 0.740 | 0.676 | 71.9 | 3 |
| hetero | 201 | 30 | 1.000 | 0.584 | 0.810 | 0.664 | 76.7 | 3 |
| hetero | 201 | 100 | 0.914 | 0.531 | 0.834 | 0.774 | 97.2 | 3 |
| hetero | 201 | 300 | 0.826 | 0.585 | 0.774 | 0.748 | 95.6 | 3 |
| hetero | 301 | 30 | 0.995 | 0.567 | 0.861 | 0.670 | 60.3 | 2 |
| hetero | 301 | 100 | 0.813 | 0.570 | 0.762 | 0.655 | 95.6 | 3 |
| hetero | 301 | 300 | 0.753 | 0.545 | 0.753 | 0.759 | 75.3 | 3 |

The joint score is not uniformly best. At heterogeneous F_H, 300 ms, the three seeds give joint/filter-only/margin-only/mean/shuffle concordances of 0.775/0.740/0.581/0.740/0.676, 0.826/0.768/0.562/0.774/0.748 and 0.753/0.690/0.560/0.753/0.759. Guaranteed-separation bins [1,1.25), [1.25,2), [2,∞) contain 136/321/417 resolved pairs and give micro agreement 0.676/0.729/0.863. These micro summaries do not replace the target-level macro endpoint; small guaranteed ratios only weakly bound the true separation. Frozen G_B direction choices against random at 300 ms have four wins, three losses and two unresolved outcomes across nine graph/target groups; against mean they have four wins, one loss, three same-direction ties and one unresolved outcome. Neither a uniform transfer failure nor a uniform advantage follows.

N.2 Coverage and zero lower bounds

All 4,308 positive intervals contain at least three distinct realized weight/event vectors and three different continuous-state paths at −L,0,+L. The twelve zero lower bounds all occur at seed301, heterogeneous F_H, target pair18, 300 ms. Their bracket is $[0,2^{-20}]$, with an actual upper failure and a passing zero certificate; the search stops at its preassigned absolute tolerance. The minimum reference margin in that group is approximately $2.03\times10^{-10}$ V, below the score floor. These observations do not prove an empty positive safe region. Across all 4,320 conditions, 3,477 stop at width tolerance, 832 at the 25% certified cap and eleven at the refinement limit; none stops at a proof gap or failed zero guard. The twelve zeros belong to the tolerance category.

N.3 Independent implementation checks

LIF direction selection uses pair indices 0, 9, 18 on each reused graph, with 16 directions, five inputs, two parameter variants and three prefixes, totaling 4320 conditions. Thirty original 300 ms Atlas references are replayed independently for all 500 units; 135000000 recorded voltage/current state words and all spikes agree bitwise. Ninety repeat/short-prefix runs agree with the long-run prefixes; one early partial technical repeat is retained separately. At certified lower bounds, −L, 0, +L replay and envelope checks cover 445824000 endpoint state values. Fixed first-pair/direction 0 audits across every graph, variant, input and horizon check 688770 rounded operations using exact rational arithmetic and 261360 exact affine inclusions. Ninety fixed score cases are independently checked by event-stream impulses, closed-form convolution and affine derivatives.

Full Atlas checks include 60 safe endpoints and 29 available failure witnesses on the fixed first-pair/direction 0 cases. They agree with local replay equality and first-discrepancy timing. The C implementation is compiled with clang -O3 -std=c11 -fno-fast-math -ffp-contract=off on Apple arm64

Darwin; Atlas version and runner binary hash are archived. Legacy source reproduction explicitly uses rustc 1.98.1, edition 2024, optimization 3, with earlier 1.86.0 build logs retained as provenance. These checks establish implementation consistency for the specified contracts, not independent physiological validation or hardware acceleration.

The sixty safe endpoints and twenty-nine available failure witnesses selected before inspection of their outcomes agree with complete Atlas-network checks in spike equality and first-discrepancy timing. They are selected implementation checks, not 710 separate full-network tests of the post hoc contraction count. The archived 24,464 bracket-resolved scoring comparisons are recomputed independently. Numerical operation checks use exact rationals for the specified cases and do not constitute machine verification of every compiler or library operation.

N.4 Post hoc witness and order calculations

Matching graph, variant, input, target and direction, we retain every 30-to-100, 30-to-300 and 100-to-300 ms comparison, including non-witnesses. The heterogeneous and homogeneous 30-to-300 counts satisfying $U_{long}<L_{short}$ are 710/720 and 630/720. At heterogeneous 300 ms, seeds100/201/301 contribute 120/244, 154/327 and 120/258 changed/common-resolved G_B/F_H orders: 394/829 of 1,080 candidates across nine target groups. Unresolved pairs are not imputed. The near-half fraction is descriptive, not an equivalence test against random ordering.